\documentclass[12pt,preprint]{aastex}
\usepackage{appendix}
\usepackage{lscape}

\slugcomment{submitted to ApJ on 7 November 2008}

\begin{document}{}

\def\t0{\theta_{\circ}}
\def\muo{\mu_{\circ}}
\def\sd{\partial}
\def\be{\begin{equation}}
\def\en{\end{equation}}
\def\bv{\bf v}
\def\bvo{\bf v_{\circ}}
\def\ro{r_{\circ}}
\def\rhoo{\rho_{\circ}}
\def\etal{et al.\ }
\def\msun{\,M_{\sun}}
\def\rsun{\,R_{\sun}}
\def\lsun{L_{\sun}}
\def\msunyr{M_{\sun} yr^{-1}}
\def\kms{\rm \, km \, s^{-1}}
\def\mdot{\dot{M}}
\def\Md{\dot{M}}
\def\curf{{\cal F}}
\def\ecs{erg cm^{-2} s^{-1}}
\def \haebe{HAeBe}
\def \mum {\,{\rm \mu m}}
\def \simali {{\sim\,}}
\def \K {\,{\rm K}}
\def \Angstrom     {\,{\rm \AA}}
\newcommand \g            {\,{\rm g}}
\newcommand \cm           {\,{\rm cm}}

\title{Dust Processing and Grain Growth in Protoplanetary Disks in the Taurus-Auriga
Star-Forming Region}

\author{B.A. Sargent\altaffilmark{1},
W.J. Forrest\altaffilmark{2},
C. Tayrien\altaffilmark{2},
M.K. McClure\altaffilmark{3},
Dan M. Watson\altaffilmark{2},
G.C. Sloan\altaffilmark{4},
A. Li\altaffilmark{5},
P. Manoj\altaffilmark{2},
C.J. Bohac\altaffilmark{2},
E. Furlan\altaffilmark{6},
K.H. Kim\altaffilmark{2},
J.D. Green\altaffilmark{2}}

\altaffiltext{1}{Space Telescope Science Institute, 3700 San Martin 
                 Drive, Baltimore, MD 21218;
                 {\sf sargent@stsci.edu}}
\altaffiltext{2}{Department of Physics and Astronomy, University of 
                 Rochester, Rochester, NY 14627}
\altaffiltext{3}{Department of Astronomy, The University of Michigan, 
                 830 Dennison Building, 500 Church Street, Ann Arbor, MI, 
                 48109-1042}
\altaffiltext{4}{Center for Radiophysics and Space Research, 
                 Cornell University, 
                 Ithaca, NY 14853}
\altaffiltext{5}{Department of Physics and Astronomy, 
                 University of Missouri, Columbia, MO 65211}
\altaffiltext{6}{Spitzer Fellow; JPL, Caltech, Mail Stop 264-767, 
                 4800 Oak Grove Drive, Pasadena, CA 91109}

\begin{abstract}

Mid-infrared spectra of 65 T Tauri stars (TTS) 
taken with the Infrared Spectrograph (IRS) 
on board the {\it Spitzer Space Telescope} are modeled using 
populations of optically thin dust at two temperatures to probe the 
radial variation in dust composition in the uppermost layers of protoplanetary 
disks.  Most spectra 
with narrow emission features associated with crystalline silicates require 
Mg-rich minerals and silica, but a very small number suggest other components.  
Spectra indicating large amounts 
of enstatite at higher temperatures (400-500\,K) also require crystalline silicates, 
either enstatite or forsterite, at temperatures lower (100-200\,K) 
than those required for spectra showing high abundance of other crystalline 
silicates.  A few 
spectra show 10 $\mum$ complexes of very small equivalent width.  They are fit well 
using 
abundant crystalline silicates but very few large grains, inconsistent with the
expectation 
that low peak-to-continuum ratio of the 10 $\mum$ complex always indicates grain 
growth.  Most spectra 
in our sample are fit well without using the opacities of large crystalline silicate
grains.  If large grains grow by agglomeration of submicron grains of all dust
types, the amorphous silicate components of these aggregates must typically be more 
abundant than the crystalline silicate 
components.  We also find that the more there is of one crystalline dust species,
the more there is of the others.  This suggests that crystalline silicates are 
processed directly from amorphous silicates, whether by evaporation of the amorphous 
grains and condensation in chemical equilibrium or by annealing of the amorphous 
precursor grains.  This also suggests that neither forsterite, enstatite, nor silica 
are intermediate steps along the way to producing either 
of the other two for the majority of the crystalline dust produced.  Crystalline
silicate abundance is 
correlated tightly with disk geometry, in the sense of higher crystallinity 
accompanying more settled disks, which are commonly associated with growth and settling 
of grains.  
Large-grain abundance is also correlated with disks that are more highly
settled, but with a 
wide range of large grain abundance for a given degree of settling.  We interpret
this range to mean that the settling 
of large grains is sensitive to individual disk properties.  We 
also find that lower-mass stars have higher abundances of large grains in their
inner regions.  

\end{abstract}


\section{Introduction}

Comets are 
thought to be largely unaltered reservoirs of leftover material from the primordial 
mixture of dust in the Solar Nebula orbiting the newly-formed Sun 
\citep[e.g.,][]{wooden05}.  Spectral 
observations of Comet Hale-Bopp by \citet{crov97} using the {\it Infrared Space 
Observatory} \citep[{\it ISO};][]{kess96} allowed for detailed spectral models 
over a wide range of wavelengths \citep{lg98,wooden99,har02}.  The amorphous dust 
of pyroxene ([Mg,Fe]SiO$_3$) or olivine ([Mg,Fe]$_2$SiO$_4$) 
stoichiometry (henceforth, ``amorphous pyroxene'' or ``amorphous olivine'',
respectively) was 
found to require a stoichiometric ratio of magnesium to iron of about unity, being 
relatively iron-rich \citep{har02}; the crystalline pyroxene must be iron-poor, 
as it is cooler \citep{wooden99}; and the larger grains must be porous, fluffy 
aggregates of smaller grains \citep{lg98,har02}.  Interplanetary Dust Particles 
\citep[IDPs;][]{brad03}, believed to originate 
from comets, were often found to be highly porous aggregates of Mg-rich 
crystalline silicate dust, having qualities very similar to 
those inferred from comet spectra and 
spectra of Young Stellar Objects (YSOs; systems in the process of forming stars and 
planets).

Using increasingly powerful telescopes and sensitive detectors, 
astrophysicists have obtained increasingly higher-quality spectra for decreasingly 
fainter objects.  \citet{cw85} distinguished 10 $\mum$ silicate complexes 
in absorption versus emission, while \citet{honda03} identified amorphous 
silicates, forsterite, enstatite, and silica using their spectrum of the 10 $\mum$ 
complex of Hen 3-600A.

Approaches toward analysis of the dust composition of protoplanetary disks in Herbig 
Ae/Be and T Tauri (SED Class II YSO) systems have taken two forms: computation of 
indices from flux ratios over 
the 10 $\mum$ (and sometimes longer wavelength) complex, or detailed modeling using 
opacities of minerals and amorphous dust measured on Earth.  Studies taking the former 
approach include \citet{bouw01}, \citet{prz03}, \citet{vb03}, \citet{ks05}, 
\citet{ks06}, and \citet{wat08}.  Studies taking the latter approach include 
\citet{bouw01}, \citet{bowad02}, \citet{mol02}, 
\citet{lilu03a}, \citet{lilu03b}, \citet{li03c}, \citet{bouw03}, \citet{vb04}, 
\citet{uch04}, \citet{vb05}, \citet{schutz05}, \citet{sarg06}, \citet{scheg06}, 
\citet{honda06}, and \citet{sarg08}.

The Infrared Spectrograph \citep[IRS;][]{hou04} on board the {\it Spitzer Space 
Telescope} \citep{wer04} allows good S/N spectra to be obtained at 5-37
$\mum$ 
wavelength for objects with flux densities of a few tens of mJy.  Dust at a
considerable 
range of temperatures in protoplanetary disks gives rise to the emission seen
from such 
systems in IRS spectra.  Our approach has been to model this using two temperatures 
\citep[see][]{chen06,kast06,sarg08}, though others have used more temperatures 
\citep[see][]{bouw08}.  
Here we model the dust emission over the 7.7-37 $\mum$ wavelength range for 65 
Class II YSOs in the 
1-2 Myr old Taurus-Auriga star-forming region \citep{kh95} using dust at two
temperatures, in 
order to analyze the dust composition in the inner and outer disk regions.

``Crystalline indices'' computed from flux ratios that characterize crystalline 
silicate emission have the advantage of being quickly computed and 
measured repeatably for a wide range of objects.  However, they do not make full use
of the 
information on dust emissivity available in a high-quality mid-infrared spectrum;
for instance, 
they measure the combined effect of grain growth and crystallinity, but such are
difficult to 
disentangle \citep{wat08} without detailed modeling using dust opacities, as is 
done here.

\section{Data Reduction}

Much of the process of data reduction of the 65 spectra analyzed in this study has
been described by \citet{sarg08} and \citet{fur06b}.  Here we give a brief overview 
and provide details 
specific to this study.  We note that the spectra of ZZ Tau and ROXs 42C used here 
are exactly the same as used and described by \citet{sarg08}; the 
following discussion of data reduction applies to all spectra except those of ZZ 
Tau and ROXs 42C.

\subsection{Observations}

Sixty-three of the 65 objects in our sample come from the 85 Class II Young Stellar
Objects (YSOs) analyzed by \citet{fur06b}.  The other 22 objects whose spectra were
analyzed by \citet{fur06b} were rejected from our study due to high (here, defined 
as having A$_V$\,$>$\,6) or unknown extinction, the presence of Polycyclic Aromatic 
Hydrocarbon bands, saturation of the detectors, and confusion of Class II YSOs with 
Class I YSOs in the same spectrograph slit.  To the 63 from \citet{fur06b} we add 
GG Tau B and 
Haro 6-28.  For opacity calibration (see Section 3), we reduced data for ROXs 42C 
(see \citet{sarg08}), HBC 656, and TW Cha.  All of these objects were observed using 
the combinations of IRS modules Short-Low and Long-Low or Short-Low, Short-High, 
and Long-High in either Staring Mode or Mapping Mode; for more information, see 
\citet{sarg08}.

\subsection{Extraction and Calibration of Spectra and Rebinning of High Resolution
Spectra}

We obtained Basic Calibrated Data products from the {\it Spitzer} Science Center 
(SSC) for each of the targets in our sample.  BCD data are 
corrected for the dark current, stray light, and flatfield 
variations.  The spectrum of HBC 656 was reduced using data 
from the S16.1.0 data calibration pipeline.  The spectrum of ZZ Tau was reduced 
using data from the S14.0.0 IRS data calibration pipeline, using the methods 
described by \citet{sarg08}; the present discussion of extraction and calibration 
of spectra excludes the spectrum of ZZ Tau.  For the spectra of 
TW Cha and the 64 T Tauri stars in Taurus-Auriga excluding ZZ Tau, the data are from the 
S13.2.0 data calibration pipeline.

Next we identified and fixed by interpolation in the spectral direction bad pixels,
as described by \citet{wat07} and \citet{sarg08}.  We then extracted spectra using
the Spectral Modeling, Analysis, and Reduction Tool \citep[SMART;][]{hig04}.  We
sky-subtracted data from the low spectral resolution modules SL and LL using the
off-order observation.  We did not subtract sky from high spectral resolution (SH and
LH) observations, as explained by \citet{sarg06}.  Multiple Data Collection Events 
(DCEs; detector samplings) at the same telecope position, same order, and same 
modules were averaged, and the uncertainties were calculated from the standard 
deviation of the mean.  Tapered column extraction was used for SL and LL data,
and full-column extraction was used for SH and LH data \citep{sarg06}.  We used
Relative Spectral Response Functions (RSRFs) to calibrate the flux
\citep{sarg06}.  For both orders of SL and for the both orders 
of LL (for all observations except for HBC 656), a spectral 
template of $\alpha$ Lacertae (A1\,V; M. Cohen 2004, private communication) of higher 
spectral resolution than the templates described by \citet{coh03} was used.  For the 
LL flux calibration of HBC 656 over second, bonus, and first 
orders out to 36 $\mu$m wavelength, RSRFs generated from data of $\xi$ Dra and the 
template for $\xi$ Dra from \citet{coh03} were used.  Past 36 $\mu$m, a LL first-order 
RSRF generated from data of Markarian 231 and the template for Markarian 231 \citep[J. 
Marshall, private communication;][]{marsh07,armus07} were used.  For SH and LH, the 
spectral template for $\xi$ Dra (K2\,III) by \citet{coh03} and its IRS data were 
used to generate RSRFs.  Except for GG Tau A, GG Tau B, ZZ Tau, ROXs 42C, HBC 656, 
and TW Cha, all RSRF-calibrated spectra are the same as presented by \citet{wat08}.

We rebin our high-resolution spectra (see Table A1 for whether the spectra 
were obtained with high or low spectral resolution) to low spectral resolution.  The 
width of each bin is determined by requiring spectral resolution 
R=$\lambda$/$\Delta\lambda$ 
$\simali$ 90, and rebinned spectrum is sampled such that each bin is $\Delta\lambda$
from the next bin.

\subsection{Correction for Extinction, Mispointing, Uncertainties, and Stellar
Photosphere}

We corrected our spectra for extinction using the same method described by
\citet{sarg06}, except we assume a ratio of V-band extinction to total optical depth 
at 9.7 $\mum$, A$_V$/$\tau_{9.7}$ of 25.  This value is suggested by data plotted by
\citet{chiar07}.  We use values of A$_V$ given by \citet{fur06b}.  We do not correct 
for extinction if A$_V$ $<$ 1.4, as we want to minimize any artifacts due to 
overcorrection of extinction \citep[see][]{sarg06}.  CoKu Tau/4, GG Tau A, GG Tau B, 
and Haro 6-28 are not corrected, but V410 Anon 13 is, assuming A$_V$ = 5.8 
\citep[see][]{sarg06}.  The extinction correction for ROXs 42C is the same as made by 
\citet{sarg08}.  The average (V-R) colors as measured 
by \citet{batal98} of HBC 656 (also known as AS 216) are $\simali$ 0.64, which is 
slightly bluer than the intrinsic (V-R) color of a K2 star \citep[the spectral type 
assigned to HBC 656 by][]{batal98} of 0.74 given by \citet{kh95}.  This is consistent 
with HBC 656 being of slightly earlier spectral type (e.g., K0) and having A$_V$ 
$\simali$ 0, so we assign A$_V$ = 0 to HBC 656.  Finally, we compute A$_V$ = 1.78 
for TW Cha using I and J magnitudes from DENIS \citep{camb98} and intrinsic colors for a 
K4 main-sequence star from \citet{kh95}, assigning a spectral type of K4 to TW Cha despite 
\citet{luh04} adopting K8 for TW Cha after having found various classifications for this 
star in the literature of K0, K8, and M0.

Next, we corrected for slight mispointing of the telescope from the
standard nod positions by scaling the mispointed nods, as described by
\citet{sarg08}.  The scalars used are given in Table A1 and further described in 
the Appendix.  After applying these scalars, the spectral uncertainties 
were computed.  For all objects except ZZ Tau, the flux density uncertainties were 
half the difference between the spectra obtained at the two nod positions 
\citep[see][]{sarg06}.  The uncertainties were derived from only two independent 
measurements of flux (except for ZZ Tau), so sometimes the relative uncertainty 
was less than 1\%.  Because of this and because of uncontrolled systematic errors, 
when the relative uncertainty was less than 1\%, it was set to 1\% as 
described by \citet{sarg08}.  The spectra plotted by \citet{fur06b} for CoKu Tau/4, 
DM Tau, and GM Aur showed very little excess emission above the stellar photosphere 
at $\lambda$ $<$ 8 $\mum$; for these, we subtracted Planck functions using solid 
angles and effective stellar temperatures given by \citet{sarg06} for the same 
purpose.  Photospheres were subtracted for ZZ Tau and ROXs 42C in exactly the 
manner described by \citet{sarg08}.  For all other spectra, the stellar photosphere 
contribution to the spectra appears minor, and stellar 
photospheres are not subtracted before modeling them.

\section{Analysis}

\subsection{Models}

To characterize the silicate dust composition of the upper levels of the T Tauri disks
giving rise to the emission features in our spectra, we construct detailed models 
\citep{sarg08}.  Briefly summarized, each model is a sum of emission from
blackbodies at temperatures T$_w$ (warm) and T$_c$ (cool) and emission from optically 
thin dust grains proportional to their opacities multiplied by Planck functions at 
T$_w$ and T$_c$:

\begin{eqnarray}
F_{\nu}(\lambda)^{\rm mod} & = & 
       B_{\nu}(\lambda,T_{c}) \left[{\Omega}_{c} + 
       \sum_i a_{c,i}\kappa_{i}({\lambda})\right] + \nonumber \\
& & B_{\nu}(\lambda,T_{w})
\left[{\Omega}_{w}+\sum_j a_{w,j}\kappa_{j}({\lambda})\right] ~~,
\end{eqnarray}

\noindent where B$_{\nu}$($\lambda$,T) is the Planck function evaluated at
wavelength $\lambda$ and
temperature T, $\Omega_c$ ($\Omega_w$) is the solid angle of the blackbody at
temperature T$_c$ (T$_w$), a$_{c,i}$ (a$_{w,i}$) is the mass weight of component j at
temperature T$_c$ (T$_w$), and $\kappa_j$($\lambda$) is the opacity (effective
cross-section per unit mass) at wavelength $\lambda$ of dust
species j.  The mass weights, a$_{c,i}$ (a$_{w,i}$), are equal to $m_{c,i}/d^2$ 
($m_{w,i}/d^2$), where $m_{c,i}$ ($m_{w,j}$) is the mass of dust species $i$ at 
$T_c$ ($T_w$) and $d$ is the distance to the T Tauri star, which is assumed to be 
140 parsecs for T Tauri stars in the Taurus-Auriga star-forming region \citep{ken94}.  
The blackbody components of each model, $\Omega_c$B$_{\nu}$($\lambda$,T$_c$)
and $\Omega_w$B$_{\nu}$($\lambda$,T$_w$), represent optically thick continuum emission,
unsubtracted stellar photosphere emission (F$_{\nu}$ proportional to 1/$\lambda^{2}$,
and it should be small for spectra for which we have not already subtracted a Planck
function at the stellar effective temperature to represent it), and emission from
grains in the optically thin disk regions whose opacities are featureless continua,
like amorphous carbon and very large silicate grains.  The model components 
proportional to opacity multiplied by 
a Planck function represent emission from submicron- to few-micron-sized dust
grains, located in the optically thin disk regions, with strong resonances giving
rise to detectable features in their opacities.

We will assume in the following that all grain species in the same location in the 
disk have the same temperature, the thermal coupling provided by gas.  
\citet{kadull04} find that the layer with A$_V$ $\leq$ 0.1 (where A$_V$ is the 
visual extinction of star into the disk), 
the layer above the layer where dust emission features arise (what they call the
``superheated surface layer''), has gas and dust temperatures within 10\% of each
other, suggesting the gas would bring dust grains of differing optical properties to
roughly the same temperature.  In general, settling of dust towards the disk midplane 
\citep{daless06} should enhance the gas particle density for the optically thin 
regions in the disk atmosphere that gives rise to the 10 $\mum$ silicate 
feature because the gas density is higher towards the disk midplane \citep[see 
Figure 1 of][]{gni04}, so dust and gas temperatures should be better coupled in 
more settled disks.  However, \citet{cg97} found that, for a fiducial disk at 1 AU, 
the dust grains should remain at their radiative equilibrium temperatures,
undisturbed by gas-grain collisions, for the vast majority of plausible gas
particle densities, 10$^{8}$ to 3$\times\,10^{14}$ cm$^{-3}$, at this distance from
the star.  They found that only between about 10$^{14}$ and 3$\times\,10^{14}$
cm$^{-3}$ would the dust grain temperatures be affected by collisions with gas
molecules.  At 8 AU, \citet{gh08} find that gas and dust temperatures are only
equal for A$_V$ slightly greater than 1, suggesting greater difference between dust
and gas temperature in the ``superheated surface layer'' than \citet{kadull04} 
predict.  For the purposes of dust modeling of T Tauri stars, Planck functions and 
opacity-weighted Planck functions at two temperatures (at most) will be used, 
while keeping in mind the results of these sophisticated studies of dust and gas 
coupling.  We allow for the two temperatures in order to study compositional 
differences of disks between inner disk regions and outer disk regions.  If, 
however, there is not sufficient coupling of dust and gas heating in the optically 
thin surface layers giving rise to emission features in our TTS spectra, this 
means dust that absorbs starlight poorly at optical and near-infrared 
wavelengths with respect to mid-infrared wavelengths, like enstatite and forsterite, 
will be located closer to the star than dust like iron-rich amorphous pyroxene and 
olivine at the same temperature.  Such iron-rich 
amorphous silicates absorb starlight more efficiently at visible and near-infrared 
wavelengths with respect to mid-infrared wavelengths \citep{har02}.

To determine the best fit of a model to a spectrum, we minimize $\chi^{2}$ between the
model and the 7.7-37 $\mum$ region of the spectrum, with respect to the two
temperatures, the solid angles, and the mass weights.  We choose ranges of T$_c$ and
T$_w$ to explore, and, picking a single pair of [T$_c$,T$_w$], we minimize $\chi^{2}$,

\begin{eqnarray}
\chi^{2} & = & \sum_k 
\left[\frac{F_{\nu}(\lambda_k)^{\rm irs}
 - F_{\nu}(\lambda_k)^{\rm mod}}{\Delta F_{\nu}(\lambda_k)^{\rm irs}}\right]^{2} ~~,
\end{eqnarray}

\noindent with respect to all solid angles and all mass weights.  When this results 
in mass 
weights or solid angles that are negative, we eliminate from the model the component
whose flux integrated from 7.7-37 $\mum$ is most negative, and minimize $\chi^{2}$ of
the model without this component.  This is repeated at a given temperature pair
[T$_c$,T$_w$] until there are no negative mass weights\footnote{The only solid angle 
that was found to be negative and in the subsequent iteration zeroed out was that 
of the cool blackbody component of CoKu Tau/3}.  This entire 
process is carried out for each temperature pair.  The temperature pair for
which $\chi^{2}$ per degree of freedom (where the number of degrees of freedom is the
number of data points ($\simali$200) minus the number of temperatures used, 2, minus 
the number of nonzero solid angles minus the number of nonzero mass weights) is 
lowest is the best-fit model for a spectrum.

We also determine uncertainties for our model parameters.  We define the 
uncertainty in a parameter to be the increment of the parameter from its best-fit
model value that changes by 1.0 the value of $\chi^{2}$ per degree of freedom, or 
$\chi_{\nu}^{2}$, of the model with the incremented value of the parameter.  We use 
a Taylor series expansion to approximate this.

We note that our definition of uncertainty in a parameter is similar to
what \citet{wooden99} call ``independent point uncertainty'' in that all model
parameters are kept at their best-fit values except for the one whose uncertainty is
being determined.  However, our uncertainties are determined differently from how
others have been determined in the literature.  \citet{press92} suggest taking the
uncertainty to be that increment of the parameter that increases $\chi^{2}$ by a 
number computed from the incomplete gamma function for a given level of significance.  
\citet{avni76} found that uncertainties computed using the increment in $\chi^2$ 
determined in this manner are quite consistent with uncertainties computed from 
Monte Carlo analysis.  In such an analysis, the standard deviation in the set of 
parameter values obtained by fitting multiple versions of the same spectrum, each 
manipulated by adding noise according to Poisson statistics, is taken to be the 
uncertainty in that parameter.

For example, the model for DG Tau has only eight constraints.  The incomplete 
gamma function implies that an increment of $\chi^2$ of 9.31 would determine the 
68.3\% (one sigma) uncertainty level on parameters.  This would result in a one 
sigma uncertainty on the mass weight of cool small amorphous pyroxene of 
$1.68\times10^{-18}$ $\frac{gm}{cm^2}$.  This is nearly identical to the 
uncertainty on this component obtained {\it via} the Monte Carlo analysis 
described previously of $1.62\times10^{-18}$ $\frac{gm}{cm^2}$.  Using the method 
adopted in this study, an uncertainty of $8.21\times10^{-18}$ $\frac{gm}{cm^2}$ 
is obtained, which is almost 5 times higher than the one sigma uncertainties 
obtained using the incomplete gamma function to determine the increment of 
$\chi^2$ or the Monte Carlo approach.  We note that, since the number of data 
points (about 200) in one of our spectra is much greater than the number of model 
constraints (by a factor of $>$20), our uncertainty level involves increasing 
$\chi^{2}$ by much more than suggested by \citet{press92}.  By so doing, we 
conservatively bias the uncertainties in our parameter values to many times one 
sigma.

Our models include the opacity of annealed silica from \citet{fab00}.  This was
shown to provide the best fit to emission features at 9, 12, 16, and 20 $\mum$ in
astronomical spectra of TTS by \citet{sarg08}.  We now justify our choice of
opacities for submicron amorphous and crystalline silicates and large
(few-micron-sized) silicate dust grains.  In order to choose the best opacity curves
to use to model emission from TTS, we assume that all disks in our sample have the
same major kinds of dust, but in different relative abundances.  Assuming all disks
have basically the same major kinds of dust - amorphous silicates of olivine and
pyroxene composition, crystalline olivines, crystalline pyroxenes, and crystalline
silica - our goal is to find the best realistic opacities to represent these major
kinds of
dust, not necessarily to constrain the specific properties of the grains themselves
(e.g., sizes, porosities, composition).

\subsection{Forsterite Opacity}

The first opacity for our models with which we concern ourselves is that of 
forsterite. Olivines have been used before in modeling dust emission to fit 
emission (and sometimes absorption) complexes at 10.0, 11.2, 16, 19, 23, 28, 
and 33 $\mum$ \citep[e.g.,][]{mol02}.  One
spectrum for which these features appear at very high signal-to-noise ratio (S/N) is
that of HBC 656, which we include as our first and most important forsterite exemplar. 
Others with these features in their spectra at fairly high S/N ratios that we
include as forsterite exemplars are DK Tau, GN Tau, IS Tau, ROXs 42C \citep[which has
already been identified as a silica exemplar; see][]{sarg08}, V836 Tau, and V955 Tau.

To fit the forsterite features in these exemplars' spectra, we have compiled opacities
of crystalline olivine from a wide range of sources.  From \citet{jag98}, we include
opacities
from transmission measurements of the following olivine powders pressed into
potassium bromide (KBr) pellets: forsterite (Mg$_2$SiO$_4$), magnesium-rich olivine
(Mg$_{1.8}$Fe$_{0.2}$SiO$_4$), hortonolite (MgFeSiO$_4$), and fayalite
(Fe$_2$SiO$_4$). 
To explore further the effects on opacity of the full range of iron content in the
olivine solid series (olivines of all Mg/Fe ratios), we include
8 opacities measured by \citet{koi03} for the range of 0.77$<$x$<$1.0 for olivines
of composition Mg$_{2x}$Fe$_{2-2x}$SiO$_4$, and also x=0.218.  In addition, we
created a
hybrid forsterite opacity using the x=1 (``Fo100'') opacity from \citet{koi03} for
$\lambda$ $<$ 15 $\mum$ and room-temperature forsterite opacity measured by
\citet{chi01} for $\lambda$ $>$ 15 $\mum$.  We also include the opacities of
forsterite powder measured by \citet{fab01} and annealed forsterite by
\citet{fab00}.  Finally,
we compute opacities for the Continuous Distribution of Ellipsoids 
\citep[see][]{bh83} and CDE2 \citep[see][]{fab01}.  CDE2 and CDE are distributions 
of ellipsoids whose size can be described as being in the ``Rayleigh-limit'', or 
characteristic size (of the semimajor axes of the ellipsoid) 
much less than $\frac{\lambda}{2\pi}$, with ellipsoids described by every possible
set of axial ratios weighted more toward near spherical shapes (CDE2) and equally
(CDE), respectively.  For both shape distributions of forsterite, we compute
opacities using optical constants for the three crystallographic axes of forsterite
provided by \citet{sog06}.  To account for this anisotropy of forsterite, we
compute opacities for forsterite grains in a given shape distribution (CDE2 or CDE)
three times, once for each of the 3 crystallographic axes' optical constants; we
average these three resulting opacities to account for forsterite grains oriented
randomly \citep[see][]{bh83}.

As a starting mixture (not the same as the standard mixture determined and used 
later) of opacities for other components to use in modeling our
forsterite exemplars, we use the non-forsterite opacities from \citet{sarg08}; 
that is, CDE2 opacities for amorphous olivine and amorphous pyroxene using 
optical constants from \citet{dor95} for MgFeSiO$_4$ and 
Mg$_{0.7}$Fe$_{0.3}$SiO$_3$, respectively, ``En90'' enstatite opacity 
(Mg$_{0.9}$Fe$_{0.1}$SiO$_3$) from \citet{chi02}, and annealed 
silica opacity from \citet{fab00}.  For large grains, we use the aforementioned
amorphous silicate optical properties and Mie Theory \citep{bh83} to compute the
opacity of 5 $\mum$ radius spheres that are 60\% vacuum by volume, computing
effective complex dielectric functions for this porous amorphous silicate material
using Bruggeman Effective Medium Theory \citep[EMT;][]{bh83}.

The scattering 
efficiencies of small amorphous and crystalline silicate grains are negligible 
because they are in the Rayleigh limit for {\it Spitzer} IRS wavelengths and are 
not included in their opacities.  The scattering component of opacity is not 
negligible for 5 $\mum$ radius grains.  The scattering cross section is 
comparable to the absorption cross section in the 10 $\mum$ region for both 
amorphous pyroxene and olivine dust grains.  Since we model the emission from 
an optically thin region (the upper disk layers), only the absorption component 
contributes to our model.  Full radiative transfer treatments of the emergent 
intensity from these optically thick disks must include the effects of 
scattering on the 5-37 $\mum$ IRS spectra.  We note the detailed radiative 
transfer models of \citet{daless01}, who cite earlier work by \citet{mn93}, show 
that scattering by large grains does not much change the flux from optically thin 
regions and makes the flux from optically thick regions slightly smaller.  For 
this reason, we deem the lack of inclusion of scattering in opacities not of 
great concern.

We give in Table 1 
the mass weights in units of our conservative uncertainties, {\~a}$_k$ (a measure 
of significance), of the olivine components for each of the seven forsterite 
exemplars.  Those olivine opacities whose {\~a}$_k$ are highest contribute
most significantly to the fit of a given forsterite exemplar.  A given entry in 
a column in Table 1 is the mass weight of the component named in the first entry 
on the same row divided by its uncertainty, giving the significance of a 
particular opacity used in a given model.  A prefix of 
``C'' (or ``cool'') means the opacity was used at the lower temperature, 
and a prefix of ``W'' (or ``warm'') means the opacity was used at the 
higher temperature.  Of the entries whose names include ``J98'' (meaning 
those opacities were presented by \citet{jag98}), ``fo'' denotes forsterite, 
``ol'' denotes Mg-rich olivine, ``ho'' denotes hortonolite, and ``fa'' denotes 
fayalite.  The subsequent 9 opacities come from \citet{koi03}, with the number 
after ``Fo'' indicating the stoichiometric abundance, expressed as a percentage, 
of Mg with respect to Fe in the ratio Mg/(Fe+Mg) for olivine [Mg,Fe]$_2$SiO$_4$.  
Entries for Fo100, Fo90.7, and Fo77, and the opacity of forsterite powder 
presented by \citet{fab01} are not included at either temperature because 
no mass of any of them was ever used in any of the models of the forsterite 
exemplars.  ``annfor'' denotes the opacity of forsterite presented by \citet{fab00} 
annealed for the longest duration.  ``fohyb'' is the hybrid opacity (explained in 
the text).  S6 means the optical constants come from \citet{sog06}, and CDE and CDE2 
are the shape distributions for which opacities are computed using these optical 
constants.  We find that a few opacities recur from model to model, and some of 
those at both model dust temperatures.  Arrows in the final column indicate recurring 
opacities.  Forsterite powder measured by \citet{jag98} shows up in 4 models as
cool dust, but not as warm dust.  Our forsterite ``hybrid'' opacity shows up in 3
models as cool dust, but only once as warm dust.  Annealed forsterite occurs in 4
models as a warm dust component, but not as a cool component.  Fayalite from
\citet{jag98} occurs as warm dust with frequently high {\~a}$_k$ for the 6
exemplars, but as cool dust only once.  The opacity that most repeatedly occurs as
both warm and cool dust components in the models of the exemplars is the CDE2
opacity using optical constants for forsterite given by \citet{sog06}.  Frequently
associated with this is also the CDE opacity for forsterite from the same optical
constants.

The best fits seem to come from a combination of CDE2 and CDE shape distributions of
forsterite using optical constants from \citet{sog06}, suggesting that a shape
distribution intermediate in weighting between CDE2 and CDE is indicated.  
Further, we note that the CDE shape distribution is somewhat unphysical, requiring 
ellipsoids either infinitely long or infinitesimally thin.  We therefore computed 
an opacity curve for a custom-designed shape 
distribution of ellipsoids intermediate in weighting between that of CDE2 and of
CDE.  We note that a similar attempt was made by \citet{zub96} resulting in what
those authors called ``modified Continuous Distribution of Ellipsoids''.  As the
features for forsterite grains in the CDE shape distribution peak at wavelengths
slightly longward of the peaks in the data and those of the CDE2 shape
distribution of forsterite grains peak at wavelengths shortward of the peaks in
the data, we expect the intermediate shape distribution to give opacity peaks at
more optimal wavelengths
than do the CDE or CDE2 shape distributions.  Not knowing exactly how to weight
one ellipsoidal shape relative to another, like CDE we weight all allowed shapes
equally, but unlike CDE we do not allow ellipsoidal shapes with L$_j$ parameter
\citep[see][]{bh83} equal to zero in the shape distribution\footnote{The 
L$_j$ parameters are inversely proportional to the semimajor axis of one of the 
three ellipsoidal axes, there is one L$_j$ parameter for each of the three axes, 
and the sum of the L$_j$ parameters equals 1.  Therefore, two L$_j$ parameters, 
L$_1$ and L$_2$, suffice to describe completely the shape of the ellipsoid.  This 
can be represented graphically with L$_1$ as the horizontal axis and L$_2$ as the
vertical axis.  Any conceivable ellipsoidal shape is then represented on this
graph as a point in the first quadrant satisfying L$_2$ $<$ 1-L$_1$}.  Our 
restricted shape distribution, which we call the ``truncated Continuous 
Distribution of Ellipsoids'', or tCDE, weighs all
points equally within the bounds of a right triangle on the graph bounded by the
three vertices (L$_1$, L$_2$) of (0.005, t), (0.005, 0.99), and (0.995-t, t).  We
chose not to search for the best forsterite opacity using the tCDE shape distribution
of forsterite and an untested enstatite opacity (En90), but, instead, to search
jointly for the best value of the bound, t, on the tCDE shape distribution and the
best enstatite opacity.  This is because enstatite opacities have features of
similar widths and at many of the same wavelengths (10, 19, 23, 28, 33) as the
opacities of forsterite.

\subsection{Enstatite Opacity}

Now we focus on choosing the best opacity of enstatite, also searching for 
the best bound in the tCDE shape distribution for forsterite.  \citet{sarg06} 
found FN Tau to require enstatite to fit narrow
features in its 10 $\mum$ complex at 9.3, 10.6, 11.2, and 11.6 $\mum$.  In this
study, we are interested in modeling all silicate dust features found in IRS
spectra, so we look for guidance longward of the 10 $\mum$ complex to the spectrum
of FN Tau.  FN Tau has a very weak 16 $\mum$ feature (unlike the forsterite 
exemplars), a strong 28 $\mum$ feature relative to both the 23 and 33 $\mum$ 
complexes, and a double-peaked 33 $\mum$ complex.  We look for these patterns 
elsewhere in our Taurus-Auriga sample to find other enstatite exemplars.  DH Tau 
also has a narrow 9.3 $\mum$ feature, strong 28 $\mum$ complex, and double-peaked 
33 $\mum$ complex.  We also include Haro 6-37 and HK Tau as enstatite exemplars 
because both have narrow features at 9.3, 10.6, and also between 11 and 12 $\mum$, 
though they lack prominent crystalline silicate features at wavelengths longward of 
the 10 $\mum$ complex.

As with the olivines, numerous lab measurements of pyroxenes exist.  We include
opacities computed from transmission spectra of submicron pyroxene powders in
potassium bromide (KBr) pellets for orthoenstatite (MgSiO$_3$ crystals in
orthorhombic crystal structure), clinoenstatite (MgSiO$_3$ in monoclinic crystal
structure), the aforementioned En90, ``En80'' (Mg$_{0.8}$Fe$_{0.2}$SiO$_3$), ``En70''
(Mg$_{0.7}$Fe$_{0.3}$SiO$_3$), ``En60'' (Mg$_{0.6}$Fe$_{0.4}$SiO$_3$), ``En50''
(Mg$_{0.5}$Fe$_{0.5}$SiO$_3$), and ``En00'' (FeSiO$_3$, or ferrosilite) by
\citet{chi02}.  We also consider similar measurements of synthetic orthoenstatite and
clinoenstatite \citep{koi00}; natural clinoenstatite from Akita, Japan,
orthoenstatite from Norway, and synthetic clinoenstatite and orthoenstatite 
\citep{koi00}; and hypersthene (Mg$_{0.5}$Fe$_{0.5}$SiO$_3$), bronzite
(Mg$_{0.8}$Fe$_{0.2}$SiO$_3$), and clinoenstatite \citep{jag98}.  We also compute
opacities of orthoenstatite grains in the CDE2 shape distribution using 
optical constants for the three crystallographic axes provided by \citet{jag98}.  At
the same time, for the rest of the dust types in the models, we use the same
non-forsterite and non-enstatite opacities as were used in the models of the forsterite
exemplars in the previous section, and we explore values of the bound, t, on the tCDE
shape distribution of forsterite ranging from 0.04 to 0.1.  We also include the
opacity of the CDE2 shape distribution of forsterite grains.  We use all of these
opacities to model the spectra of the enstatite exemplars FN Tau, DH Tau, Haro 6-37,
and HK Tau, in addition to the forsterite exemplars.

Table 2 gives the significance of the mass weights of the opacities of crystalline 
pyroxene and crystalline olivine in the models of the enstatite and forsterite 
exemplars.  The opacities named ``en90'', 
``en80'', ``en70'', ``en50'', and ``en00'' were presented by 
\citet{chi02}.  The ones named ``J98CDE2'' use the optical constants for 
the three crystallographic axes of orthoenstatite presented by \citet{jag98} to 
compute opacity for grains in a CDE2 shape distribution.  ``K0so'' is the synthetic 
orthoenstatite, whose opacity is presented by \citet{koi00}.  ``J98b'' and ``J98c'' 
are bronzite and clinoenstatite opacities, respectively, presented by \citet{jag98} 
obtained via transmission measurements of crushed grains pressed into pellets.  
Entries for orthoenstatite, clinoenstatite, and ``en60'' opacities from 
\citet{chi02}, synthetic clinoenstatite from \citet{koi00}, and hypersthene from 
\citet{jag98} are not included at either temperature because 
no mass of any of them was ever used in any of the models of the 
exemplars of both forsterite and enstatite modeled jointly.  
``S6'' in the last five rows refers to optical constants presented by \citet{sog06} 
used to compute opacities for the CDE2 shape distribution (the first of the five) 
and for the tCDE shape distributions with bounding parameter (see discussion in 
Section 3.4), t, of 0.10, 0.08, 0.06, and 0.04 for the second through fifth of the 
five opacities.

Among the pyroxene opacities, CDE2 orthoenstatite from
optical constants provided by \citet{jag98} shows up at significant levels as a cool
dust component but only weakly as a warm dust component.  Ferrosilite (En00) by
\citet{chi02} often contributes significantly as a warm dust component but not as a
cool component.  The one opacity that contributes significantly most frequently as
both a warm and a cool dust component is the clinoenstatite (enstatite having 
monoclinic crystalline structure) En90 from \citet{chi02}.  Though we use the opacity 
of enstatite of composition Mg$_{0.9}$Fe$_{0.1}$SiO$_3$ having slight iron content, 
the fit at long wavelengths is not ideal and suggests caution in the interpretation 
of finding crystalline silicates with any iron content.  This opacity 
was used by \citet{sarg06} in modeling the 10 $\mum$ complexes of 12 TTS and by
\citet{sarg08} in modeling silica exemplars.  Among the various opacities of
forsterite, the tCDE opacity with bound t=0.1 gives the best fit to both the
enstatite and forsterite exemplars.  We note the improvement of fit of the spectrum of
ROXs 42C reported by \citet{sarg08} when using tCDE instead of CDE forsterite
\citep[optical constants by][]{sog06}, with $\chi^{2}$ per degree of freedom of 4.3 
and 5.1, respectively.

\subsection{Amorphous Silicate Opacities}

Now we turn to amorphous silicate opacities.  \citet{sarg06} showed that the
transitional disks (disks whose outer parts are optically thick to mid-infrared
radiation but whose inner regions are very optically thin) CoKu Tau/4, DM Tau, and
GM Aur were fit well using opacities of mostly submicron amorphous silicates having
smooth 10 and 20 $\mum$ features.  LkCa 15 was shown by \citet{esp07} to be a
pre-transitional disk, which are like the aforementioned transitional disks but with
more dust in the inner regions.  As with the transitional disks, LkCa 15 shows 
smooth 10 and 20 $\mum$ features.  FM Tau, TW Cha, and UY Aur also show
similarly smooth 10 and 20 $\mum$ features.  We consider these seven objects to be 
amorphous silicate exemplars.

\citet{dor95} gives complex indices of refraction for amorphous dust of pyroxene
composition (``amorphous pyroxene'') MgSiO$_3$, Mg$_{0.7}$Fe$_{0.3}$SiO$_3$, and
Mg$_{0.5}$Fe$_{0.5}$SiO$_3$ and also for amorphous dust of olivine composition
(``amorphous olivine'') MgFeSiO$_4$.  We use complex indices of refraction for 
samples of amorphous forsterite (Mg$_2$SiO$_4$) and amorphous enstatite (MgSiO$_3$) 
obtained by laser ablation of crystals of forsterite and enstatite, respectively 
\citep{sd96}.  We also use complex indices of refraction for samples of amorphous 
forsterite and amorphous enstatite prepared by sputtering of samples of MgSi and 
Mg$_{2}$Si in an atmosphere of 50:50 argon-oxygen \citet{day79}.  Additionally, we 
use optical constants for a sample of amorphous bronzite obtained by quenching a 
melt of a natural sample of bronzite (Mg$_{0.9}$Fe$_{0.1}$SiO$_{3}$) from 
Paterlestein, Germany analyzed by \citet{dor88} and for amorphous pyroxene of 
cosmic composition (Ca$_{0.03}$Mg$_{0.52}$Fe$_{0.45}$SiO$_{3}$; sample ``1S'') by 
\citet{jag94}.  For the same reasons provided by \citet{sarg06}, we assume the 
amorphous silicate dust grains are in the CDE2 shape distribution.

We also compute opacities of 5 $\mum$ radius, 60\% vacuum porous grains.  The narrow
features in the opacity curves of crystalline grains are very
sensitive to the exact details of these grains' properties such as shape,
composition, porosity, size, etc, as we have already indicated in our exploration of
the opacities of submicron crystalline grains used to fit silica, forsterite, and
enstatite exemplars.  For this reason, we, as \citet{bouw01}, do not include 
opacities of large crystalline grains.  The opacities for large amorphous silicate 
grains are less sensitive to these details.

Large grains (greater than 1 $\mum$ in characteristic size) in protoplanetary disks 
should not be homogeneous.  Instead, they should be like Interplanetary Dust
Particles, which are heterogeneous aggregates whose components are $\simali$ 0.1 
$\mum$ in size \citep[see][]{har02} and composed chiefly of either amorphous
silicate, forsterite, or enstatite \citep[see the review by][]{brad03}.  How 
should the opacities of such large heterogeneous grains appear?  The
recent study by \citet{min08} concluded that the opacity of a large
heterogeneous aggregate is equal to the sum of opacities of its constituents,
the opacities resembling those of homogeneous grains of characteristic size
corresponding to the abundance of the particular constituent in the large aggregate.  
As an example in their study, they compute the opacity of a large porous aggregate
of amorphous silicate, forsterite, and enstatite, the amorphous silicate being the
most abundant component of the aggregate.  The opacity of the amorphous component of
the aggregate resembled that of homogeneous grains of amorphous silicate of size
similar to that of the heterogeneous aggregate, while for the lowest abundance of
crystalline grains in the aggregate, the forsterite contribution to the aggregate
opacity resembled the opacity of much smaller grains of homogeneous forsterite.  
Very similar conclusions were also reached by \citet{bouw08}, finding that the 
typical grain size of crystalline silicate grains used in their models of seven 
spectra were submicron, but typical amorphous silicate grain sizes were up to 6 
$\mum$ radius (solid grains).

Because the grains in protoplanetary disks grow primarily by sticking together, 
it is likely large grains will be heterogeneous, composed of various sub-micron 
components, similar to the IDPs.  The small amorphous components are believed to 
come directly from the ISM, with sizes $<$ 0.25 $\mum$ in radius.  Since nearly 
all our objects show evidence for this amorphous component, one expects a 
significant amorphous component in the large, porous grains.  Furthermore, the 
amorphous components were present at the initial formative stages of the 
protoplanetary disks.  On the other hand, the crystalline components were produced 
later, by as yet unknown processes in the protoplanetary disks 
\citep[e.g.,][]{vb04,sarg06,wat08,bouw08}.  Thus one would 
expect amorphous components to dominate the heterogeneous fluffy grains.  In this 
case, the large grain spectra will resemble the sum of completely amorphous large 
grains with an admixture of sub-micron crystalline grains.  The models presented 
here will correctly detect the large grains, through the broadening of the silicate 
features to longer wavelengths.  Any crystalline components of large grains will 
be modeled as small crystalline grains.  The opacity of a large porous heterogeneous 
grain with a significant abundance of crystalline components, however, will not be 
modeled well by our standard set of opacities, especially at longer wavelengths 
($\lambda$ $>$ 20 $\mum$) in the IRS spectra (see Appendix B).  Because of optical 
depth effects, this will somewhat underestimate the crystalline mass fraction 
contributed by large grains.

To model our amorphous silicate exemplars, we use annealed silica, enstatite,
and forsterite in the tCDE shape distribution (with bound t=0.1) as our crystalline
silicate opacities.  We use the aforementioned amorphous silicate optical properties
for grains in the CDE2 shape distribution to compute opacities of submicron grains of 
amorphous pyroxene and amorphous olivine and also of 5 $\mum$ radius, 60\% vacuum
grains of amorphous pyroxene and amorphous olivine.  In testing the amorphous
pyroxene and amorphous olivine opacities, we chose the material whose opacities of both
submicron and 5 $\mum$ grains showed up most frequently as a significant contributor
as both warm and cool dust.  In Table 3, we list the significance of the mass
weights of the amorphous pyroxenes and amorphous olivines used in the models of our
seven amorphous silicate exemplars.  ``D95'' refers to optical constants 
presented by \citet{dor95}, ``D79'' refers to optical constants presented by 
\citet{day79}, ``SD96'' refers to optical constants presented by \cite{sd96}, 
and ``J94'' refers to optical constants presented by \citet{jag94}.  The 
amorphous olivine ``Ol'' from \citet{dor95} is MgFeSiO$_4$, and the amorphous 
olivine from \citet{day79} is amorphous forsterite.  The amorphous 
pyroxenes with ``Py5'' and ``Py10'' in their names refer, respectively, to 
Mg$_{0.5}$Fe$_{0.5}$SiO$_3$ and MgSiO$_3$ presented by \citet{dor95}.  The 
amorphous pyroxene from \citet{jag94} is amorphous pyroxene of cosmic 
composition, and the amorphous pyroxenes from \citet{day79} and \citet{sd96} 
are amorphous enstatite.  ``Sm'' denotes small Rayleigh-limit (here, submicron) 
size grains in the CDE2 shape distribution.  ``Lg'' denotes large 5 $\mum$ 
radius 60\% vacuum porous spheres.  Entries for opacities computed from optical 
constants for amorphous forsterite from \citet{sd96}, amorphous pyroxene of 
composition Mg$_{0.7}$Fe$_{0.3}$SiO$_3$ from \citet{dor95}, and amorphous 
bronzite from \citet{dor88} are not included at either temperature because no 
mass of any of them was ever used in any of the models of the amorphous silicate 
exemplars.

Although it is never used as a cool dust component, the amorphous olivine MgFeSiO$_4$ 
from \citet{dor95} is the best amorphous olivine because it is a very significant 
warm dust component in the models of all seven amorphous exemplars.  LkCa 15 
provides one of the most stringent tests of the amorphous silicate features, as its 
silicate features are very prominent and have high S/N ratios.  Amorphous olivine 
\citep{dor95} almost exclusively fits this spectrum's 10 $\mum$ feature.  Amorphous 
pyroxene can dominate as a cool dust component over amorphous olivine; for 
instance, the amorphous pyroxene of cosmic composition \citep{jag94} exclusively fits 
the 20 $\mum$ feature of LkCa 15.  Although the rest of
the amorphous pyroxenes are fairly equal contenders for best amorphous pyroxene 
opacity, the amorphous pyroxene of cosmic composition by \citet{jag94} shows up as 
the most significant amorphous pyroxene opacity in the most exemplars, so we choose 
this as our amorphous pyroxene opacity.  The minor modification to the amorphous pyroxene 
opacity usually improved (but sometimes made worse) the $\chi_{\nu}^2$ by $\simali$ 
0.2 over the amorphous pyroxene opacity used by \citet{sarg06,sarg08}.

The amorphous pyroxene of cosmic composition used in our models has a 
magnesium-to-iron ratio of 52:45, while the amorphous olivine used has a Mg-to-Fe 
ratio of 50:50.  These are not far from the Mg-to-Fe ratio for the
``cosmic'' abundances of these two elements \citep[Mg:Fe $\simali$ 4:5, 
measured by][]{hhk90,sw95,cox01}.  We also 
note the imaginary parts of the complex dielectric functions in the visible and
near-infrared regions (0.2-8 $\mum$ wavelengths) of the best amorphous olivine
\citep{dor95} and the best amorphous pyroxene \citep{jag94} are very similar to that
of ``astronomical silicate'' \citep{dl84}, whose near-ultraviolet, visible, and
near-infrared imaginary part of the dielectric function was specifically constructed
so that astronomical silicate grains would heat to the correct temperatures in the
presence of stars.  This was important in attempting to obtain a self-consistent
model of the {\it Spitzer} IRS spectrum of IP Tau by \citet{sarg06}.  We use the
best opacities that fit our various forsterite, enstatite, and amorphous silicate
exemplars to fit our sample of 65 spectra of TTS from the Taurus-Auriga 
star-forming region.

The first real test of this ensemble of dust opacities to be used as a standard 
mixture for dust emission models was to fit the mid-infrared spectrum of the 
Trapezium nebulosity.  The 8-13 $\mum$ spectrum of the Trapezium \citep{fgs75} was 
used by \citet{dl84} to derive the emissivity of ``astronomical silicate'' grains, 
which represent the dust in the Interstellar Medium.  The 8-13 $\mum$ spectrum of 
the Trapezium presented by \citet{fgs75} was combined with the 16-38 $\mum$ 
spectrum of the Trapezium obtained by subtracting 3\% of the spectrum of the 
Kleinman-Low nebula from the spectrum obtained pointed at the Trapezium 
\citep[see][]{for76}.  This accounted for contamination by emission from the KL 
nebula of the wide beam of the detector while pointed toward the Trapezium 
\citep{for76}.  The best-fit model for the Trapezium spectrum is shown in Figure 1.

As can be seen, the dominant components used to fit the spectrum are 
submicron amorphous pyroxene and submicron amorphous olivine.  This is 
consistent with more recent analyses of dust composition of the ISM along the 
line-of-sight to the Galactic Center by \citet{kemp04}, who found their dust models 
required negligible large grains and very small amounts of crystalline silicates 
(less than 2.2\% by mass).  The parameters for the model shown in Figure 1 are 
given in Table 4.

We describe in the Appendix our tests of Bruggeman effective medium theory used to
compute effective complex dielectric functions and Mie Theory used to compute
opacities of large grains.

\subsection{Degeneracy of Model Components}

We desire to measure the degeneracy between model components.  This is the extent 
to which one component could be replaced by other components and achieve a 
similarly good model fit.  After computing the best-fit models (to be described 
in the next chapter) for the spectra of FN Tau (an enstatite exemplar), IS Tau 
(a forsterite exemplar), ZZ Tau (a silica exemplar), and DM Tau (an amorphous 
silicate exemplar), $\chi^{2}$ was minimized with respect to the 
blackbody solid angles and dust mass weights over 7.7-37 $\mum$ wavelengths at the 
two dust temperatures (found by the best-fit models of these exemplars), not 
eliminating components with negative solid angle or mass weight.  In the process of 
minimizing $\chi^{2}$, a 16$\times$16 element covariance matrix was computed, each 
of the rows and columns of which belong to one of the 7 dust species or the 
blackbody at one of two temperatures.  Each of the elements of this matrix is a 
sum over all concerned wavelengths (7.7-37 $\mum$) of a product.  The product is 
opacity times the Planck function (or just the Planck function in the case of a 
blackbody component) specific for a given row times the opacity times the Planck 
function (or, again, just the Planck function in the case of a blackbody component) 
specific for a given column, all divided by the square of the flux density 
uncertainty.  The inverse of this matrix gives the covariance matrix \citep[see 
discussion in Chapter 14 of][]{press92}.  The diagonal elements of this matrix are 
the variances of each of the dust components in the model, and the off-diagonal 
elements are the covariances of the dust component of a given row with the dust 
component of a given column.

The correlation coefficient, r, is computed for each off-diagonal element by 
dividing the covariance of that element by the square root of the product of 
the variances of the two dust components in question.  One variance is the 
diagonal element of the same column, the other is that of the same row.  
Two highly degenerate components will be highly anticorrelated, with a correlation 
coefficient very near -1.  A correlation coefficient near zero means the two 
components are not correlated.  In addition, for each off-diagonal element, 
the probability, P, of a coefficient of equal or greater magnitude being found 
for a non-correlated data set \citep[which is the probability of the correlation 
coefficient having been drawn from a random distribution; see][]{taylor82} is 
computed.  P near 0\% indicate significant correlation 
coefficients, and P near 100\% indicate insignificant correlation coefficients.

Here we discuss the most significantly degenerate component pairs, which are 
shown in Table 5.  The most 
negative r's are for the pair of cool large amorphous olivine and cool 
large amorphous pyroxene.  For this pair, r is between -0.87 and -0.89 (and 
P=0.0\%) for the four representative exemplars FN Tau (enstatite exemplar), IS 
Tau (forsterite exemplar), ZZ Tau (silica exemplar), and DM Tau (amorphous 
silicate exemplar), indicating the most degeneracy between components.  Warm 
large amorphous olivine and warm large amorphous pyroxene are also fairly highly 
anticorrelated (degenerate); for this pair, r is usually around {-0.72} over the 
sample of 65 spectra.  Similarly, cool small amorphous pyroxene and cool small 
amorphous olivine are fairly significantly degenerate with each other, with 
-0.82$>$r$>$-0.89 over the four representative exemplars (P=0.0\%).  The warm 
small amorphous pyroxene and amorphous olivine are also highly degenerate, with 
r usually near -0.73 over the sample of 65 spectra.  Cool enstatite is typically 
fairly significantly degenerate with cool forsterite, with r usually near 
r=-0.46 over the sample of 65 spectra, as enstatite and forsterite share very 
similar features in their opacities for wavelengths longward of the 10 $\mum$ 
complex.  Silica also shares a 20 $\mum$ feature close to 19 $\mum$ complex of 
enstatite, explaining the spread of -0.29$\geq$r$\geq$-0.50 (P$\leq$0.1\%) in this 
pair of components for the four representative exemplars.  At warm temperatures, 
enstatite is only significantly degenerate with silica, with r for the 65 spectra 
in the sample usually being about -0.50, as warm silica and warm enstatite have 
strong, narrow features at 9.3 and 9.1 $\mum$, respectively, while warm forsterite 
is not degenerate with either, not sharing a strong feature with silica or 
enstatite in the 10 $\mum$ region.

For the highly degenerate pairs, the dust emission models accurately determine 
the sum of the masses of the two components, but not so accurately the individual 
masses.  This indicates the model finding that the inner disk (warm component) 
tends to be dominated by amorphous olivine while the outer disk (cool component) 
shows more amorphous pyroxene is probably not physically real.  This, at least 
partially, explains the negative masses often found to give the very best fit.  
By zeroing the negative mass, the complementary component's mass is increased to 
give nearly as good a fit.  Figure 2 demonstrates this effect by showing the 
degeneracy between cool small amorphous pyroxene and cool small amorphous olivine 
for the model of the Trapezium shown in Figure 1.  For the best-fit model 
temperature pair, the model is recomputed but without setting any of the 
components in the model to zero, allowing mass weights to be negative if that 
results in the lowest $\chi^{2}$ per degree of freedom (reduced $\chi^{2}$).  
All resulting model parameters are held at their new best-fit values except for 
the mass weights of cool small amorphous pyroxene and cool small amorphous 
olivine, which are varied over the ranges indicated on the plot.  The levels of 
reduced $\chi^{2}$ resulting from exploring this range of mass weights of these 
two dust components are shown as contours on the plot, with the levels of the 
first three contours indicated.  This contour plot suggests that replacing some 
amount of one of the cool small amorphous silicates with the same mass of the 
other gives a very similarly good fit, changing reduced $\chi^{2}$ only very 
slightly.

Histograms of the correlation coefficient between four representative pairs of 
dust components for all 65 spectra in the sample are given in Figure 3.  The 
histograms all show single, well-defined peaks.  This indicates the correlation 
coefficients are measuring real degeneracy between model components.

\section{Results}

Dust emission models are fit to the 65 Taurus-Auriga spectra, which we show 
in Figures 4-14.  Table 6 gives the parameters of these models 
(temperatures, blackbody solid angles, mass weights, reduced $\chi^{2}$).  
$\chi^{2}$ is minimized over 7.7-37 $\mum$, so both dust temperatures used in the 
models are well constrained by relative uncertainties of 10\%.  The median high and 
low model dust temperatures are 545\,K and 127\,K, respectively.  In the DR Tau 
model, which has dust at precisely these temperatures, most of the 10.0 $\mum$ 
wavelength flux above that from the blackbodies originates from 
545\,K dust, while most of the 20.0 $\mum$ flux above that from the blackbodies 
originates from 127\,K dust.  Assuming a distance of 140pc to DR Tau, assuming all 
of the optically thin dust flux 
at 10.0 $\mum$ comes from 545\,K dust in a disk of total optical depth at 10.0 
$\mum$ of 0.1, and assuming all of the optically thin dust flux at 20.0 
$\mum$ comes from 127\,K dust in a disk of total optical depth at 20.0 $\mum$ of 
0.1, the radii of these two disks in the case of DR Tau are {\bf 0.75 AU} and 
{\bf 11.5 AU} for the {\bf warm} and {\bf cool} optically thin dust, respectively.  
We list at the end of Table 6 the best opacities that fit our various forsterite, 
enstatite, and amorphous silicate exemplars to fit our sample of 65 spectra of
TTS from the Taurus-Auriga star-forming region.

\subsection{Extreme Inner Disk Grain Growth}

The spectra of DM Tau, DO Tau, UZ Tau/e, XZ Tau, and ZZ Tau IRS in our sample show 
evidence of high abundances of large grains.  Figure 4 shows our models of these
five spectra, breaking into components the model of UZ Tau/e as an
example of the models.  Warm large amorphous olivine dominates the 10 $\mum$ 
complexes of these five objects' spectral models.  The 20 $\mum$ features also are 
partly fit by this same component, but usually more of the flux required to fit the 
20 $\mum$ features comes from cool submicron amorphous pyroxene.  The overall shape 
of the spectrum of DM Tau differs from that of the other spectra in Figure 4 because 
it is a transitional disk \citep{cal05}, lacking significant dust closer to the 
central star than $\simali$ 3 AU.  We note that these five spectra support our use 
of opacities of large grains of amorphous silicates and do not require those of 
large crystalline grains.  That 
our models fit our spectra well is consistent with amorphous silicates being more
abundant both as small grains and as constituents in larger grains (see Section 3.4).  
We note that significant abundances of large grains have been found for SR20, a TTS 
in the Ophiuchus star-forming region \citep{mcc08}, CS Cha \citep{esp07cs}, and 
CVSO 224 \citet{esp08}.  \citet{fur07} found a large abundance of large grains 
around HD 98800 B, a somewhat older ($\simali$ 10 Myr old) YSO in the TW Hydrae 
association.  Large abundances of large grains around other somewhat older stars 
were found by \citet{bouw08} and \citet{ks06}.

\subsection{Prominent Forsterite Spectral Features}

In Figure 5 we show the spectra of six spectra with prominent forsterite features,
five of which are our forsterite exemplars, with the model of IS Tau broken down
into components.  We note in support of the forsterite opacities used in our models
that the fits at all wavelengths to the spectra of these stars, DK Tau, F04147+2822,
GN Tau, IS Tau, V836 Tau, and V955 Tau are quite good.  All major features and
complexes of forsterite at 10.0, 11.2, 16, 19, 23, and 33 $\mum$ are fairly well
fit.

The spectrum of F04147+2822 (Figure 5) is especially interesting.  Its 10
$\mum$ complex lacks any of the distinctive narrow emission features characteristic
of the forsterite, enstatite, or silica and is fit well by a combination of emission
from submicron amorphous silicate grains and large amorphous pyroxene.  At longer
wavelengths, however, emission features at 19, 23, and 33 $\mum$ characteristic of
forsterite dominate and, correspondingly, are fit well by our forsterite profile for
grains in a tCDE shape distribution.  This suggests a greater abundance of
forsterite in the outer disk than the inner disk.  One possible explanation is that
this spectrum is a sum of emission from two disks, one with very little
crystallinity throughout and blue in continuum color and another disk with a lot of
crystallinity throughout the disk and red in continuum color, but this system is not
known to be multiple.  If this spectrum is from only one protoplanetary disk,
this would contrast with the finding by \citet{vb04} of greater crystallinity within 2
AU than outside of 2 AU in disks around 3 Herbig Ae/Be stars.  However, the two are 
not inconsistent, as T Tauri stars are less luminous than Herbig Ae/Be stars.  Two 
AU in HAeBe stars should correspond to a much lesser radius in disks around T Tauri 
stars.  It 
suggests that, on average, the regions to which we refer as ``inner'' and ``outer''
disk regions in the population of 65 T Tauri stars in Taurus-Auriga are both outside
of the regions in T Tauri star disks analogous to the region inside 2 AU of Herbig
Ae/Be disks.

In some spectra, however, the long-wavelength side of the 10 $\mum$ feature is not
fit well.  DD Tau, DE Tau, and V710 Tau all share this problem, as is seen in
Figure 6, with the model of FX Tau broken into its components.  There are many ways
by which forsterite opacity features can be centered at longer wavelengths than
those in our tCDE opacity curve.  Grain shape distributions more heavily weighted
toward extreme shapes (extremely flat, extremely elongated), larger grains (or large
aggregates of small grains, with a large fraction of these small grains being
forsterite), olivine grains with nonzero Fe/Mg ratios, and forsterite grains with
greater porosity would all result in forsterite emission features at longer
wavelengths than resulted from the forsterite tCDE curve \citep[solid spheres of
forsterite give rise to opacity features at shorter wavelengths than those of
forsterite in the CDE shape distribution, and the opacity curve of 60\% porous
spheres of forsterite is almost identical to that of forsterite in the CDE shape
distribution][]{sarg06}.  Also, both CoKu Tau/3 and FX Tau (Fig. 4) have 23 and 33 $\mum$
emission complexes with emission that extends to longer wavelengths than provided
by the submicron forsterite grains in the tCDE shape distribution in our models. 
That both spectra lack significant 28 $\mum$ complexes indicates the dominant form
of crystals giving rise to 20-37 $\mum$ emission for these two systems is forsterite
and not enstatite.  It should be noted, though, that the 10 $\mum$ complexes in the
spectra of both CoKu Tau/3 and FX Tau are fit fairly well.  We therefore conclude
there is variation in the exact details of the olivine grain populations in the
protoplanetary disks in our sample, in terms of average grain shape, grain size,
grain composition, and grain porosity.  For DD Tau, DE Tau, and V710 Tau, this
variation occurs for the warmer forsterite grains, while for CoKu Tau/3 and FX Tau,
this variation occurs for the cooler forsterite grains.  We note that most of these 
five also seem to have emission features at 14 $\mum$ wavelength (see below, Section 
4.8).  Whatever gives rise to these 14 $\mum$ features is not necessarily related 
to the poor fits to the forsterite features because the poor fitting happens over 
very limited wavelength ranges of 11-11.5 $\mum$ and 23-25 $\mum$.

\subsection{Variation in Silica}

In addition to the olivine component of our models, the silica component may also
vary in details of grain properties among all disks in our sample.  While the silica
features at 9, 12.6, and 20 $\mum$ in the spectrum of ZZ Tau (and the other silica 
exemplars) are fairly fit well
with the annealed silica opacity used, others like the 12.6 $\mum$ features in the
spectra of DN Tau and FZ Tau are not quite so well fit (see Figure 7).  FZ Tau has a
single-peaked feature in this wavelength range, but it peaks at 12.45 $\mum$,
shortward of the wavelength at which annealed silica peaks.  This could suggest the
polymorph of silica giving rise to this feature is $\alpha$-quartz at somewhat
elevated temperatures of 500-625\,K \citep[see discussion by][]{sarg08}.  DN Tau
appears to have a double-peaked feature, which would suggest $\alpha$-quartz at
$\simali$ 300\,K \citep{sarg08}, but the S/N at these wavelengths is low for this
spectrum so its reality is more doubtful (note the discussion on the effect of
unresolved emission lines in this section).  According to \citet{sarg08}, the
presence of silica argues for high-temperature processing, as it is not present in 
the interstellar medium and therefore should not be present as the starting dust
mixture as the protoplanetary disk formed from its protostellar envelope. 
If $\alpha$-quartz can be confirmed, the silica, once formed, must cool slowly enough 
to allow phase transition from a higher-temperature polymorph, like tridymite or 
cristobalite, to $\beta$-quartz, then $\alpha$-quartz.

\subsection{Enstatite Exemplars}

We show the model fits to our enstatite exemplars in Figure 8, with the model of FN
Tau broken into components.  Other than a mismatch between model and spectrum for FN
Tau at 11.6 $\mum$, the fits to the 10 $\mum$ complexes are generally good. 
However, we note the poorness of fit of our models to the longest wavelength
complexes (28 and 33 $\mum$) of the spectra of FN Tau and DH Tau, our two best
enstatite exemplars.  The height-to-continuum ratios in our models of these two
exemplars for the 28 and 33 $\mum$ complexes are unacceptably low compared to those
in the spectra.  We note the progressively increasing height-to-continuum ratio of
the 19, 23, 28, and 33 $\mum$ complexes in the spectra of both DH Tau and FN Tau. 
The rest of our sample are sufficiently fit using crystalline dust at two 
temperatures.  This suggests that a population of submicron enstatite or forsterite 
dust at a third (very low) temperature would be required to give acceptable fits to 
these spectra.  We note this lack of a third model temperature is not a problem for 
spectra requiring abundant 
forsterite (Figures 5 and 6).  Note in Figure 6 the problem is not insufficient
peak-to-continuum ratio for the model forsterite features; rather, the problem is 
insufficient width of the features, which is a problem with the forsterite opacity.

The LH part of the spectrum of Haro 6-37 (19.3-37 $\mum$ wavelength)
suffers from an artifact, in which each of the 10 spectral orders are tilted such
that the flux density of their short-wavelength end is higher than it should be and
the flux density of their long-wavelength end is lower than it should be.  These
tilted spectral orders, which could be interpreted to resemble narrow crystalline
silicate features longward of 19.3 $\mum$ wavelength, result from an artifact of the
reduction of the LH spectrum of this object.  This artifact could be due either to
unsubtracted sky emission or to mispointing of the telescope with respect to Haro
6-37 along or perpendicular to the length of the LH slit.

\subsection{Amorphous Silicate Exemplars and Transitional Disks}

Spectra requiring abundant submicron amorphous silicate grains and model fits to 
these spectra are shown in Figures 9 and 10.  Figure 9 shows the
spectra and corresponding models of five of our amorphous silicate exemplars (see
Section 3.4) and the model breakdown into components for FM Tau.  In general, from
7.7 to slightly longward of 30 $\mum$ wavelength, the fits of the models to the data
are excellent.  For GM Aur and especially LkCa 15, the each model fits the continuum 
at wavelengths greater than 30 $\mum$ with emission from a weak 33 $\mum$ complex 
of forsterite.  This is an artifact, as these two 
spectra require a third model temperature to fit the data using emission from a 
blackbody or amorphous silicates at a very low temperature.  Figure 10 shows 
spectra whose models also require relatively high abundances of submicron amorphous 
silicates, though they require more crystalline silicates and large grains than 
those shown in Figure 9; the model of HQ Tau is broken into its components as an 
example.  Again, the model fits to the data are very good, with the exception of a 
slight insufficiency in the peak-to-continuum ratio of the forsterite 33 $\mum$ 
complex in the model of DP Tau.

\subsection{Small-Equivalent-Width 10 $\mum$ Complexes}

The spectra of CY Tau, DF Tau, DG Tau, DL Tau, DQ Tau, IT Tau, and V807 Tau show 10
$\mum$ complexes of small equivalent width (see Figure 11; the model of DF Tau is
broken into components).  CY Tau, DF Tau, DL Tau, and IT Tau also have essentially 
flat spectra over all IRS wavelengths, being consistent with the 
models presented by \citet{fur06b} and their interpretation of settling of dust
grains.  Settling of dust in a disk results in the disk being flatter and less
flared, which gives rise to a bluer spectral continuum color in the mid-infrared and
also less equivalent width in the distinctive silicate features.  \citet{aw05} noted 
CY Tau for its flat spectral colors over infrared wavelengths.

These spectra are well fit mostly by emission from submicron dust grains and the two
blackbody components in the models.  CY Tau requires a small amount of silica to fit
in the 10 $\mum$ complex and a moderate amount of cool forsterite to fit longer
wavelength complexes, especially the 33 $\mum$ feature.  The emission required by
large grains in the model for DF Tau to fit the longer wavelengths of its spectrum
is the largest of the five low 10 $\mum$ equivalent width spectra, in addition to
small amounts of emission from submicron dust grains.  Little emission
from forsterite is required to give rise to the 10 $\mum$ feature of DG Tau, while a
fair amount of submicron amorphous pyroxene is required to fit its mild 20 $\mum$
feature.  The 10 $\mum$ feature of DL Tau requires modest amounts of emission from
forsterite and silica and lesser amounts from other kinds of dust.  The LH spectrum
(19.3-37 $\mum$) of DL Tau suffers from an order-tilt artifact as does Haro 6-37 
(see above), limiting the precision with which we can determine the crystallinity 
and large grain content of the cooler outer disk, but it appears to be optimally 
fit using a mixture of both amorphous and crystalline grains.  DQ Tau has 
subtle features at 9.3, 9.8, 10.6, 11.1, and 11.6 $\mum$ that require a moderate 
amount of enstatite.  IT Tau requires small amounts of many of the submicron amorphous 
and crystalline grains to fit its miniscule 10 $\mum$ complex, but its broad 20 
$\mum$ feature is fit well by submicron amorphous pyroxene.  The same applies to the 
fit of the 20 $\mum$ feature of V807 Tau, though it requires a modest amount of 
large amorphous pyroxene to fit its 10 $\mum$ feature.  The abundance of crystalline
grains with respect to amorphous grains in the models of these five spectra is
consistent with the finding by \citet{wat08} using all (and more) of the spectra
analyzed in this study that increased crystallinity accompanies more advanced
settling of dust in disks as measured by continuum indices n$_{6-13}$ and
n$_{13-31}$.  It is also noted that these seven spectra with small equivalent width 
10 $\mum$ features do not indicate large abundances of large grains 
(see Table 6).  We explore these issues more in Section 5.

\subsection{Mixed Dust Compositions}

In Figures 12-14, we show the spectra and corresponding models to the rest of our
sample, with eight spectra in each figure arranged top-to-bottom alphabetically by
their names.  For all of these spectra, the abundances of the various dust species
required by the models are mixed.  Typically, no one dust type has an abundance
required by its model much more than the abundances of any of the other dust types
in that model.  The spectrum of AA Tau (and others) have what appear to be a broad 
emission feature at 14 $\mum$, but this probably originates from gas emission 
\citep{carr08}; also, SH and LH spectra of AA Tau and a few other objects show 
unresolved emission lines from H$_2$O and OH (see next subsection).  The models of 
FP Tau, GI Tau, and VY Tau are broken into components in Figures 12, 13, and 14, 
respectively.  Note that FS Tau and FV Tau in Figure 13 and GK Tau, HN Tau, and Haro 
6-28 in Figure 14 suffer from the same LH order-tilt artifact as did Haro 6-37 (see 
Section 4.4).  Also note the high point-to-point noise in the 20-25 $\mum$ parts 
of the LL spectra of FT Tau, GG Tau B, and GH Tau in Figure 13 and IQ Tau in
Figure 14.  This ``fringing'' artifact is due to the delamination of the LL 
order-sorting filter 
and shows up when the telescope has been mispointed with respect to a source in
the direction perpendicular to the slit for a LL observation, with increasing
severity of the artifact for greater mispointing of the telescope.

\subsection{Unresolved Emission lines}

There are some very narrow gaseous emission lines in our spectra.  The spectra of 
DG Tau (Figure 11), DM Tau (Figure 4), and FS
Tau (Figure 13) all show 12.8 $\mum$ features arising from [NeII].  This can affect
identification of the silica component, as noted by \citet{sarg08} with respect to
the modeling by \citet{sarg06} of TW Hya.  The presence of this feature
in addition to a narrow feature at $\simali$ 12.4 $\mum$ in the SL spectrum of TW
Hya led \citet{sarg06} to conclude the presence of $\alpha$-quartz, which has a
double-peaked feature matching these two wavelengths.  Investigation of 
high-resolution spectra of TW Hya showed that these two features belonged not 
to $\alpha$-quartz but to the Humphreys-$\alpha$ Hydrogen line (HI n=7-6) at 12.37 
$\mum$ and [NeII] at 12.81 $\mum$.

C$_2$H$_2$ appears in the spectra of CoKu
Tau/3 (Figure 6), DF Tau (Figure 11), and DL Tau (Figure 11).  This emission line is
located at the more innocuous wavelength of 13.7 $\mum$ \citep{carr08}, and does not
appear to have significantly affected the dust models of the spectra of these three
objects.  However, emission from HCN centered just longward of 14.0 $\mum$ 
\citep{carr08} is more problematic.  Examples of spectra with this feature are AA
Tau (Figure 12), BP Tau (Figure 12), and IT Tau (Figure 11).  First, the feature occurs
often in spectra taken with the SL and LL modules, which we splice together at 
precisely 14.00 $\mum$.  Often there is a flux level mismatch between SL and LL at this
wavelength, making establishing the reality of this feature difficult.  Second, 
the width of this feature varies.  We have no
dust component with an emission feature centered around 14 $\mum$, so when the feature 
is wider, more data deviate from our model, worsening the
model fit to the data.  Thirdly, observations of evolved stars in the Large Magellanic 
Cloud \citep{sloan06,sloan08} sometimes show a feature at $\simali$ 14 $\mum$ 
probably belonging to dust related to silicates.  Lastly, there are emission features 
near these wavelengths in spectra of Calcium-Aluminum Inclusions in meteorites 
\citep{posch07} and in laboratory spectra of the melilite solid series \citep{chi07}, 
making identification of all of the 14 $\mum$ features with HCN emission more 
ambiguous.

Finally, emission lines of H$_2$O have been discovered in high resolution {\it
Spitzer} IRS spectra of classical TTSs in Taurus, such as AA Tau \citep{carr08}. 
These lines are found in its SH and LH spectra, especially past 25 um.  
Many of these lines were found in the {\it Spitzer} IRS spectrum of the Class 0 YSO
protostar NGC 1333 IRAS 4B \citep{wat07}.  The lines are generally spread out over
wavelength, but there are clumps of lines around 24.5-25.5, 27.5-28.5, and 30-31
$\mum$ wavelength.  These water lines are unresolved, and our SH and LH spectra (R
$\simali$ 600) have been rebinned to SL/LL resolution (R $\simali$ 90), so the
effect of the lines generally being distributed over all wavelengths in SH and LH is
to raise slightly the level of the continuum above the continuum underneath the 
water lines in the high resolution spectra.  The clumps at 25, 28, and 30.5 $\mum$ 
in the  original high-resolution spectra, when rebinned, will show up as very slight
$\simali$ 0.5
$\mum$ wide bumps in the rebinned spectra.  We typically see no prominent
crystalline silicate features centered at either 25 or 30.5 $\mum$ in our spectra
\citep[though DH Tau has a small feature centered around 30.5 $\mum$ wavelength 
that has been attributed to enstatite before; see][]{mol02}, but we do see many 28
$\mum$ features in our data.  Both the opacities of forsterite and enstatite have
28 $\mum$ complexes, the 28 $\mum$ complex of enstatite being more prominent with
respect to its 23 and 33 $\mum$ complexes, so the net effect of not accounting for
the water emission might be obtaining a enstatite abundance slightly too large.

\subsection{Imperfect Extinction Correction}

Though we attempted to correct for extinction as explained in Section 2, two spectra
proved to be difficult in this regard, those of FV Tau (Figure 13) and CoKu Tau/3
(Figure 6).  For FV Tau, A$_V$=5.33, and for CoKu Tau/3, A$_V$=5.  Both were just
under the upper limit of A$_V$=6 above which we would not correct
for extinction and therefore not model, and both spectra show absorption at 15.2
$\mum$ from CO$_2$ ice; additionally, CoKu Tau/3 has ice absorption features at 6.0 
and 6.8 $\mum$ wavelengths as do heavily embedded Class I YSOs in Taurus-Auriga
\citep{wat04,zas07,boo08}.  Though the model fit to the 10 $\mum$ complex of CoKu 
Tau/3 is pretty good, the same is not true for FV Tau.  We suspect this to result 
from imperfect extinction correction.  At longer wavelengths, however, the fit to FV 
Tau (excluding the region around the 15.2 $\mum$ CO$_2$ ice absorption feature) is 
quite good, while the fit to the crystalline silicate complexes at 23 and 33 $\mum$ 
is only adequate.  As any extinction correction at longer wavelengths is less than 
that for the 10 $\mum$ silicate complex, we attribute this not to a problem with 
extinction correction, but rather to the opacity used to fit forsterite, which is 
discussed in the next subsection.

\section{Discussion}

\subsection{Inner versus Outer Disk Crystallinity and Grain Growth}

From the models for all 65 spectra, we computed the percentage of mass in a 
given dust grain species at one temperature out of all mass in dust at that 
temperature.  With these mass percentages, we compute histograms displaying 
both warm and cool dust mass percentages.  The histograms for total warm
and cool large grain mass fraction, total warm and cool crystalline grain mass
fraction, warm and cool forsterite mass fraction, warm and cool enstatite mass
fraction, and warm and cool silica mass fraction are given in Fig. 13.

The test described and shown in Appendix B of fitting a 5 $\mum$ radius 
heterogeneous grain profile with our standard model suggests that the abundance 
of crystalline grains could be underestimated by our standard models by up to 
50\%.  In addition, it also suggests the large grain abundance could be 
slightly underestimated by 23\%.  The net effect of this 
bias is to underestimate the true crystalline and large grain abundances.

Figure 15a shows that the inner disk regions typically have much larger mass 
fraction in large grains than the outer disk regions.  The mean and median in 
the histogram for warm large grains are around 50\%; for cool large grains, they 
are between 0 and 10 percent, 
though with a considerable ``tail'' up to 90\%.  Noting that our 5 $\mum$ radius 
porous grains are very similar in shape of opacity to 2 $\mum$ solid grains 
\citep{sarg06}, our average and median warm large grain mass fractions of 44\% 
and 45\%, respectively, compare favorably to the finding by \citet{siag07} of 
the average grain size for 1-2 Myr old systems being between 1.5 and 2 $\mum$ 
radius (as probed by the 10 $\mum$ feature).  Our average and median cool large 
grain mass fraction were 17\% and 0\%, respectively.

Figure 15b shows histograms of warm and cool crystalline dust mass fraction.  
Both histograms peak at low percentages of about 5\% for warm crystalline dust 
and about 12\% for cool crystalline dust (each of these percentages being the 
{\it mode} of the distribution), and there is considerable overlap.  
The cool crystalline dust histogram extends slightly to higher mass fraction 
bins, but we note the greater large grain mass fractions for the warmer inner 
disk regions.  The average and median mass fractions for warm crystalline dust 
were 17\% and 11\%, respectively, and the average and median mass fractions 
for cool crystalline dust were 23\% and 15\%, respectively.  Greater large 
grain mass fractions can lead to slight underestimation of the crystalline
abundance, so the amount of crystallinity in the inner disk regions could be 
higher and closer to that of outer disk regions.  Our findings of the mean, 
median, and mode of the warm crystalline dust mass fractions of 17\%, 11\%, 
and $\simali$ 5\% are consistent with those by \citet{honda06}, but they are 
slightly less consistent with those by \citet{siag07}.  \citet{honda06} 
find typical warm crystalline dust mass fractions (as probed by the 10 $\mum$ 
feature) typically to be between 5 and 20\%, regardless of system age, while 
\citet{siag07} find typical warm crystalline dust mass fractions to be 
between 5 and 7\%.  Though crystalline 
mass fractions may be greater in the inner disk regions than outer disk regions 
(see Table 6), the total mass of the cool dust in our models is between $\simali$ 
10 and 1000 times greater than that of the warm dust in the models.  The total 
mass in crystalline silicates is often greater for the cool crystalline dust 
than the warm crystalline dust, as Table 6 shows for V955 Tau.  The crystalline 
mass fractions for warm and cool crystals are 24.7\% and 20.6\%, respectively, 
being very similar.  However, there is $\simali$ 250 times more mass is cool dust 
than warm, so there is much more mass in 115\,K crystalline dust than in 488\,K 
crystalline dust according to the model.

Figures 15c, 15d, and 15e show the enstatite, forsterite, and silica mass fraction
histograms.  In Figure 15c, the warm enstatite mass fractions are very
slightly displaced to larger values than the cool enstatite mass fractions.  The
previously discussed potential for slight underestimation of crystalline grain 
mass fraction for the inner disks due to greater grain growth reinforces our 
conclusion that the abundance of enstatite is greater in the inner disk regions.  
Figure 15d shows that both the warm and cool forsterite mass fraction histograms 
show a peak in the lowest bin; however, a large fraction of the cool forsterite 
distribution extends to greater mass fractions.  First, we note the larger 
uncertainties in cool forsterite mass fractions than warm forsterite mass 
fractions (Table 6).  Second, we once again note the potential to 
underestimate inner disk crystallinity.  Together, these two considerations 
suggest the warm forsterite abundance is similar to that of cool forsterite for 
our sample.  Figure 15e shows the same is true for silica as for forsterite; for 
the same reasons as forsterite, the warm and cool silica abundances in our sample 
are similar.  The higher relative abundance of enstatite over forsterite in inner 
disk regions than for outer disk regions was also found by \citet{bouw08}.

\subsection{Correlations}

In order to search for correlations between pairs of model parameters, stellar
properties, and disk properties, we have computed the linear correlation
coefficient, r \citep{bev69}, and its corresponding probability, P, of finding a
linear correlation coefficient of magnitude greater than or equal to these 
coefficients on a null data set, as above (Section 3.5).  The correlation 
coefficient is weighted by uncertainties which are, for one data point, the square 
root of the sum of the squares of the uncertainties of the two quantities for 
which the correlation coefficient is being computed \citep[see discussion on weighted 
least-squares fitting with uncertainties in both x and y by][]{bev69}.  Note that 
scaling all dust mass weights by the same scalar less than unity would not 
change the correlation coefficient between two dust mass fractions.  The uncertainty 
weighting of the correlation coefficient shows up in both numerator and denominator 
of the correlation coefficient \citet{bev69} to the same power, so the scalar 
applied in the denominator would divide the scalar applied in the numerator to give 
1 times the original correlation coefficient.  Note that if r and the number, N, of 
data points that are being tested for correlation do not change, P will 
not change either because P depends only upon r and N.  Values of r 
that significantly deviate from zero and values of P close to zero suggest significant 
correlation or anticorrelation.  We deem as significant P$\leq$2.0\%, resulting in
$|$r$|$ $\geq$ 0.29.  Stellar masses, disk-to-star mass ratios, n$_{6-13}$ and 
n$_{13-31}$, and their uncertainties, plus multiplicity counts, stellar luminosities, 
and mass accretion rates used in this correlation analysis come from \citet{wat08}.  
Submillimeter slopes were obtained from the study by \citet{aw05}.

First we explore trends between dust mass fractions.  All trends between crystalline 
components are positive.  Figure 16 shows a positive correlation 
between warm enstatite and warm forsterite with correlation coefficient r=0.62 and 
probability P of having been drawn from a random distribution of 0.0\%.  This trend 
is fairly clear, with a heavy concentration of points at low mass 
fractions and a collection of points with higher enstatite and forsterite mass 
fractions.  Warm enstatite also has a correlation with warm silica, with 
r=0.31 and P=1.2\%.  Figure 17 shows a positive correlation between cool 
forsterite and cool silica, with r=0.29 and P=2.0\%.  Cool forsterite also 
correlates with warm large grain mass fraction (r=0.32 and P=0.9\%) and warm 
crystalline grain mass fraction (r=0.32 and P=1.0\%).  The correlation 
between warm silica and cool crystalline (cool enstatite plus cool forsterite 
plus cool silica) mass fractions is shown in Figure 18, with r=0.32 and P=0.9\%.  This 
correlation is likely related to the correlation between warm silica and cool 
forsterite (r=0.36 and P=0.4\%).  Warm enstatite also correlates with cool 
forsterite (r=0.29 and P=1.9\%).  Cool forsterite mass fraction and 
submillimeter slope \citep[$\alpha$ reported by][]{aw05} correlate, with r=-0.40 
and P=2.0\%.  Lower $\alpha$ means flatter submillimeter slope (flux density at 
submillimeter wavelengths, F$_{\nu,smm}$, is proportional to $\nu^{\alpha}$), 
implying that more growth of grains to millimeter sizes (instead of sizes of a 
few microns) accompanies a greater abundance of cool forsterite.

The general sense of these trends is that the crystalline dust abundances all
correlate positively with each other.  If the abundance of one type of crystal is
high, it is likely that the abundances of the other two types will also be
high.  The general finding that crystal abundances track other crystal abundances
has been noted before.  \citet{bouw01}, {\it via} detailed modeling of the 10 $\mum$
complexes of Herbig Ae/Be stars, noted correlation between forsterite and silica
abundance in Herbig Ae/Be stars, and \citet{vb05} (using a similar analysis) noted
correlation between enstatite abundance and total crystalline abundance in a similar
population of Herbig Ae/Be stars.  \citet{wat08} measured crystalline abundances
using indices computed from ratios of flux integrated over small wavelength bands
characteristic of emission from pyroxene, olivine, and silica at 10 and 33 $\mum$
(they call these bands P$_{10}$, O$_{10}$, S$_{10}$, and O$_{33}$, respectively); in
their study, they found warm pyroxene correlated with warm silica and warm olivine,
and that warm olivine correlated with warm silica and cool olivine.  From this we
conclude, as did \citet{wat08}, that whatever produces the different kinds of
crystals (forsterite, enstatite, and silica) produces them at a rate faster than
the crystals can transform from one to another.  As a specific example: if some
silica forms by incongruent melting of enstatite, that amount of
silica formed must be much less abundant than any silica formed by another
means \citep[e.g., {\it via} incongruent melting of amorphous pyroxene;][]{sarg08}.  
There is no significant correlation of crystalline silicate masses with disk-to-star 
mass ratio.  A positive correlation of 
disk-to-star mass ratio might be expected for crystalline dust produced by shock 
annealing \citep{hd02}.  The lack of this correlation does not necessarily mean 
shock annealing is not connected to dust processing, however, as disk mass estimates 
are based on assumed opacities of dust at submillimeter wavelengths.  The amount by 
which these assumed opacities differ from actual opacities varies according to the 
extent of grain growth to millimeter sizes \citep[see discussion by][]{daless06}.

\citet{fur06b} computed and interpreted the SED continuum indices n$_{6-13}$ and
n$_{13-25}$ as primarily measuring the degree of flaring of disks.  Indices that are
increasingly positive indicate increasingly flared disks, and more negative indices 
indicate increasingly flatter disks.  The flattening of these disks was interpreted 
to mean that dust had settled from high in the disk atmosphere towards the disk 
midplane.  \citet{wat08} used n$_{13-31}$ instead of n$_{13-25}$ to reduce 
contamination of the indices by the 20 $\mum$ amorphous silicate feature.  We use
the n$_{6-13}$ and n$_{13-31}$ indices, interpreting them as primarily measuring 
the degree of flaring of disks \citep{fur06b,wat08}.  We omit from our searches of
correlation of parameters with disk continuum indices the points for the
transitional and pre-transitional disks CoKu Tau/4 \citep{daless05}, DM Tau and GM
Aur \citep{cal05}, and LkCa 15 \citep{esp07}.  These four objects have spectra whose
continuum colors are very red due not to highly flared disks but, rather, to the
clearing of almost all small dust grains in the innermost regions of the disks.  The
disk indices for these four systems would measure how clear they are of such
dust in their inner regions, so they are not included in the search for trends of
parameters with disk indices.

Figure 19 illustrates a tight anticorrelation between warm
crystalline grain abundance and n$_{6-13}$, with r=-0.50 and P=0.0\%.  This can be
explained by correlations between each of the three warm crystalline types of
grains and n$_{6-13}$.  Warm forsterite and n$_{6-13}$ are anticorrelated with r=-0.42
and P=0.1\%, warm enstatite and n$_{6-13}$ are correlated with r=-0.41 and 
P=0.1\%, and warm silica and n$_(6-13)$ are correlated with r=-0.39 and P=0.2\%.  
Cool enstatite correlates with n$_{13-31}$ (r=-0.31 and P=1.9\%).

These findings are similar to those of \citet{wat08} 
that greater warm olivine and warm silica, measured by O$_{10}$ and S$_{10}$, 
respectively, anticorrelate with n$_{6-13}$ and that greater warm enstatite, warm 
olivine, warm silica, and cool olivine (warm enstatite and cool olivine being 
measured by P$_{10}$ and O$_{33}$, respectively) anticorrelate with n$_{13-31}$.  
The spectra described in Section 1.6 (low 10 $\mum$ complex equivalent width) all 
indicate, with the exceptions of DG Tau and V807 Tau, highly settled systems with 
abundant crystalline silicates, though with relatively large uncertainties on these 
crystalline abundances.  The settling of disks is interpreted here, as did 
\citet{fur06b} and \citet{wat08}, as part of the evolution of protoplanetary disks.  
The processing of dust into crystalline silicates is also interpreted as evolution 
of such disks.  Further, the models by \citet{cies07} predict crystalline mass 
fraction should be correlated with dust sedimentation both in inner and outer disk 
regions.  Therefore, the correlation of crystalline abundance with disk 
settling is expected as a general result of protoplanetary disk evolution.

Warm and cool large grain abundances both anticorrelate with n$_{13-31}$.  Figure 
20 shows the correlation between warm large grain abundance and n$_{13-31}$ 
(r=-0.36 and P=0.5\%).  This trend is interpreted as an increase in large grain 
abundance expected as a result of evolution of protoplanetary disks.  Figure 
21 shows that cool large grain abundance also anticorrelates with n$_{13-31}$ 
(r=-0.32 and P=1.5\%).  \citet{rettig06} found other tentative evidence for dust 
settling with grain growth.  Grain growth is, in fact, expected to be the
cause of settling of dust in protoplanetary disks.  From first-principles modeling,
\citet{weid97} find in his simulations of grain growth and settling towards disk
midplane of protoplanetary disks that at 60,000 years of evolution in the disk 
(the initial diameter of their dust grains was 1 $\mum$), at 30 AU from the central
star they find particles with diameters from one $\mum$ to near one centimeter. 
The range of ages assumed for Class II YSOs in the Taurus-Auriga star-forming
region is 1-2 Myr, so substantial grain growth should have occurred for the 65
objects whose spectra are analyzed in this study.  Also, at a disk age of 0.1 Myr,
\citet{weid97} predicts that at a disk radius of 30 AU, the dust-to-gas mass ratio
at 6 AU above the midplane should be 10$^{-3}$ what it was at time zero, when the
disk was initially well-mixed, at the same location in the disk.  This ratio of the
dust-to-gas mass ratio to its initial well-mixed value is what \citet{daless06}
call the settling parameter, $\epsilon$.  As shown by \citet{fur06b}, the disk
indices of some of the Class II YSOs in the Taurus-Auriga star-forming region are
consistent with such low values of $\epsilon$ (0.01-0.001).  \citet{bouw08} also find 
that the differing morphology of IRS spectra of YSOs are consistent with dust growth 
and settling.  Large grain abundance anticorrelating with n$_{13-31}$ is consistent 
with the expectation by \citet{weid97} of increasing grain growth over time and 
also greater settling of larger dust grains towards the disk midplane.

Also significant is the result shown in Figure 15a that more large grains are 
found in inner disk regions than outer disk regions.  The observations and models by 
\citet{vb04} required more large grains for disk regions inside 2 AU than for disk 
regions outside 2AU for spectra of two of the three Herbig Ae/Be stars they studied.  
\citet{aw07} find that the surface density of protoplanetary disks decreases with 
radius in the disks as r$^{-0.5}$ or r$^{-1}$, so inner disk regions 
should be denser.  This could encourage more rapid grain growth.

However, as noted previously, the correlations between large grain
abundances and n$_{13-31}$ are not very strong.  One possible explanation for the 
weakness of correlation we find between warm large grain abundance and n$_{13-31}$ 
is that we are not measuring the true large grain abundance from each spectrum 
because we do not account for differing amorphous-to-crystalline component ratios 
in the large inhomogeneous aggregate grains from system to system.  The more
crystalline components the average aggregate large grain in a given disk has, the 
less reliable a measure our large amorphous silicate grains are of the large grain 
extent in such disks \citep{min08}.

Another possibility to explain the weakness of trend of warm large grain abundance 
with n$_{13-31}$ is that sufficiently high turbulence prevents rapid grain growth 
and settling of few-micron-sized grains to the disk midplane
\citep{dudo04,hubbard06}.  
Perhaps different disks can have the same degree of flattening but differing 
abundances of large grains in the optically thin surface layer.  \citet{dudo05} note 
that collisions replenish small grains.  However, \citet{cies07} finds that turbulence 
can also encourage collisions that facilitate grain growth.  These large grains, once 
formed, can be stored in a dead zone underneath the turbulent live zone, and can be 
protected even if the dead zone is disturbed by the live zone above it \citep{cies07}.

Warm large grain abundance also correlates with the known number of stars in the 
star system, with r=0.33 and P=0.7\%, so perhaps multiplicity contributes to the 
dispersion seen in the trend between warm large grain abundance and n$_{13-31}$.  
Figure 22 shows a comparison of the histogram of warm large grain abundance for 
single star systems and the same for multiple star systems - one immediately notices 
the higher large grain mass fractions for multiple systems than single systems.

Finally, Figure 23 shows a fairly weak anticorrelation between warm large grain
mass fraction and stellar mass, with r=-0.36 and P=0.5\%.  This may be related to the 
correlation between warm small amorphous grain mass fraction and stellar mass 
(r=0.40 and P=0.1\%).  In our sample, the spectra 
of low mass stars indicate higher levels of warm large grain abundance than for
higher-mass stars.  \citet{apai05} found in their study of six brown dwarf spectra
higher mass fractions of crystalline grains for lower mass stars.  They concluded 
that the region giving rise to the 10 $\mum$ feature was at smaller disk 
radii for less luminous, lower mass stars, and that in these innermost disk radii the
crystalline dust abundance was higher.  We interpret the trend of increasing warm
(inner disk) large grain mass fraction with decreasing stellar mass similarly, in
view of our finding of greater grain growth in inner disk regions than outer disk
regions (Figure 15).

\section{Summary and Conclusions}

We have analyzed the dust composition of 65 T Tauri stars using spectra from the 
IRS on the {\it Spitzer Space Telescope}.  These spectra show very prominent 
emission features and complexes at 10, 16, 19, 23, 28, and 33 $\mum$ wavelengths, 
which are characteristic of silica and silicate grains (both amorphous and 
crystalline) with sizes from submicron to a few microns.  We have constructed 
spectral models for each of the spectra that include, firstly, blackbodies at two
temperatures and, secondly, Planck functions at those two temperatures multiplied 
by scaled dust emissivities.  These represent inner and outer disk emission from, 
firstly, the optically thick disk midplane and blackbody grains and, secondly, 
dust in the optically thin disk atmosphere with strong infrared resonances.  
The best fit for a given set of dust opacities is given by the pair of 
temperatures for which the global $\chi^{2}$ per degree of freedom is a minimum.  
We have tested the opacities on high-quality IRS spectra to find the best 
opacities to represent emission from submicron grains of enstatite, forsterite, 
silica, and amorphous silicates and large (few micron size) grains in the 
protoplanetary disks of the Taurus-Auriga association.

We conclude the following:

\begin{itemize}
\item High S/N spectra suggest the best crystalline silicate opacities to
use in modeling are those of iron-poor pyroxene (enstatite) and iron-poor olivine
(forsterite).  The best
amorphous silicate opacities to use are those whose iron-to-magnesium ratios are
nearly 1 (cosmic).  This is consistent with the finding by \citet{har02} that
cometary amorphous silicates require significant iron in their compositions in
order to heat to inferred dust temperatures.  The adopted amorphous silicate 
closely matches the heating of \citet{dl84} ``astronomical silicates''.
\item Though we use opacities of grains with 50\% porosity and 5 $\mum$ 
radius of both 
amorphous pyroxene and amorphous olivine, calculations of the covariance between
these components in models of various high S/N spectra show they are highly
degenerate.  This means our models cannot readily distinguish between the two
species as large grains, but our models are more sensitive to the sum of their
masses.  Submicron amorphous pyroxene and submicron amorphous olivine are similarly
degenerate but slightly less so.  Cool silica is degenerate with cool submicron
amorphous silicates as they share 20 $\mum$ features.  Cool enstatite and cool
forsterite are somewhat degenerate because they share prominent 23, 28, and 33 
$\mum$ features (only somewhat because the 28 $\mum$ feature of enstatite is 
relatively stronger than that of forsterite).
\item A few spectra of T Tauri stars show evidence for extensive grain 
growth in that protoplanetary disks as indicated especially by their 10 $\mum$ 
complexes.
\item Though most spectra are fit satisfactorily using our truncated CDE 
forsterite opacity, 
a few are not.  The deviations of the 10, 23, and 33 $\mum$ features between our
models and the spectra suggest olivine grains of greater size, greater porosity,
greater iron content (namely, a nonzero iron content), or greater weighting towards
extremely elongated or flattened grain shapes in the shape distribution are 
required, or some combination of these.
\item Most spectra that require silica are fit well by the annealed silica 
opacity in the standard dust model mixture.  The spectra of FZ Tau and DN Tau 
suggest a 
silica polymorph other than annealed silica \citep[cristobalite and tridymite; 
see][]{fab00} would provide a better fit to their spectra.  This would imply 
different cooling rates for silica, once formed, than implied by the presence 
of the higher temperature polymorphs of silica \citep{sarg08}.  However, the 
12.6 $\mum$ features of these two exceptional spectra are only mildly
inconsistent with the annealed silica in the dust models, so the suggestion 
of an alternative silica polymorph is weak.
\item The models of the best enstatite exemplars, FN Tau and DH Tau, do not 
fit very well the strong crystalline silicate emission peaks at 28 and especially 
33 $\mum$, suggesting enstatite or forsterite dust at temperatures significantly 
lower than those used by our models is required by the data.  It is unknown why 
such cold material is required to fit spectra indicating high enstatite abundance.
\item Transitional and pre-transitional disks in this sample require negligible 
crystalline silicates and modest amounts of large grains (DM Tau being the 
exception, requiring significant large grains) to fit their spectra relative to the 
rest of the disks whose spectra were analyzed in this study.  A similar
(small) number of spectra of systems not known to be transitional or
pre-transitional also require high abundances of submicron amorphous silicates.
\item A few spectra have 10 $\mum$ complexes that have small equivalent
widths.  Most of these are fit well by crystalline silicates as opposed to large 
grains, which is inconsistent with the idea that decreasing peak-to-continuum ratio 
of the 10 $\mum$ complex necessarily implies grain growth.
\item The goodness of fit of our models to the majority of our spectra means
the opacities of large crystalline silicate grains are not typically required.  
This, in turn, implies that, if grains grow to larger sizes by agglomeration of 
grains of differing composition, then on average in the Taurus-Auriga YSO 
population the more abundant subcomponents of these heterogeneous aggregate 
large grains are amorphous rather than crystalline.
\item Higher abundances of large grains are found in inner disk regions
than outer
disk regions, suggesting grain growth occurs more rapidly in inner disk regions.
\item Crystalline silicate abundances are very similar in the inner and 
outer disk regions.  
This contrasts interestingly with the finding by \citet{vb04} that regions 
inside 2 AU in disks around Herbig Ae/Be stars have much more processed dust than 
regions outside of 2 AU.  F04147+2822 exemplifies the contrast, having prominent 
19, 23, 28, and 33 $\mum$ complexes but very 
little indication of crystallinity based on its 10 $\mum$ complex.  This suggests 
the inner region of this disk giving rise to the 10 $\mum$ complex has a lower 
abundance of crystals than the outer region giving rise to emission past 19 $\mum$ 
wavelength.
\item Cool forsterite also correlates with flatter submillimeter SEDs, 
suggesting grain growth to millimeter sizes is more extensive for disks with more cool 
forsterite.
\item Each crystalline silicate abundance always correlates positively and 
significantly with abundances of other crystalline silicate species.  This suggests 
that whatever produces crystalline silicates does so at a 
faster rate than that at which any of the crystalline silicate species transform into
any of the others.  This, in turn, suggests that amorphous dust is processed into 
crystalline dust at a rate greater than one kind of crystalline silicate species 
can transform into another.
\item Crystalline silicates in inner disk regions are more abundant for
bluer, flatter disks, those with more advanced settling of dust grains towards 
disk midplane.
\item Large grains in the inner disk regions are, on average, more abundant
for more settled (bluer) disks.  There is a fair degree of dispersion in the inner 
disk large grain 
abundance for a given degree of settling, suggesting different rates of settling of
large grains in different disks.  There is also an indication that the abundance 
of large grains in the inner disk correlates with known multiplicity of stellar 
system.
\item The spectra of disks around less massive stars indicate higher
abundances of large grains in their inner disk regions.
\end{itemize}

\acknowledgments This work is based on observations made with 
the {\it Spitzer Space Telescope}, which is operated by 
the Jet Propulsion Laboratory, California Institute of Technology 
under NASA contract 1407.  Support for this work was provided by 
NASA through Contract Number 1257184 issued by JPL/Caltech and 
through the Spitzer Fellowship Program, under award 011 808-001, 
and JPL contract 960803 to Cornell University, 
and Cornell subcontracts 31419-5714 to the University of Rochester.  The
authors wish to thank Harald Mutschke for sharing the opacity in tabular 
form of the annealed silica presented by \citet{fab00}.  A.L. acknowledges 
support from the Chandra theory program,
the Hubble theory programs, and the Spitzer theory programs.
SMART was developed by the IRS Team at Cornell
University and is available through the Spitzer Science Center at Caltech.  
This publication makes use of the Jena-St. Petersburg Database 
of Optical Constants \citep{hen99}.  The authors made use of the 
SIMBAD astronomical database and would like to thank those responsible 
for its upkeep.

\appendix
\section{Error Bars and Mispointing}

We corrected for slight mispointing of the telescope from the standard nod 
positions by scaling the mispointed nods, as described by \citet{sarg08}.  For 
observations taken by the low spectral resolution modules SL and LL, only 
mispointing in the dispersion direction (perpendicular to the length of the slit) 
could give rise to loss of flux.  Mispointing of the high resolution slits along
the length of the slit (spatial direction) towards the center of the slit can
actually result in obtaining more flux than is present if one does not consider the
RSRFs used to calibrate flux were constructed from observations of flux standards at
nominal nod positions.  More flux enters through the slit if the target is
positioned closer to the center of the slit than a nominal nod position (like the
flux calibrator is); in this case, one must apply a scalar factor less than 1 to the
affected nod spectrum to correct for the extra fraction of total signal from the
target not measured for the calibrator.  SH has a smaller slit and is therefore more 
sensitive to mispointing in either spectral or spatial directions; therefore, the 
relative amount of scalar correction is typically higher for SH than LH.  The scalars 
used are given in Table A1.

\section{Test of Large Grains with Heterogeneous Composition}

We conducted tests of the combination of Bruggeman EMT and Mie Theory used to
compute effective complex dielectric constants and opacities for our large grains. 
Note that \citet{sarg06} found that opacity curves of sub-micron spherical 60\% 
vacuum (porous) grains of forsterite and $\alpha$-quartz were identical to those 
computed for solid grains of the same materials in the CDE shape distribution.  We 
have computed the effective complex dielectric function for material that is, by 
volume, 60\% vacuum (complex dielectric function equal to unity), 20\% forsterite 
\citep[complex dielectric function by][]{sog06}, and 20\% amorphous olivine 
MgFeSiO$_4$ \citep[complex dielectric function by][]{dor95} as follows.  First we 
computed using Bruggeman EMT the effective complex 
dielectric function of 50\% amorphous olivine and 50\% forsterite a-axis; we also
computed the same effective complex dielectric functions for 50/50 mixtures of
amorphous olivine and each of the other two forsterite crystallographic axes, b and
c.  Then we computed using Bruggeman EMT the effective complex dielectric constants
for material that is 60\% by volume vacuum and 40\% the 50/50 mixture of amorphous
olivine and forsterite; again, the same was done for the effective complex
dielectric functions involving forsterite b and c axes.  This resulted in 3 sets
of complex dielectric functions, one for each of the 3 crystallographic axes of
forsterite, of material that is, by volume, 60\% vacuum, 20\% forsterite, and 20\%
amorphous olivine.  Next we computed opacities for each of the three sets of
effective complex dielectric functions for Rayleigh-limit-size spheres of these
materials.  Finally, we averaged the three opacities together to account for
anisotropy of optical constants \citep[see][]{bh83} for forsterite.  We then
calculated the flux density from 1 lunar mass of such grains 140 parsecs away at
300K.  1\% relative uncertainties were assumed, which are about the best we can
expect from our sample of 65 spectra.

The 7.7-37 $\mum$ part of this spectrum was then modeled using our standard
opacities, which are listed in the notes for Table 6, with the exception of using
the opacities of forsterite and amorphous olivine grains in the CDE shape 
distribution (instead of tCDE and CDE2).  We used the CDE for these two dust types 
in anticipation of the porous-CDE correspondence reported by \citet{sarg06}.  In 
Figure A1, we show the fit of the model to this test spectrum.  Although none of 
the dust opacities were precluded at the outset, the only opacities used were the 
CDE opacities for forsterite and amorphous olivine, with masses of 0.45 and 0.43 
M$_{Moon}$, respectively.  The fit is very good to all wavelengths, though we note 
deviations for the strongest features at 11.4, 19, and 23 $\mum$; we also note that
the correspondence between porous grain and CDE opacities reported by \citet{sarg06}
was not perfect.  The model used grains at 308K, only slightly different from the 
assumed 300K.  This slightly higher than assumed temperature explains the slightly 
lower (0.88 M$_{Moon}$) than assumed (1 M$_{Moon}$) total mass of dust.  We take 
this as support for the conclusion by \citet{min08} that opacities of aggregates 
can be expressed as a sum of the opacities of homogeneous grains composed of the 
same materials as the aggregate's components, as applied to heterogeneous 
aggregates in the Rayleigh size limit.

Next we conducted a similar test, but for large aggregates.  We used the same
effective complex dielectric function of vacuum, forsterite, and amorphous olivine
in a 60/20/20 volume ratio as before but this time to compute opacities for 5 $\mum$
radius grains of the material.  We computed the test spectrum assuming 1 lunar mass 
of grains at 300K located 140pc away.  Here, we used our standard opacities exactly 
as listed in the notes for Table 6, to gauge the effect of large grains in 
protoplanetary disks being aggregates of both amorphous and crystalline grains.  
Dust at only one temperature was allowed, but no dust species at this temperature 
was precluded from the model 
at the outset.  This time, the best-fit model dust temperature is lower, 264K,
and enstatite is used in addition to forsterite and large (5 $\mum$ radius porous)
grains of amorphous olivine, though at a lower significance level than either the
amorphous olivine or forsterite.  We show this model in Figure A2.  The total 
 mass of the dust used by the model, 1.25 lunar masses, is a sum of 0.96 M$_{Moon}$ 
of large amorphous olivine plus 0.20 M$_{Moon}$ of forsterite, 0.09 M$_{Moon}$ of
enstatite, and 0.001 M$_{Moon}$ of submicron amorphous pyroxene, which is somewhat 
greater than the mass assumed for the aggregates.  This increase in mass was needed 
to compensate for the lower temperature.  The model becomes an increasingly
poorer fit at increasing wavelengths.  Part of this failure may be due to the use
of the restricted shape distribution (tCDE) for the forsterite instead of the more
extreme shape distribution CDE (used previously to test the Rayleigh-limit
aggregate).  An additional population of forsterite grains at a lower temperature 
than used may improve the fit.  This suggests any serious failures of our standard 
model to fit spectra with prominent crystalline silicate features may imply a 
significant amount 
of grain growth of aggregate grains with significant crystalline silicate abundances
(here, 50\% of the solid mass of the aggregate).  This test also implies that the 
crystalline abundance could be underestimated by 50\% (0.29/1.25 instead of 0.625/1.25 
fractional abundance of crystalline silicates in heterogeneous aggregates) of the 
true value, and it implies that the large grain abundance could be 
slightly underestimated (0.96 instead of 1.25 lunar masses of large (5 $\mum$ 
radius) grains) by 100$\times$(0.29/1.25)\% = 23\%.

\section{Importance of Large Grain Opacity}

We tested the necessity of using opacities of large grains in our models. 
We used our standard dust model on IS Tau, one of the objects whose spectrum was
modeled using a significant amount of large grains, but we eliminated the opacities
of large amorphous pyroxene and large amorphous olivine in the fit.  The 
resulting model, shown in Figure A3, is a poorer fit to the 10 $\mum$ complex, not 
``filling out'' the full width of the 10 $\mum$ complex.  Reduced $\chi^{2}$ for 
this large-grains-precluded model was 4.9, which is 1.6 higher than the reduced 
$\chi^{2}$ of 3.3 for the model with large grains.  We conclude the opacities of 
large amorphous grains are required in our models.

\section{Porosity Test}

We tested how sensitive our models are to the exact combination of
porosity and grain size when we compute and use opacities of large grains in our
models to detect grain growth.  Instead of 5 $\mum$ radius 60\% porous grains, we
computed opacities for 20 $\mum$ radius grains that are 88\% porous.  We
chose this size to preserve the optical depth through the center of the
grain.  As noted by \citet{sarg06}, 5 $\mum$ radius 60\% porous grains were chosen
to preserve the optical depth through the center of the 2 $\mum$ solid grains of
amorphous olivine used by \citet{bouw01} to represent grain growth in modeling
spectra of Herbig Ae/Be stars.  This preservation of the optical depth
through the center of the grain resulted in a close correspondence of the opacities
of 5 $\mum$ radius porous grains and of 2 $\mum$ radius solid grains.  Although 90\%
vacuum (by volume) grains would be the logical extension 88\% vacuum 20 $\mum$ 
radius grains proved to result in an 
opacity most like that of 5 $\mum$ radius 60\% vacuum grains.  As a test of this
opacity, we modeled the spectrum of UZ Tau/e, noted in Section 4 to require a large
amount of large grains; here, we replace the opacity of 5 $\mum$ 60\% porous grains
of amorphous olivine with that of 20 $\mum$ 88\% porous grains of amorphous
olivine.  We show the best-fit model in Figure A4, compared to the model shown in
Figure 4 and described in Section 4.  The fit is not much different, and the
reduced $\chi^{2}$ is 5.6 instead of the 5.4 obtained with our standard opacities. 
Therefore, we cannot make precise statements on grain size and porosity individually 
for a given protoplanetary disk based only on the IRS spectra.

\clearpage

\begin{figure}[t] 
  \epsscale{0.8}
  \plotone{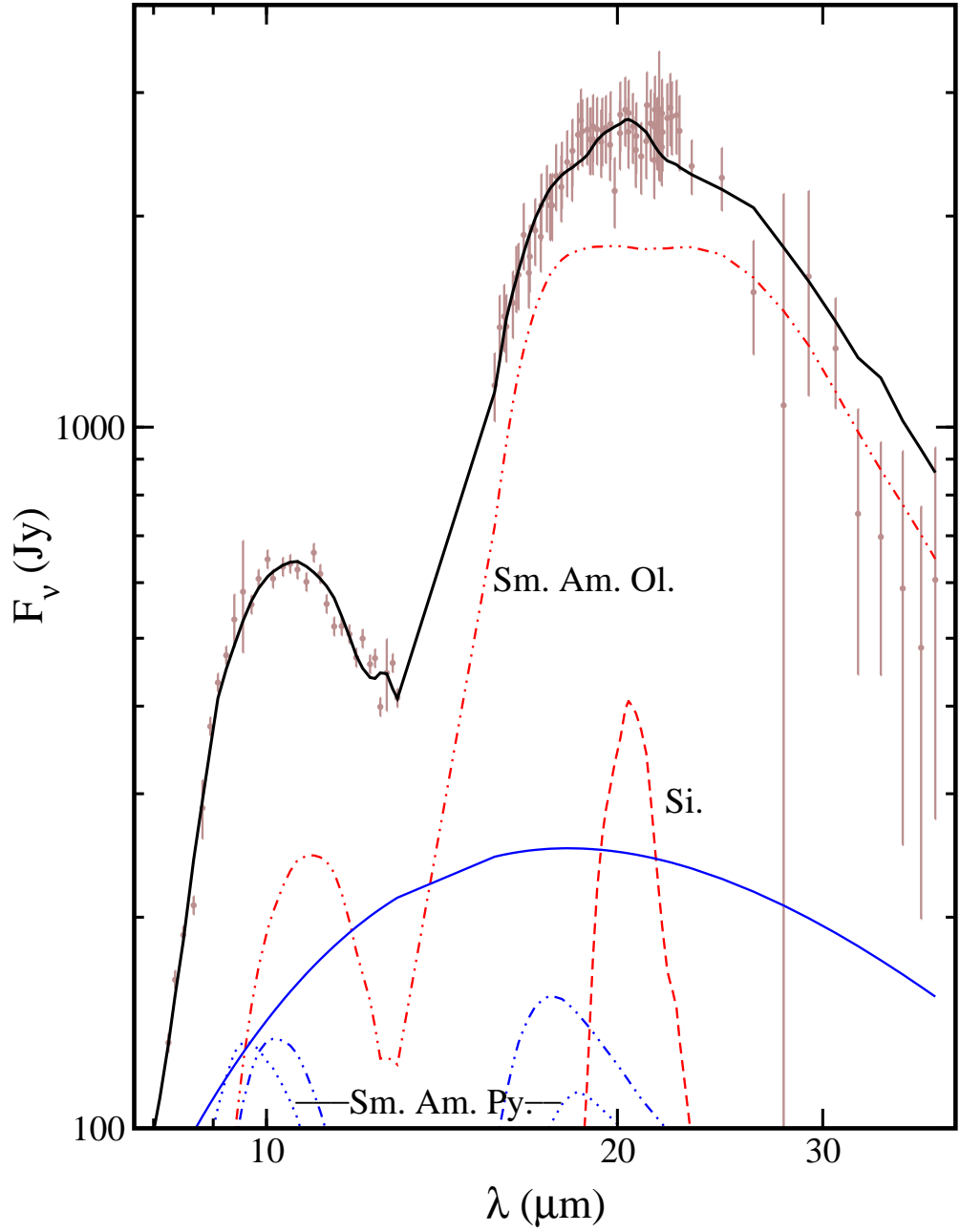}
  \caption[Model Fit to Trapezium Spectrum]{Model of the spectrum of the Trapezium.  
Model components are at bottom.  Blue lines represent model components at higher 
temperature, red lines 
represent model components at lower temperature.  Solid lines are blackbodies, 
dotted lines are submicron amorphous pyroxene, dash-double-dotted lines are 
submicron amorphous olivine, dot-long-dash lines are large amorphous pyroxene, 
dot-double-dash lines are large amorphous olivine, dot-short-dash lines are 
enstatite, long-dashed lines are forsterite, and short-dashed lines are silica.}
\end{figure}

\clearpage

\begin{figure}[t] 
  \epsscale{0.8}
  \plotone{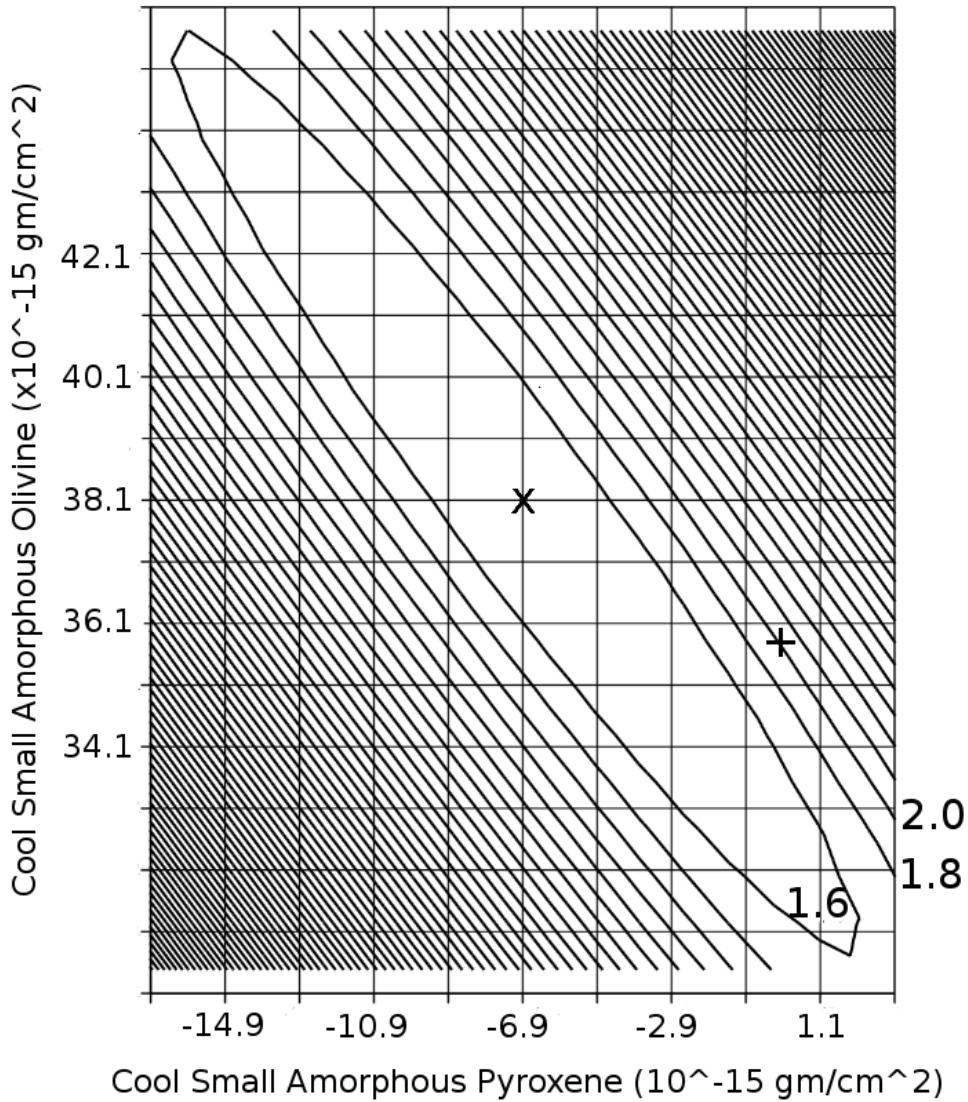}
  \caption[Cool Small Amorphous Pyroxene and Olivine Degeneracy]{Contours 
of $\chi^{2}$ per degree of freedom for models of the Trapezium spectrum 
with dust at the same temperatures as the model shown in Figure 1, but 
with dust weights allowed to be negative (the ``negative-allowed'' model).  
The first three contours of reduced $\chi^{2}$ are labeled.  The ``x'' gives 
the best-fit values of cool small amorphous pyroxene and cool small amorphous 
olivine for this model, resulting in reduced $\chi^{2}$ of 1.4.  The ``+'' 
gives the best-fit values of the same two dust components for the standard model 
(the ``negative-not-allowed'' model) shown in Figure 1, resulting in reduced 
$\chi^{2}$ of 1.6.  This value, 1.6, is lower than that indicated on the plot, 
2.0, because in the negative-not-allowed model, the mass weights of the other 
dust components were allowed to vary to obtain the global minimum in reduced 
$\chi^{2}$, while in the negative-allowed model they were not.  Note that 
nearly the same minimal level of reduced $\chi^{2}$ can be obtained by 
replacing a given amount of one of the two dust components by about the 
same mass of the other.}
\end{figure}

\clearpage

\begin{figure}[t] 
  \epsscale{0.8}
  \plotone{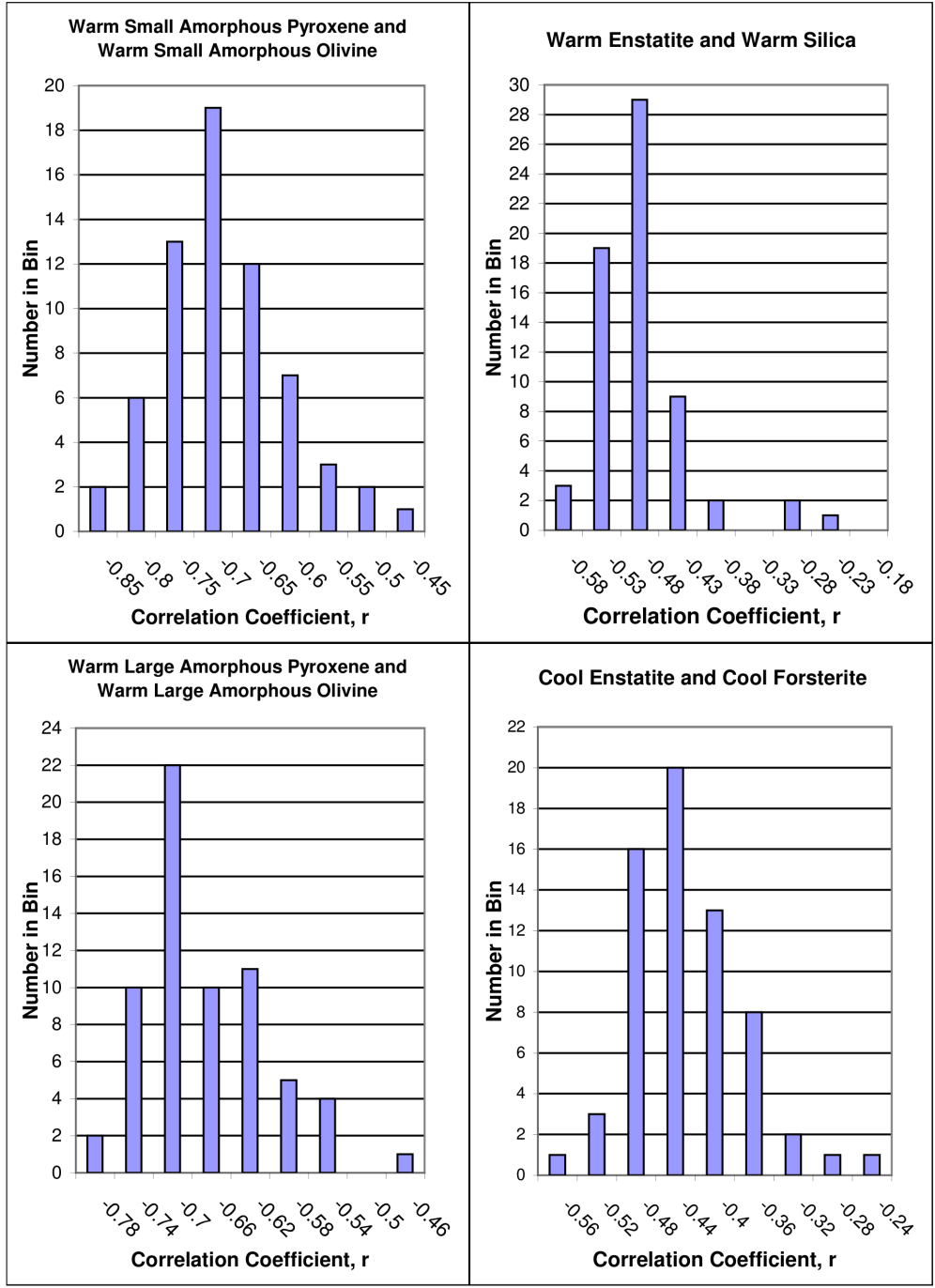}
  \caption[Dust Model Component Pair Correlation Coefficient Histograms]{Histograms 
of the correlation coefficients obtained for all 65 
spectra of Taurus-Auriga T Tauri stars analyzed in this study, for four 
representative pairs of dust components.  Varying degrees of degeneracy, 
or anticorrelation, of components are seen, with the most degeneracy 
coming from pairs of amorphous silicate grains at the same temperature.}
\end{figure}

\clearpage

\begin{figure}[t] 
  \epsscale{0.8}
  \plotone{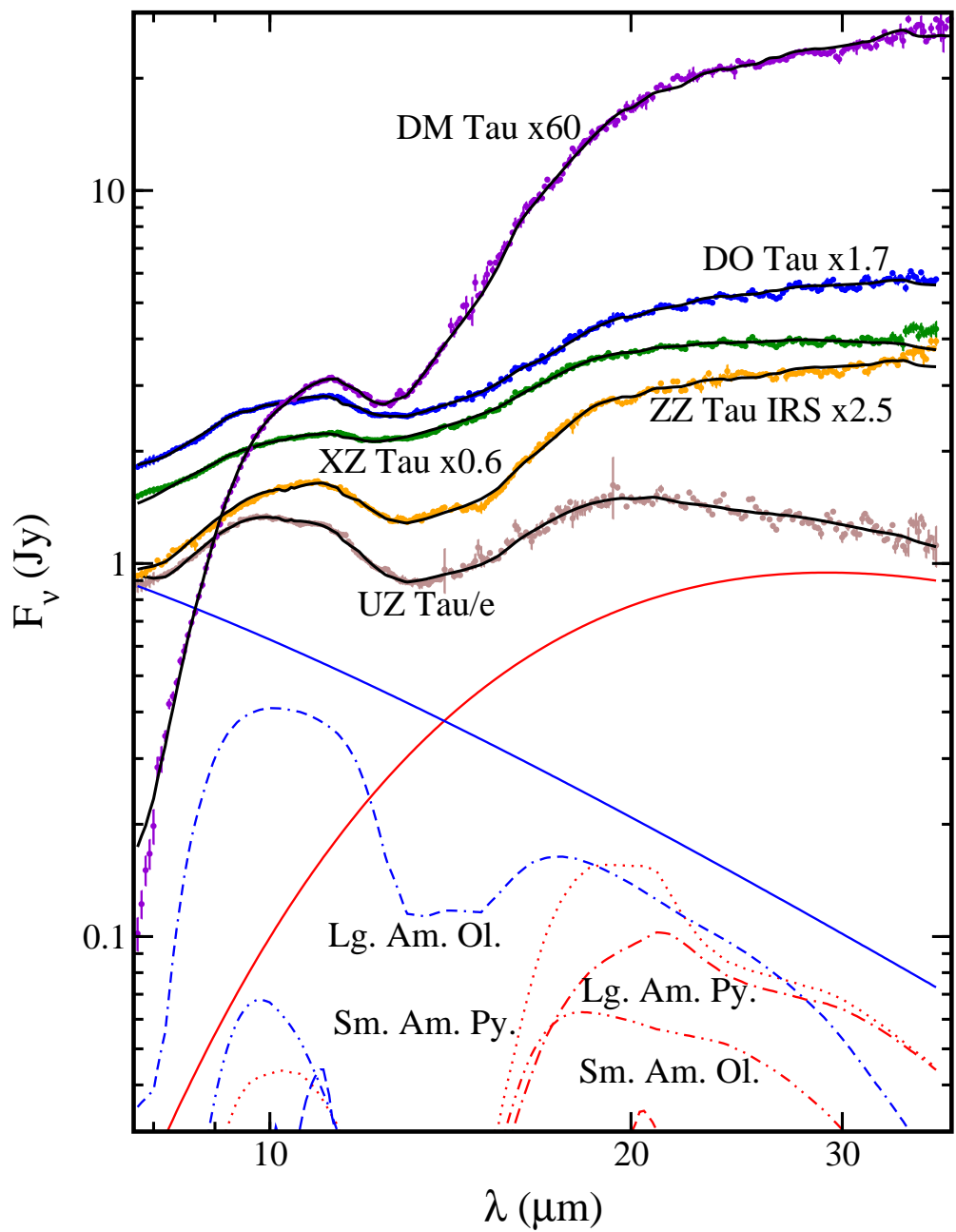}
  \caption{Spectra indicating substantial inner disk grain growth.  Model 
components for model of UZ Tau/e are at bottom; blue lines represent model 
components at higher temperature, red lines represent model components at lower 
temperature.  Solid lines are blackbodies, dotted lines are small (submicron) 
amorphous pyroxene, dash-double-dotted lines are small (submicron) amorphous 
olivine, dot-long-dash lines are large amorphous pyroxene, dot-double-dash lines 
are large amorphous olivine, dot-short-dash lines are enstatite, long-dashed 
lines are forsterite, and short-dashed lines are silica.}
\end{figure}

\clearpage

\begin{figure}[t] 
  \epsscale{0.8}
  \plotone{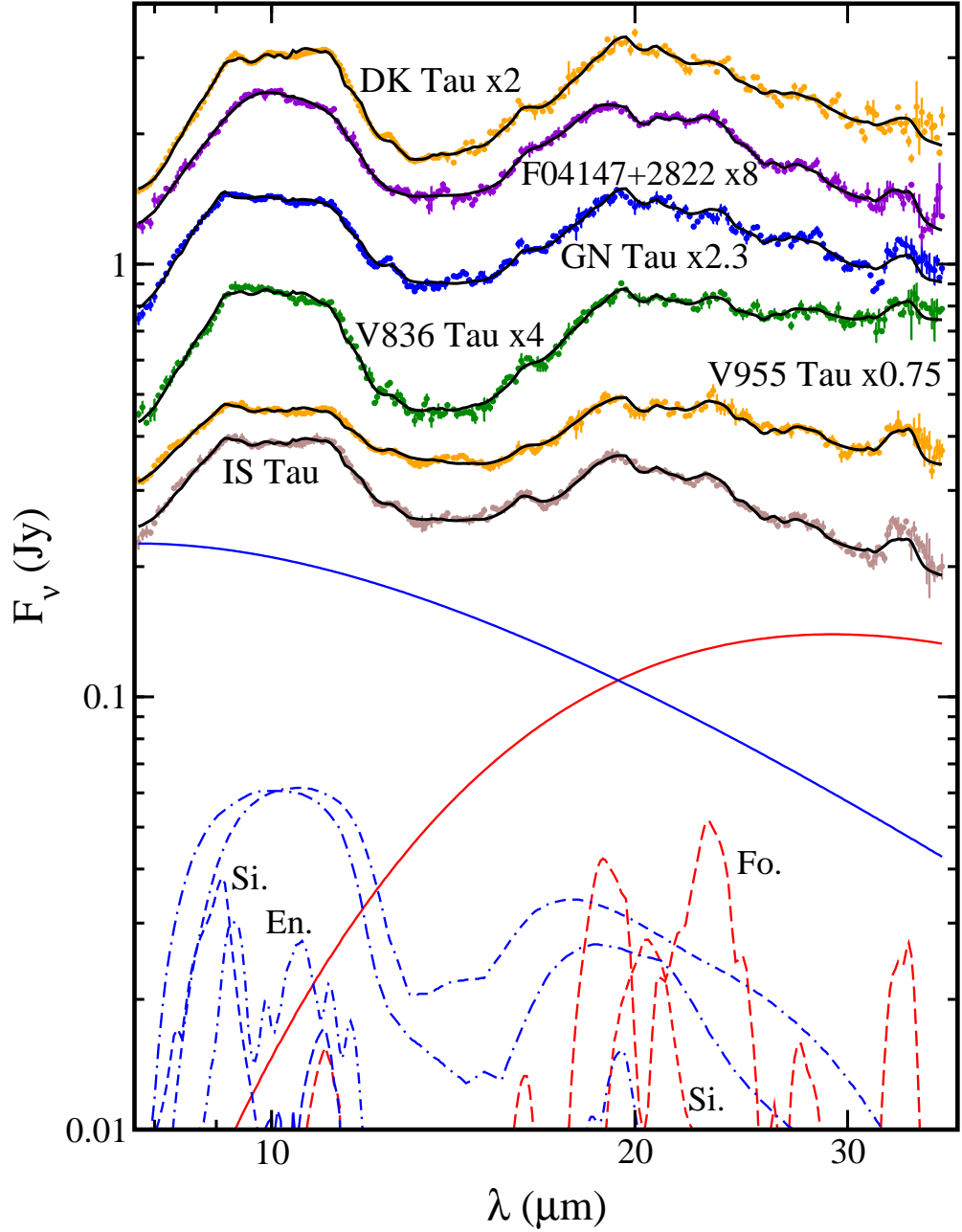}
  \caption{Spectra with prominent forsterite features.  Model components are shown
for IS Tau, same style and color convention as Figure 4.}
\end{figure}

\clearpage

\begin{figure}[t] 
  \epsscale{0.8}
  \plotone{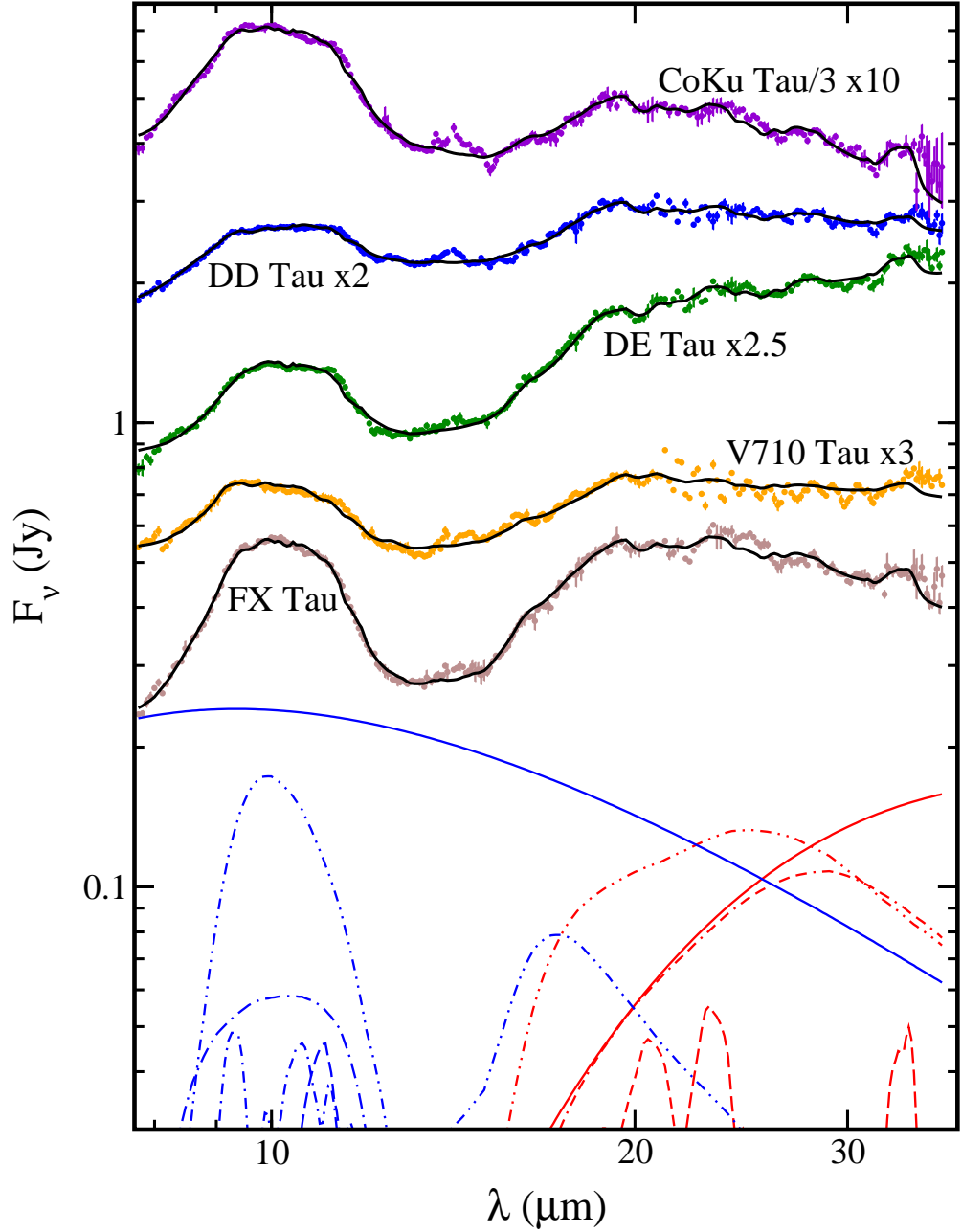}
  \caption{Spectra indicating forsterite, but not fit well by our forsterite
opacity.  Model components are shown for FX Tau, same style and color convention
as Figure 4.}
\end{figure}

\clearpage

\begin{figure}[t] 
  \epsscale{0.8}
  \plotone{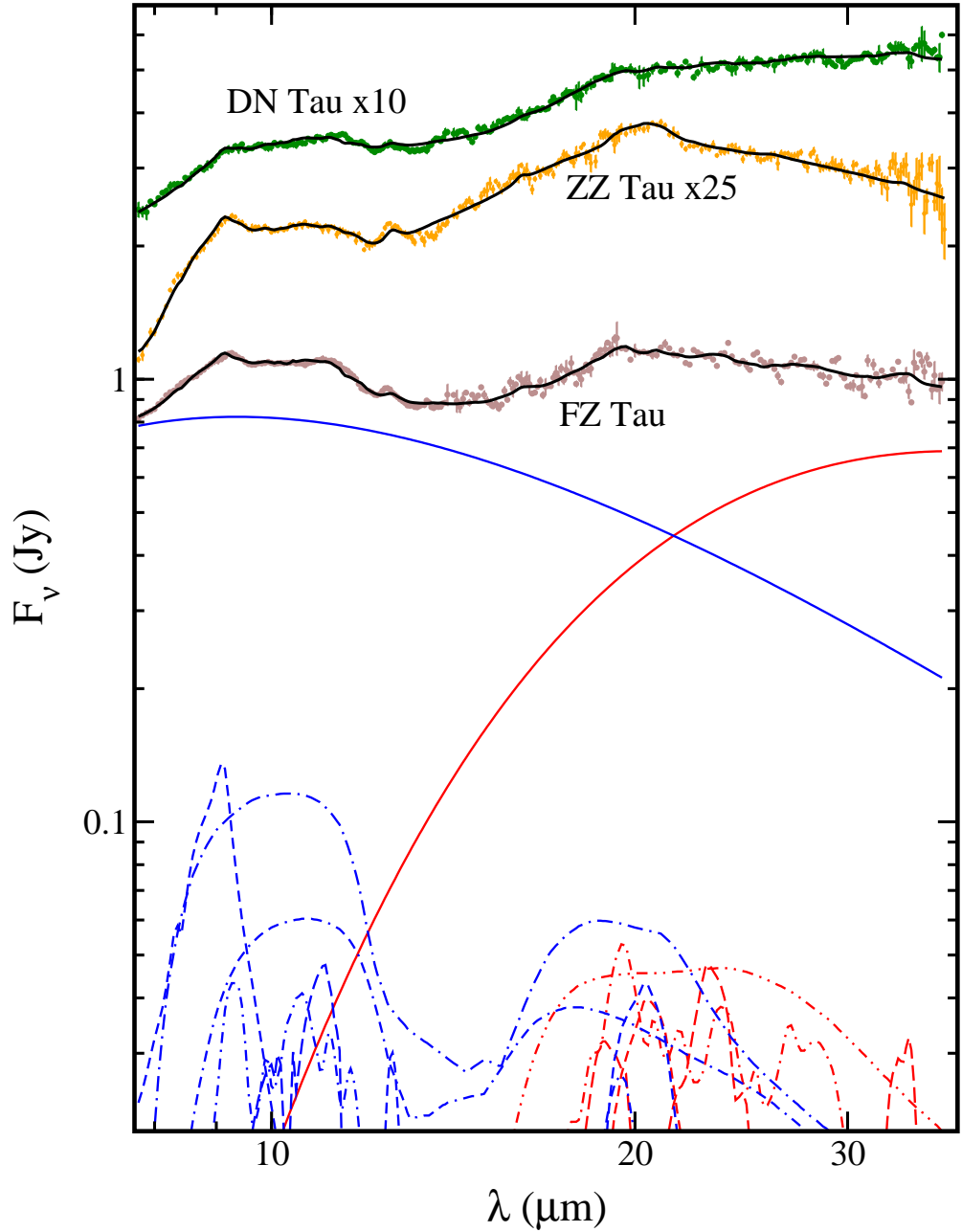}
  \caption{Spectra of DN Tau, ZZ Tau, and FZ Tau.  DN Tau and FZ Tau require silica
but are not fit well by our silica opacity, while ZZ Tau requires silica and is
fairly fit well by our silica opacity.  Model components are shown for FZ Tau,
same style and color convention as Figure 4.}
\end{figure}

\clearpage

\begin{figure}[t] 
  \epsscale{0.8}
  \plotone{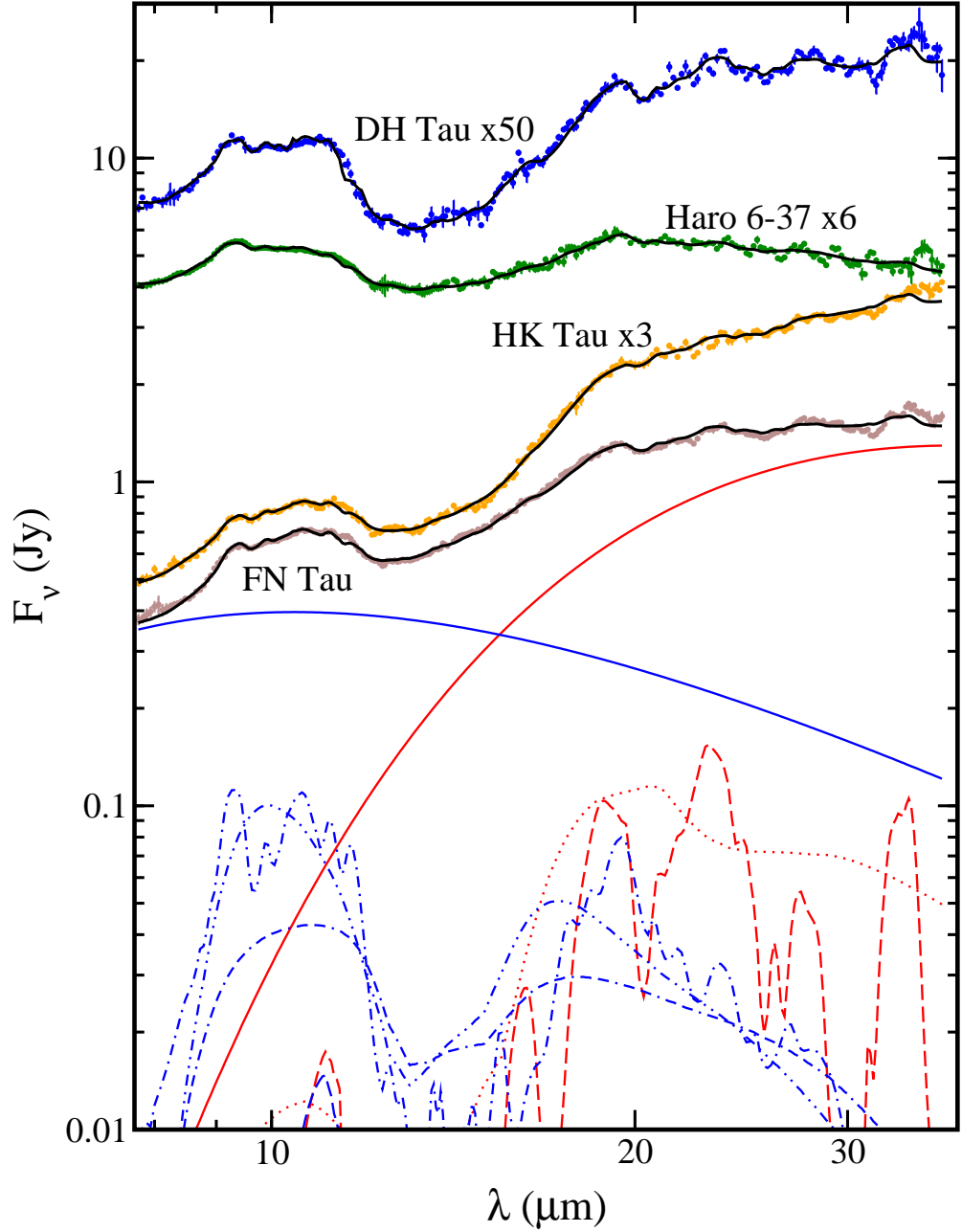}
  \caption{Spectra with prominent enstatite features.  Model components are shown for
FN Tau, same style and color convention as Figure 4.}
\end{figure}

\clearpage

\begin{figure}[t] 
  \epsscale{0.8}
  \plotone{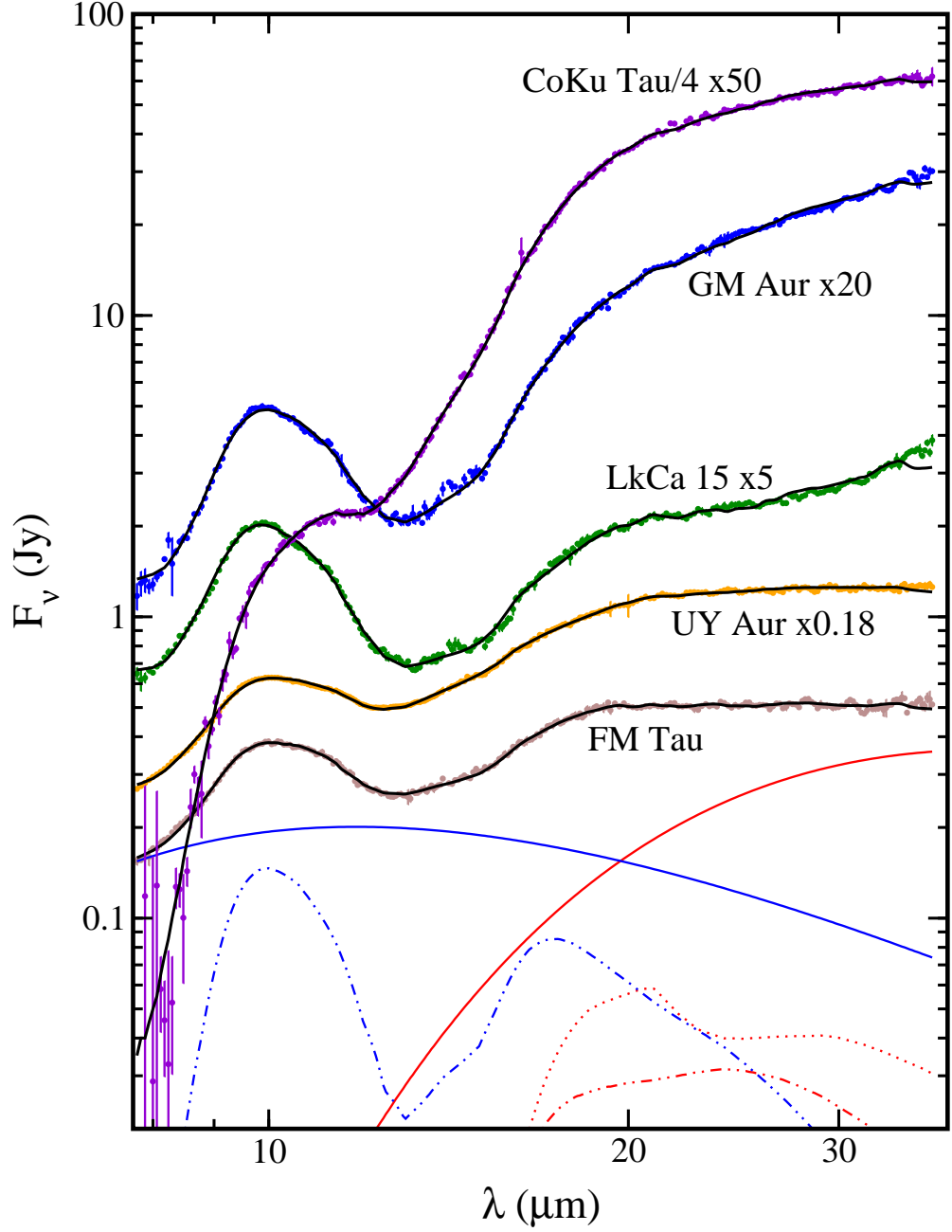}
  \caption{Spectra whose models require high abundances of submicron amorphous
silicates.  Model components are shown for FM Tau, same style and color convention
as Figure 4.}
\end{figure}

\clearpage

\begin{figure}[t] 
  \epsscale{0.8}
  \plotone{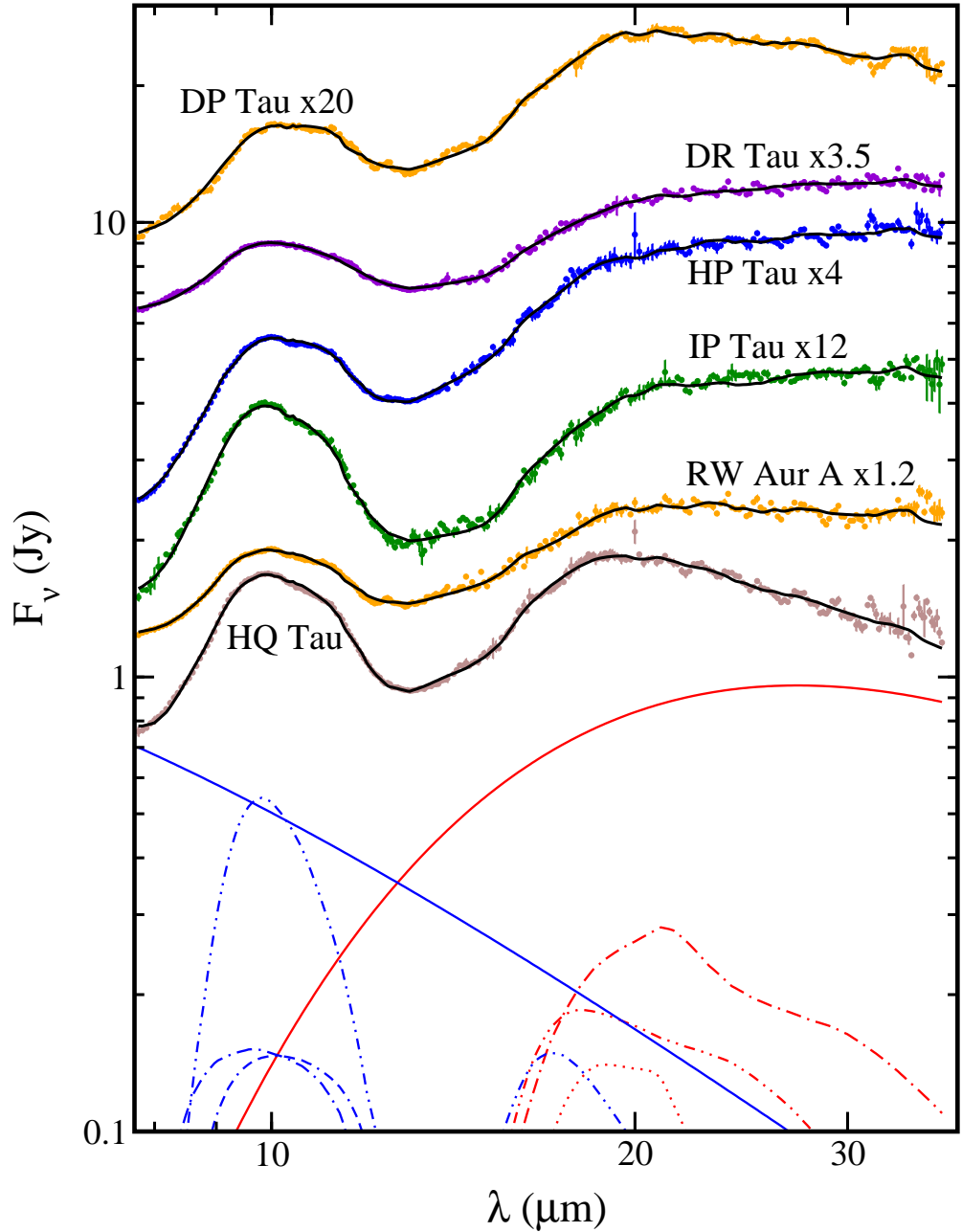}
  \caption{Spectra whose models require fairly high abundances of submicron
amorphous silicates, though not as high as those in Figure 9.  Model components
are shown for HQ Tau, same style and color convention as Figure 4.}
\end{figure}

\clearpage

\begin{figure}[t] 
  \epsscale{0.8}
  \plotone{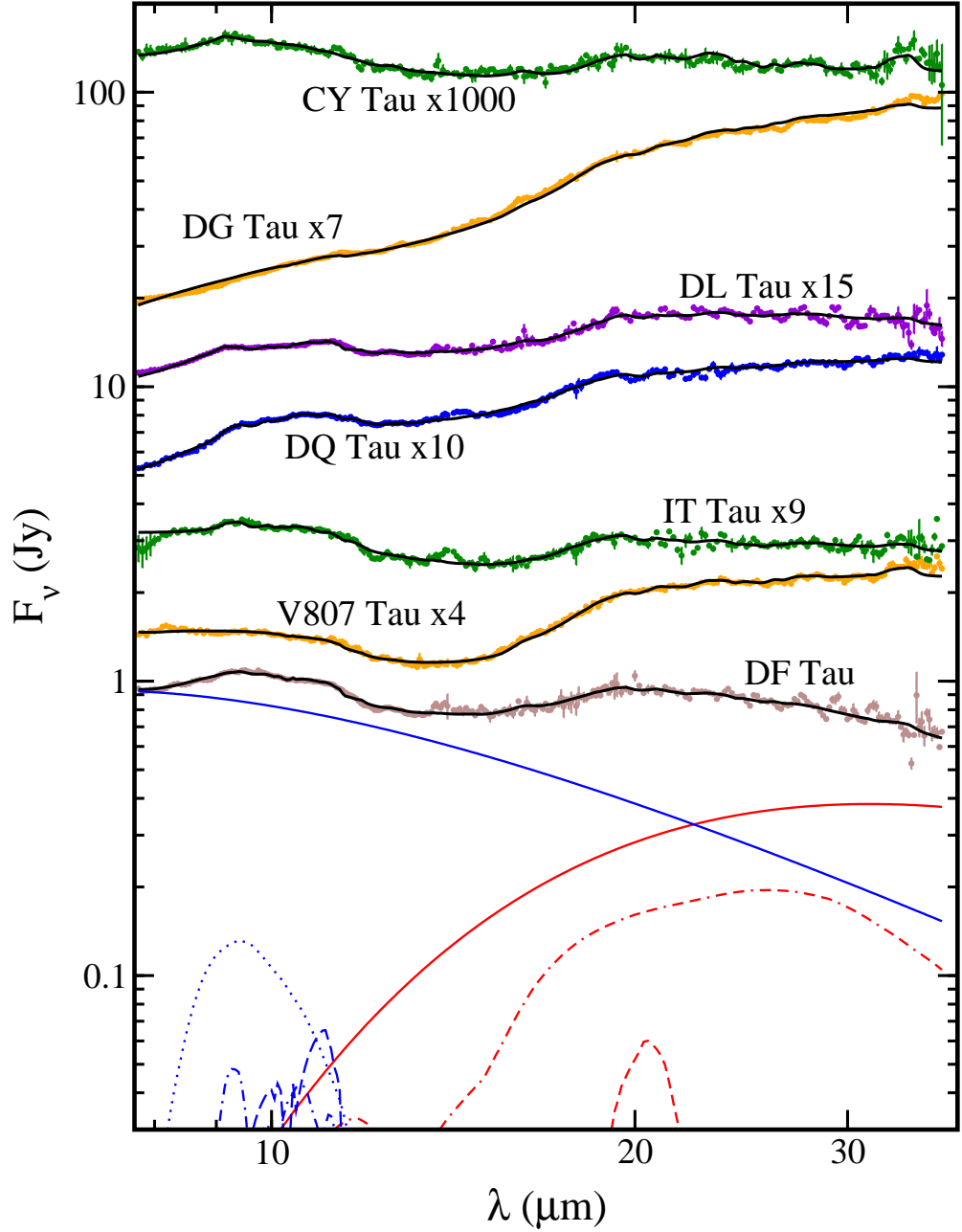}
  \caption{Spectra with low equivalent width 10 $\mum$ complexes.  Model components
are shown for DF Tau, same style and color convention as Figure 4.}
\end{figure}

\clearpage

\begin{figure}[t] 
  \epsscale{0.8}
  \plotone{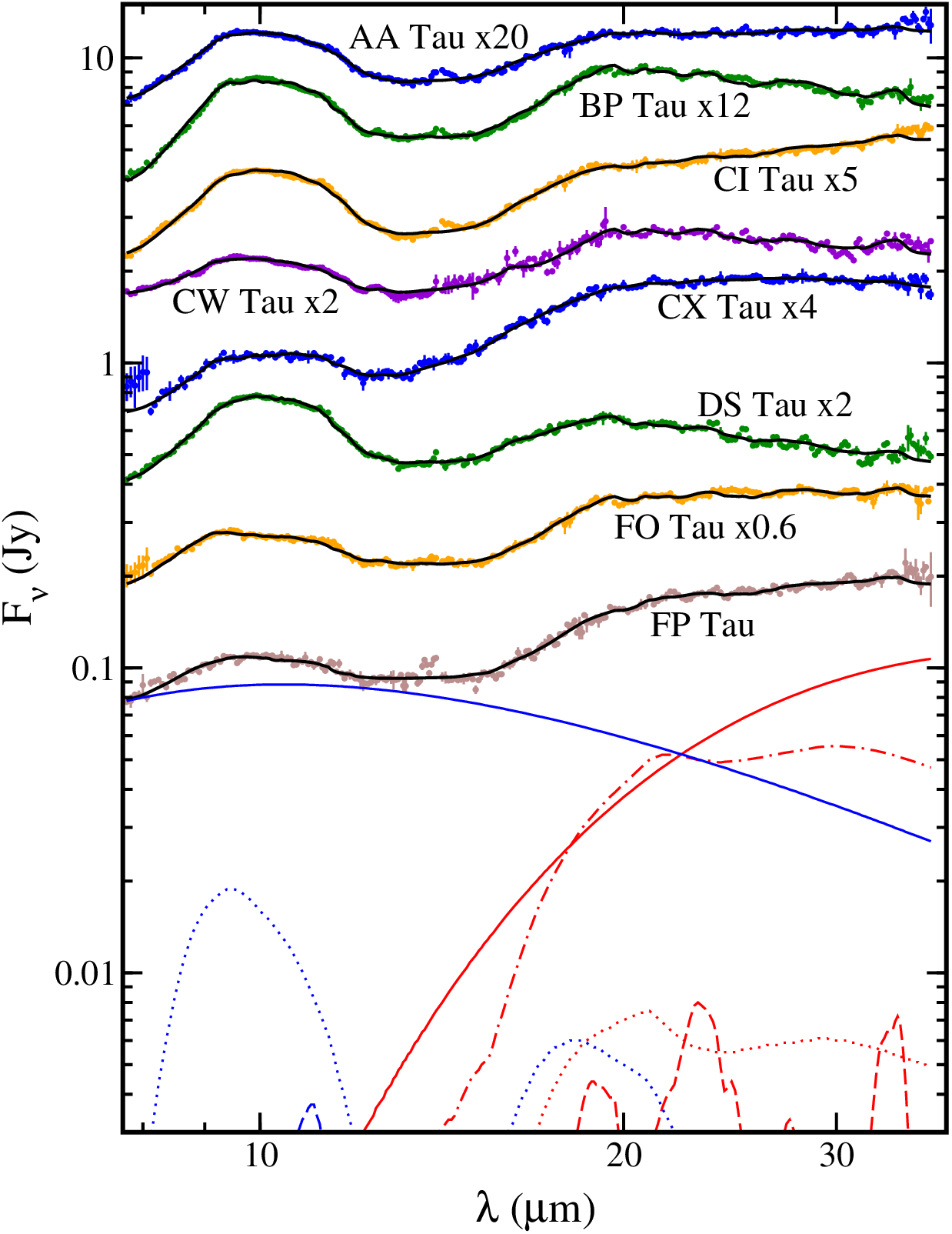}
  \caption{Spectra whose models indicate mixed compositions, with no dust component
clearly dominant over another.  Model components are shown for FP Tau, same style
and color convention as Figure 4.}
\end{figure}

\clearpage

\begin{figure}[t] 
  \epsscale{0.8}
  \plotone{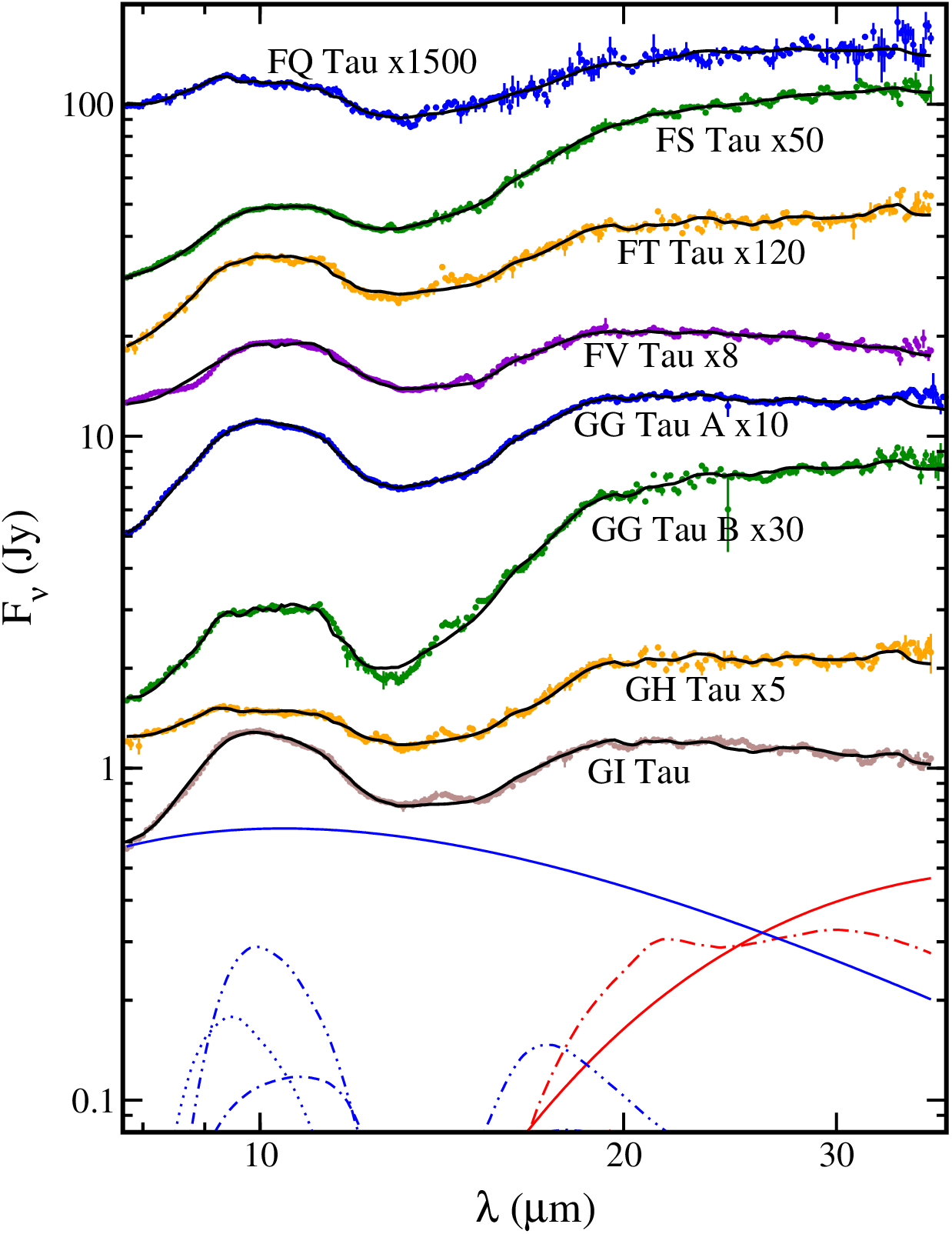}
  \caption{Spectra whose models indicate mixed compositions, with no dust component
clearly dominant over another.  Model components are shown for GI Tau, same style
and color convention as Figure 4.}
\end{figure}

\clearpage

\begin{figure}[t] 
  \epsscale{0.8}
  \plotone{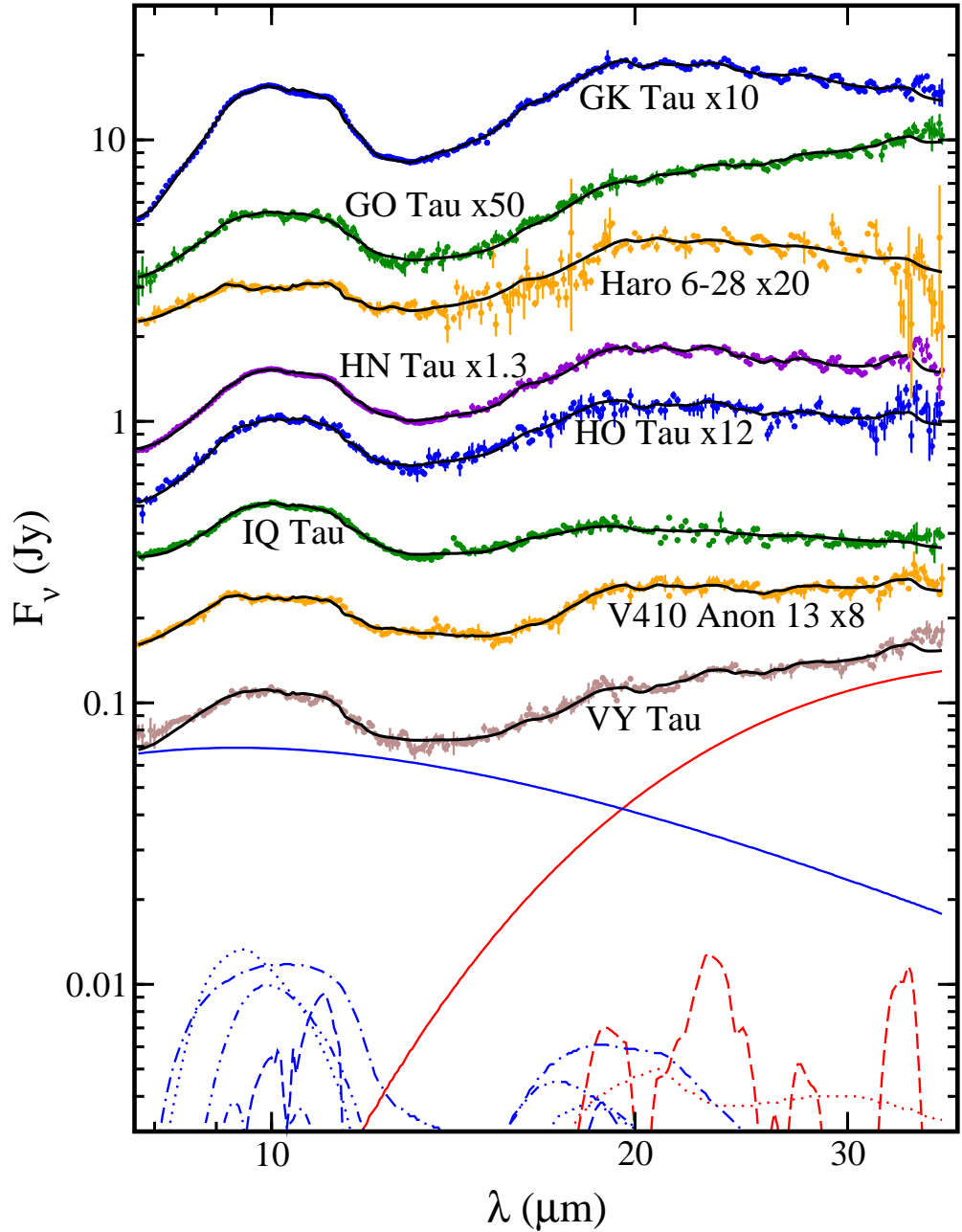}
  \caption{Spectra whose models indicate mixed compositions, with no dust component
clearly dominant over another.  Model components are shown for VY Tau, same style
and color convention as Figure 4.}
\end{figure}

\clearpage

\begin{figure}[t] 
  \epsscale{0.8}
  \plotone{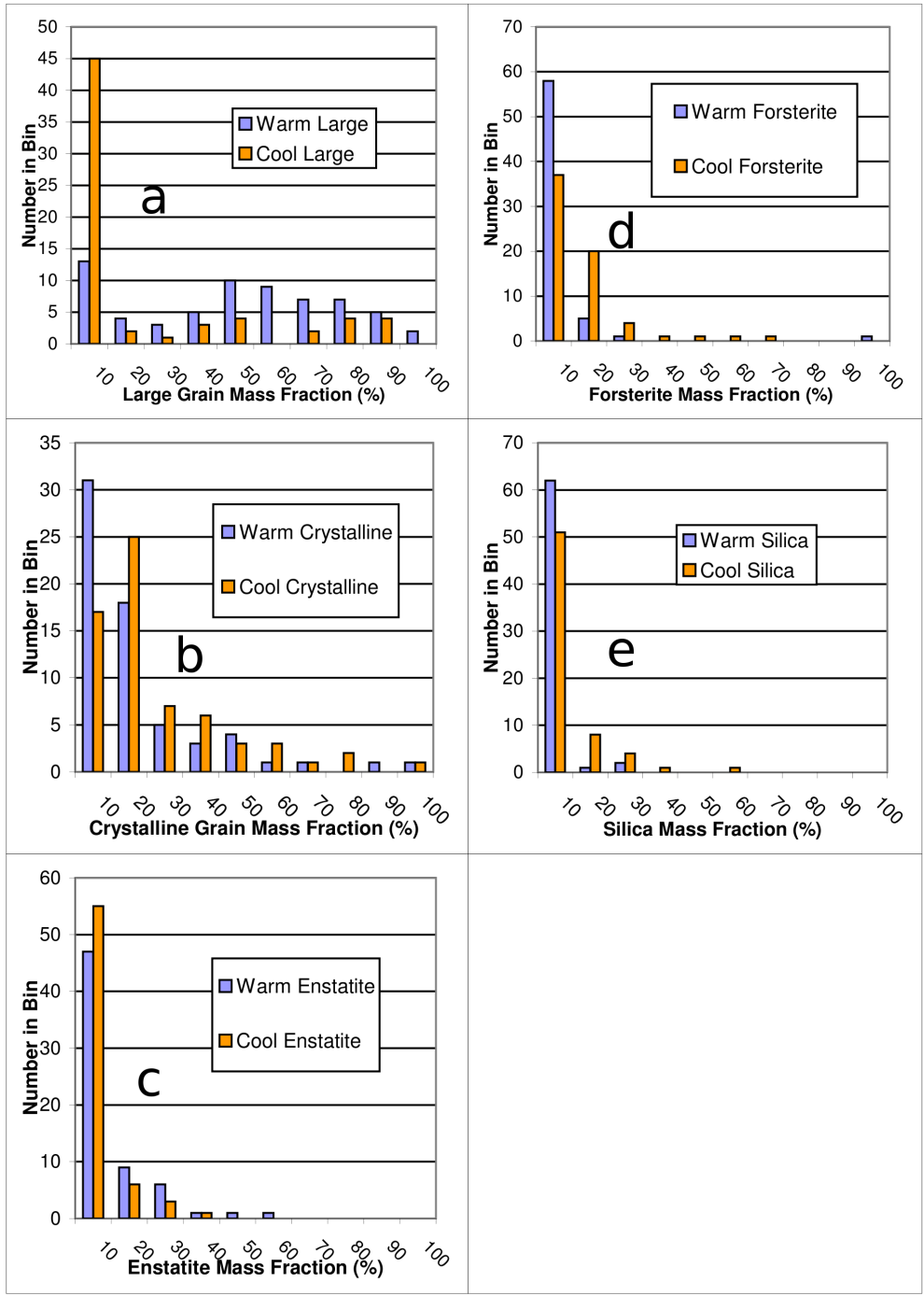}
  \caption[Histograms of Warm and Cool Dust Abundances]{Warm versus cool dust 
histograms.  In the upper left is the large 
grain mass fraction histogram.  In the middle left is the crystalline grain 
mass fraction histogram.  In the lower left is the enstatite grain mass 
fraction histogram.  In the upper right is the forsterite grain mass fraction 
histogram.  In the lower right is the silica grain mass fraction histogram.  
The number of models requiring warm grain mass fractions within a given bin are 
represented by the height of the blue rectangle to the left within that bin.  
Orange rectangles to the right are for cool grain mass fractions.}
\end{figure}

\clearpage

\begin{figure}[t] 
  \epsscale{0.8}
  \plotone{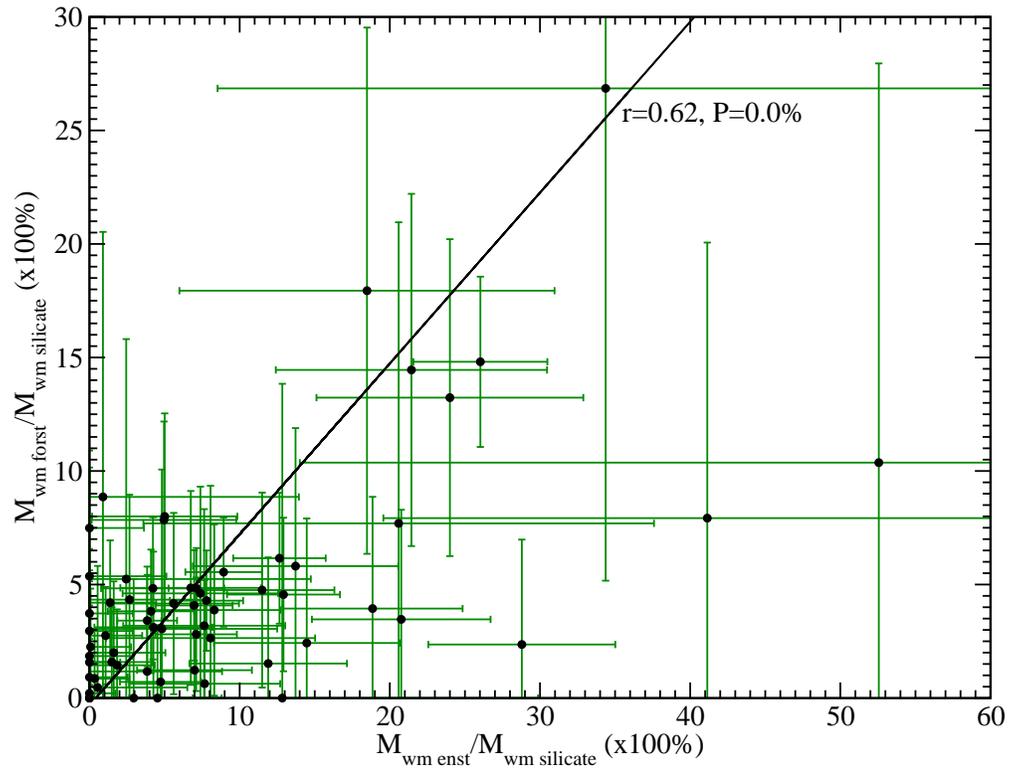}
  \caption{Warm forsterite mass fraction versus warm enstatite mass fraction trend
plot.  The thick solid line running through the points is the trendline consistent
with the computed correlation coefficient, indicated on the plot.}
\end{figure}

\clearpage

\begin{figure}[t] 
  \epsscale{0.8}
  \plotone{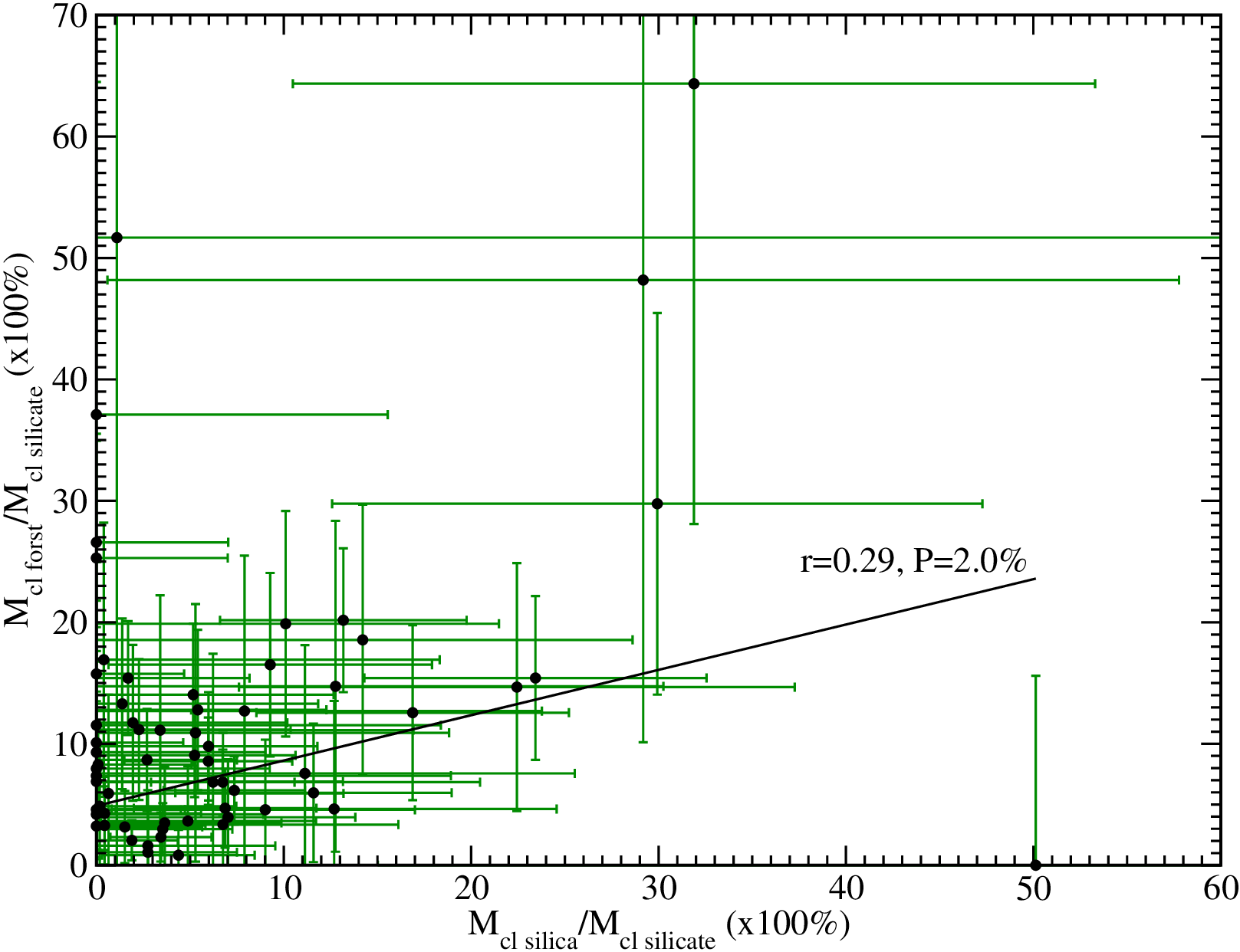}
  \caption{Cool forsterite mass fraction versus cool silica mass fraction trend
plot.  The thick solid line running through the points is the trendline consistent
with the computed correlation coefficient, indicated on the plot.}
\end{figure}

\clearpage

\begin{figure}[t] 
  \epsscale{0.8}
  \plotone{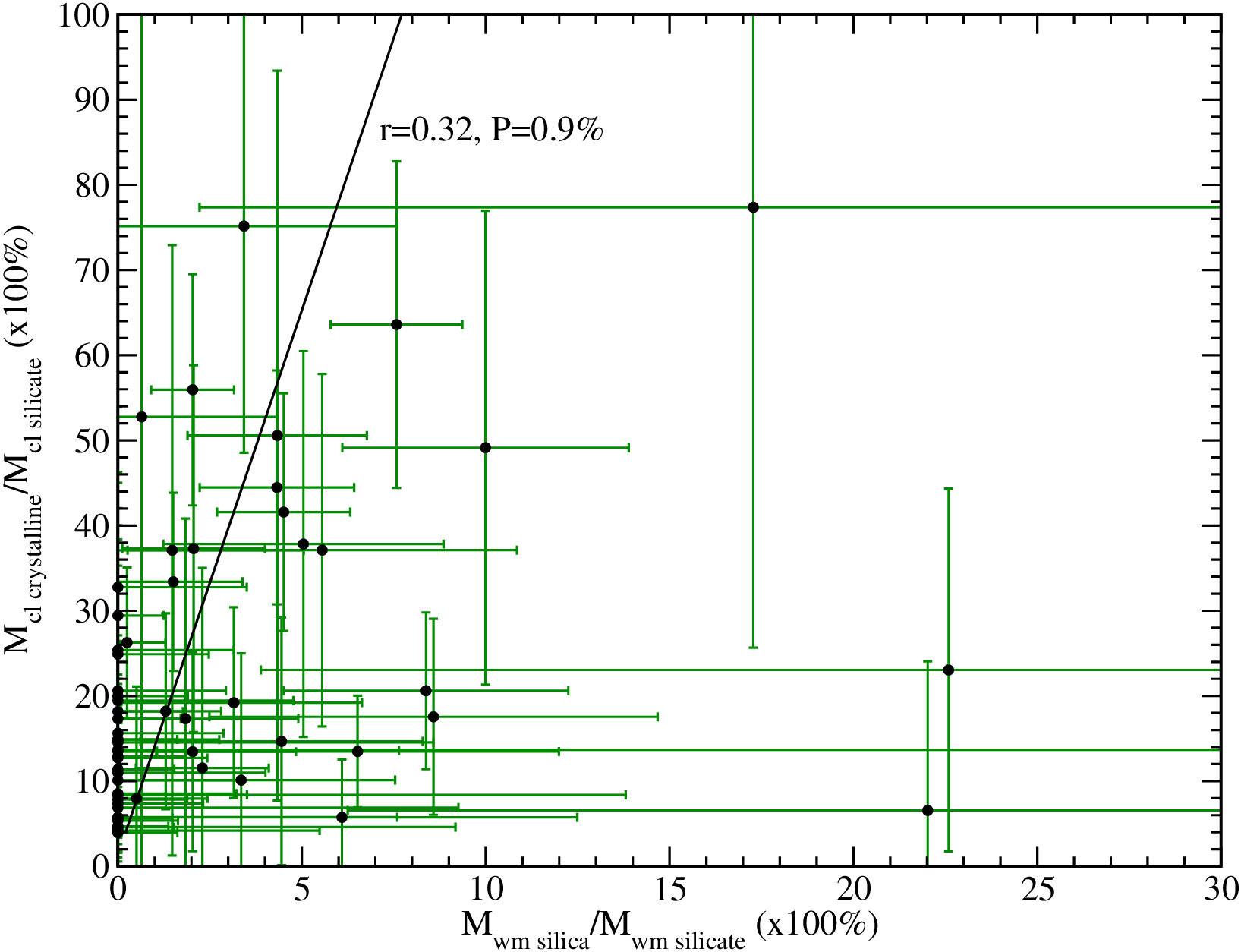}
  \caption{Cool crystalline mass fraction versus warm silica mass fraction trend
plot.  The thick solid line running through the points is the trendline consistent
with the computed correlation coefficient, indicated on the plot.}
\end{figure}

\clearpage

\begin{figure}[t] 
  \epsscale{0.8}
  \plotone{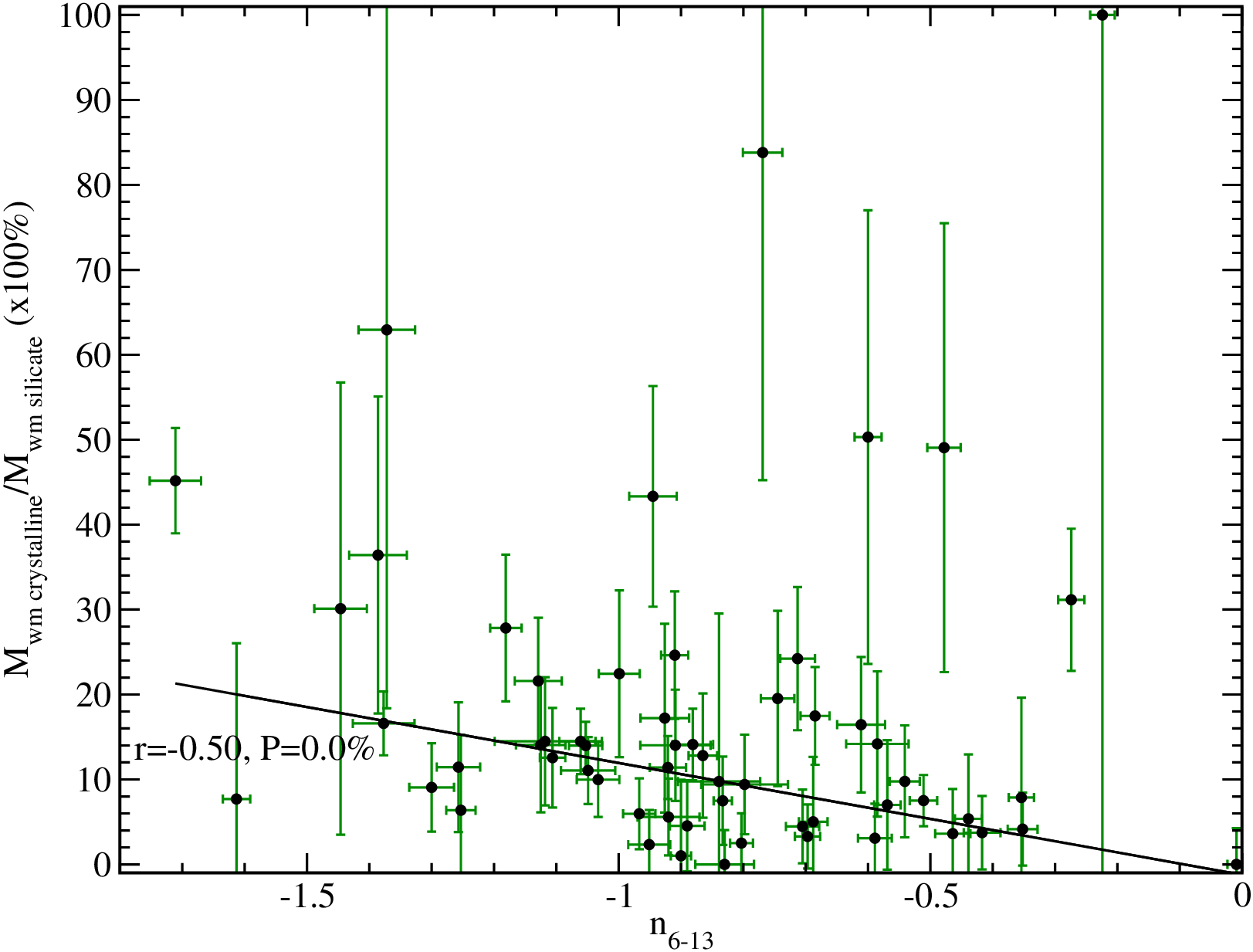}
  \caption{Warm crystalline mass fraction versus n$_{6-13}$ trend plot.  The thick
solid line running through the points is the trendline consistent with the
computed correlation coefficient, indicated on the plot.}
\end{figure}

\clearpage

\begin{figure}[t] 
  \epsscale{0.8}
  \plotone{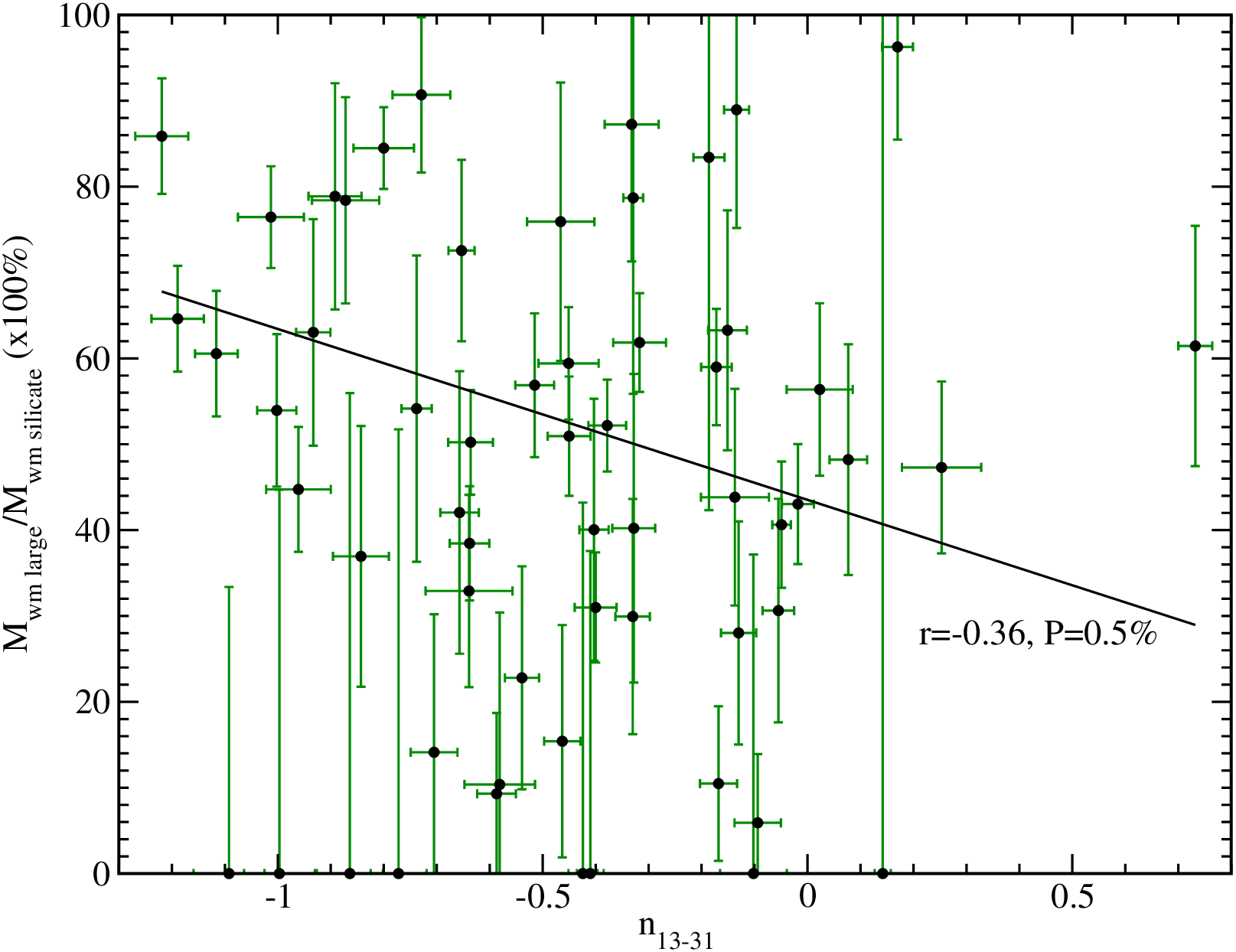}
  \caption{Cool crystalline mass fraction versus n$_{13-31}$ trend plot.  The thick
solid line running through the points is the trendline consistent with the
computed correlation coefficient, indicated on the plot.}
\end{figure}

\clearpage

\begin{figure}[t] 
  \epsscale{0.8}
  \plotone{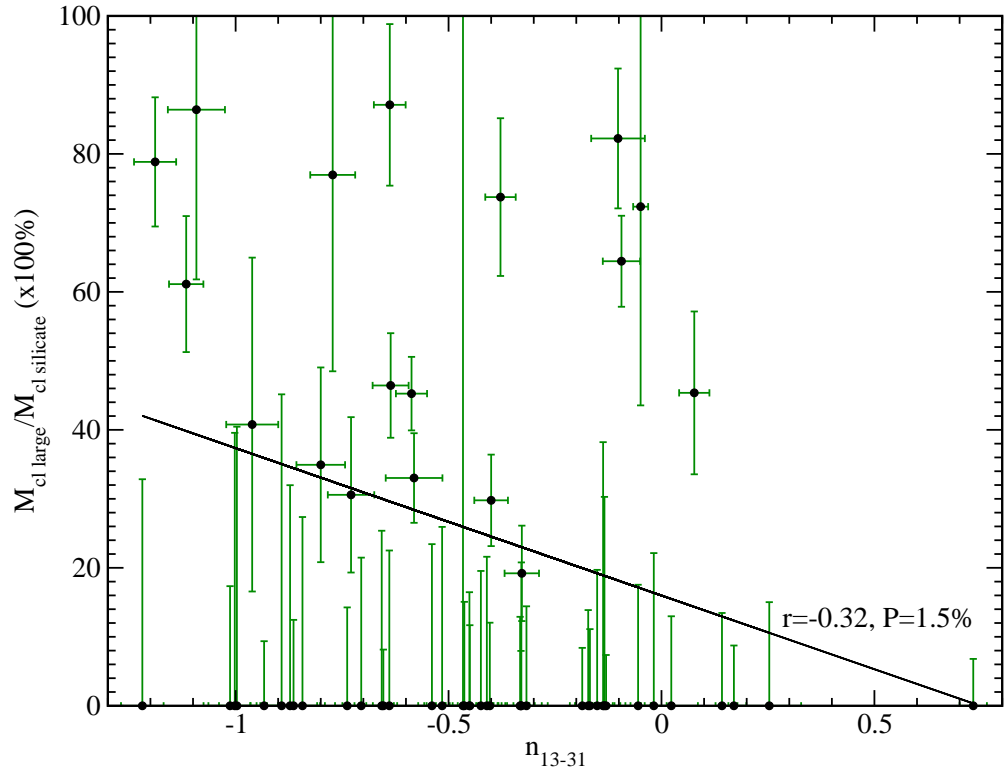}
  \caption{Warm large grain mass fraction versus n$_{13-31}$ trend plot.  The thick
solid line running through the points is the trendline consistent with the
computed correlation coefficient, indicated on the plot.}
\end{figure}

\clearpage

\begin{figure}[t] 
  \epsscale{0.8}
  \plotone{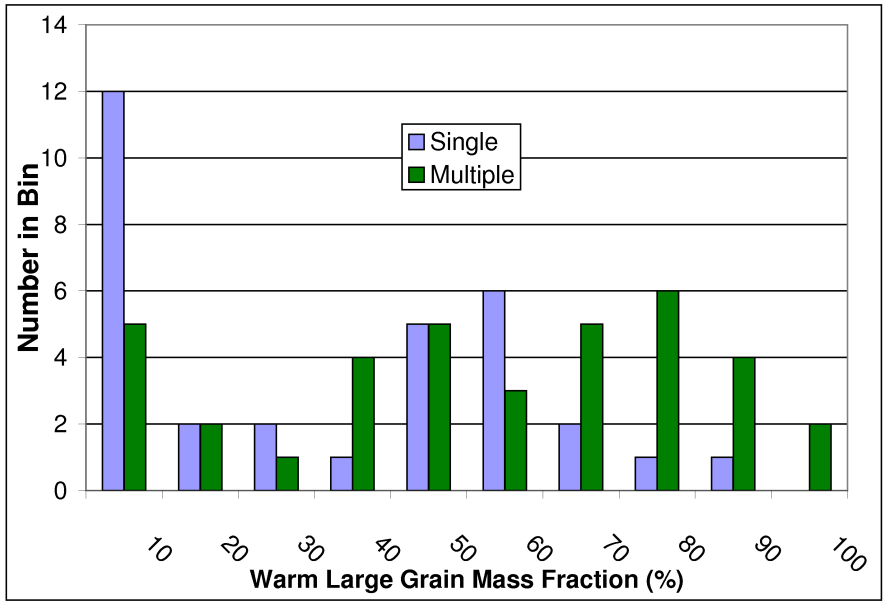}
  \caption{Warm large grain mass fraction histograms for single systems (blue) and 
multiple systems (green).}
\end{figure}

\clearpage

\begin{figure}[t] 
  \epsscale{0.8}
  \plotone{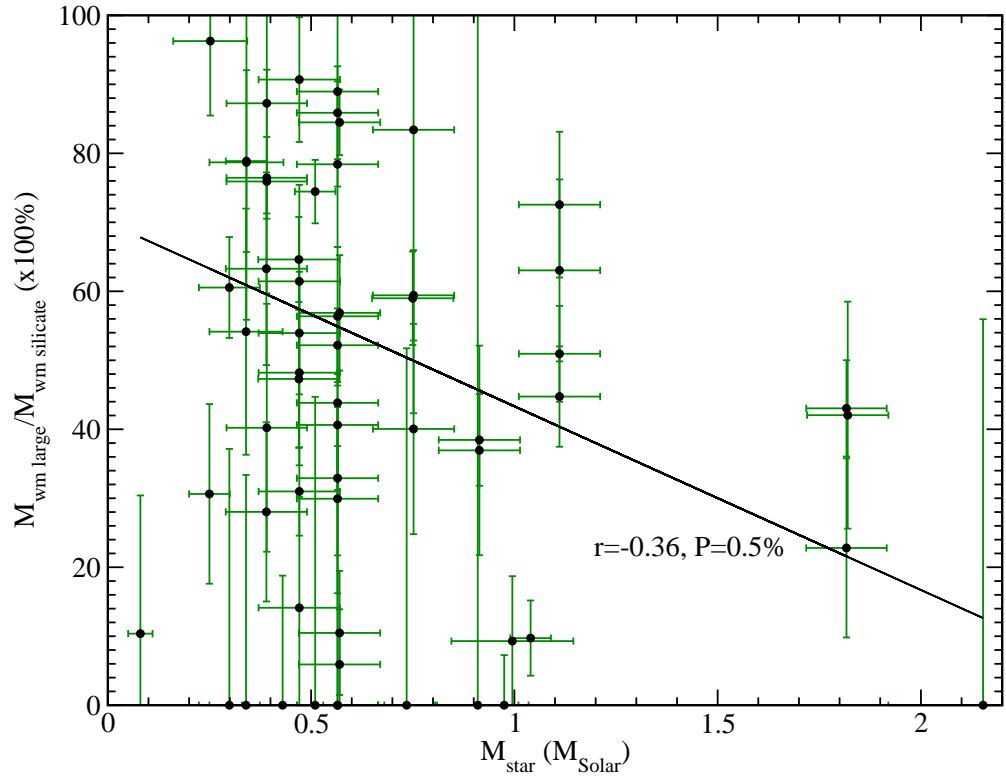}
  \caption{Warm large grain mass fraction versus stellar mass trend plot.  The thick
solid line running through the points is the trendline consistent with the
computed correlation coefficient, indicated on the plot.}
\end{figure}

\clearpage

\begin{figure}[t] 
  \epsscale{0.8}
  \plotone{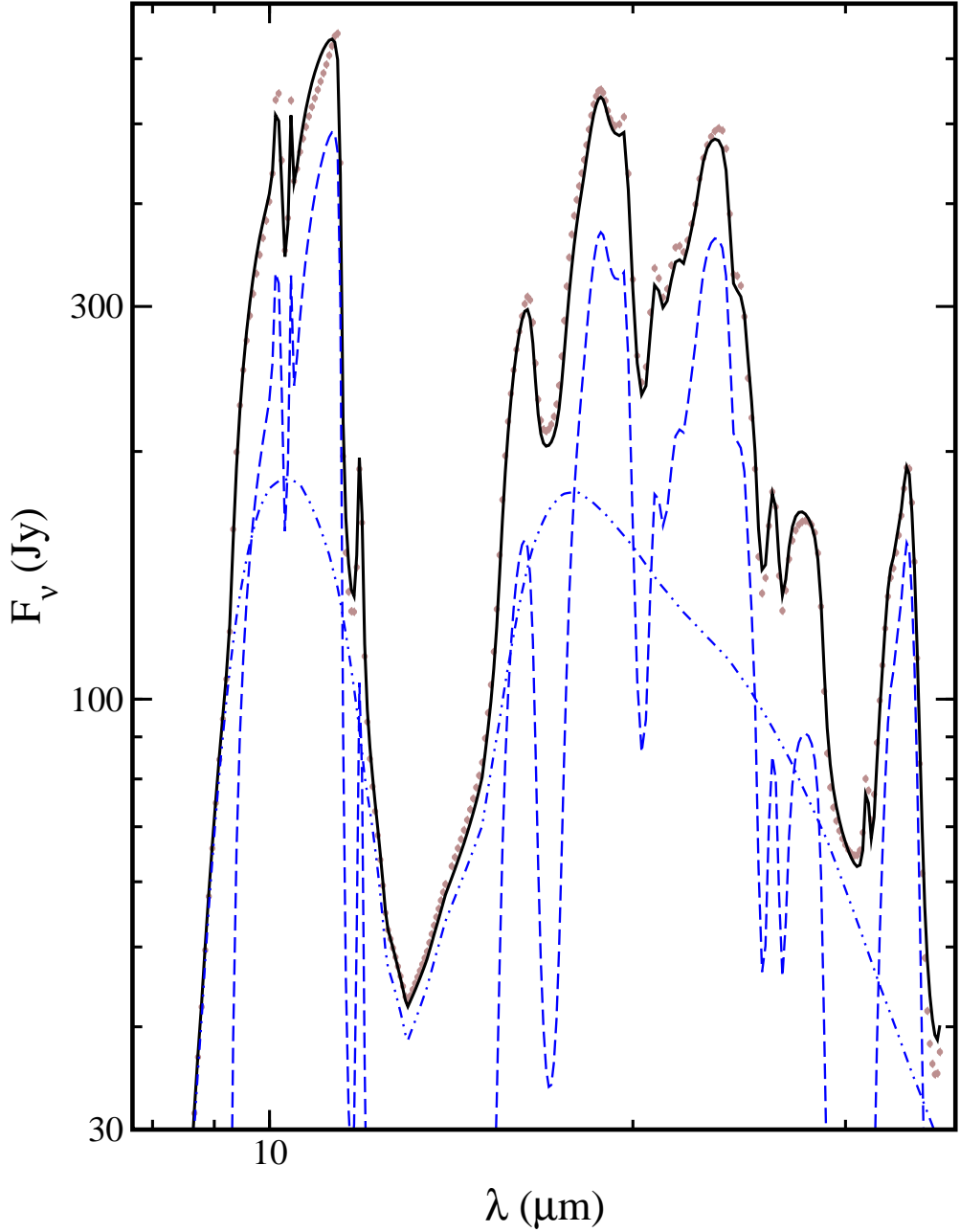}
  \figurenum{A1}
  \caption{Model of Rayleigh-limit size porous aggregate grain composed of amorphous
olivine and forsterite.  Style and color convention of model components same as
for Figure 4.}
\end{figure}

\clearpage

\begin{figure}[t] 
  \epsscale{0.8}
  \plotone{f25.eps}
  \figurenum{A2}
  \caption{Model of 5 $\mum$ radius porous aggregate grain composed of amorphous
olivine and forsterite.  Style and color convention of model components same as
for Figure 4.}
\end{figure}

\clearpage

\begin{figure}[t] 
  \epsscale{0.8}
  \plotone{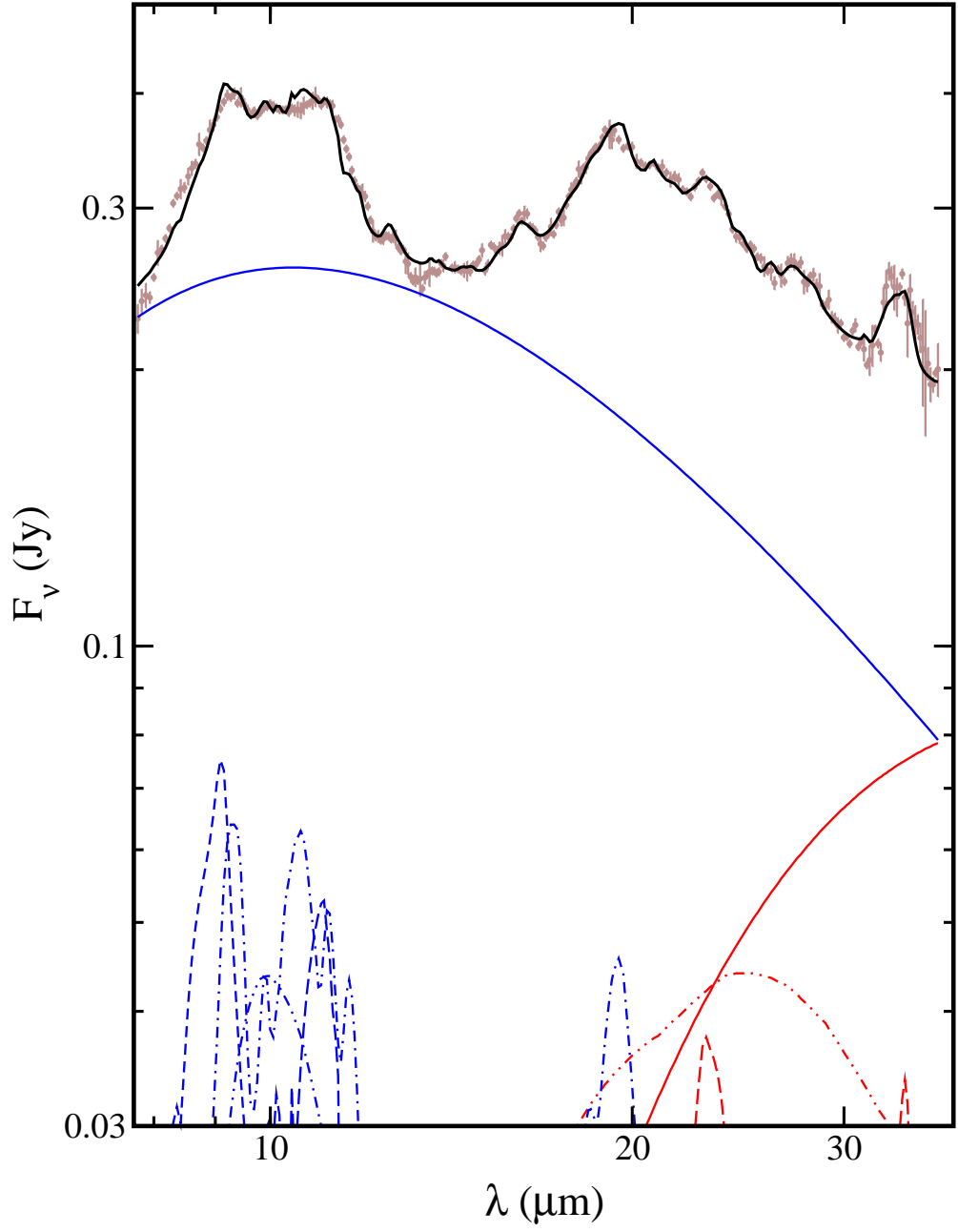}
  \figurenum{A3}
  \caption{Model of IS Tau, without using large grains at either model dust
temperature.  Reduced $\chi^{2}$ is 4.9, which is 1.6 higher than the 3.3 computed
for the model of IS Tau using large grains (see Fig. 5).  Style and color convention 
of model components same as for Figure 4.}
\end{figure}

\clearpage

\begin{figure}[t] 
  \epsscale{0.8}
  \plotone{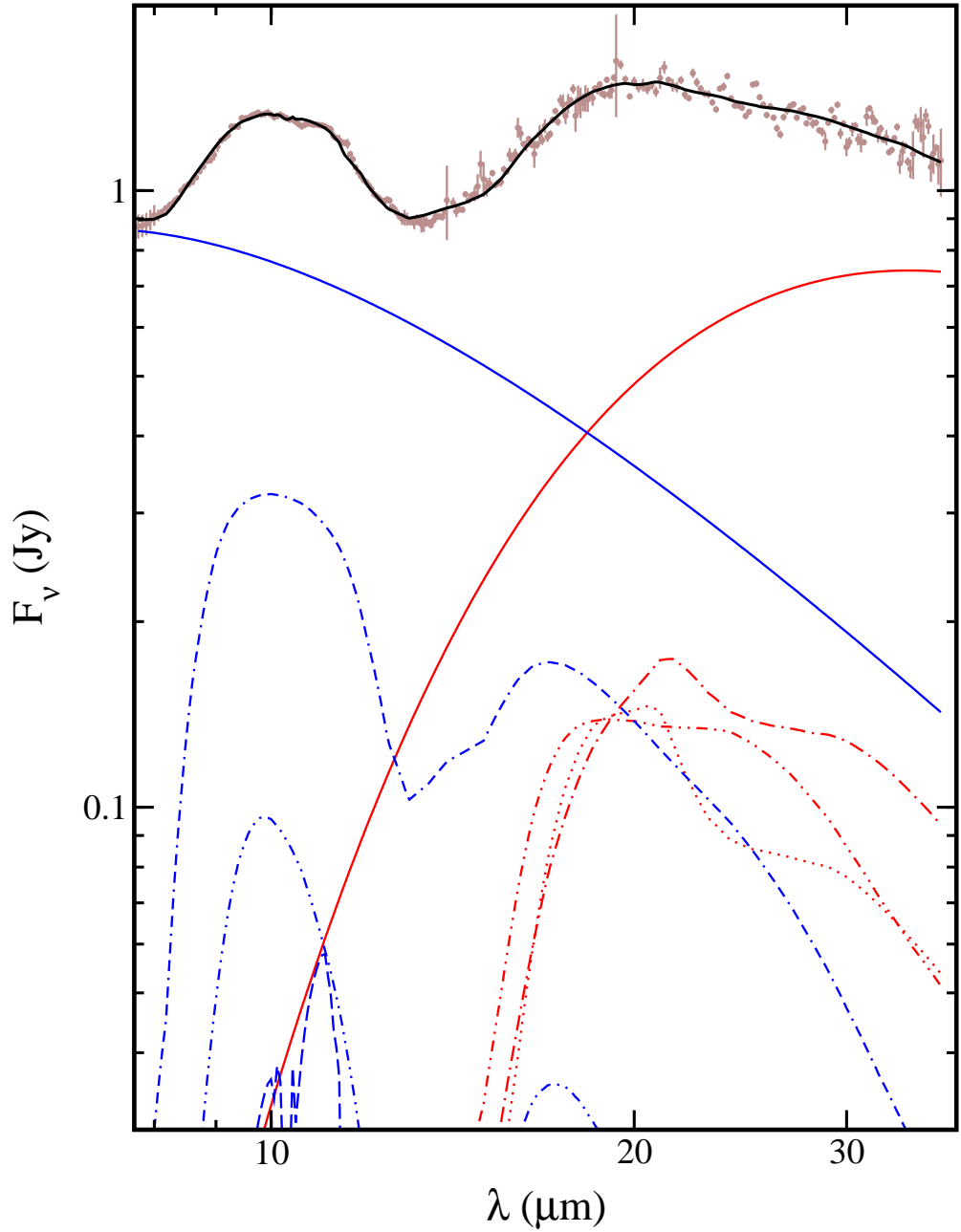}
  \figurenum{A4}
  \caption{Model of UZ Tau/e, replacing 5 $\mum$ radius 60\% vacuum porous grains
of amorphous olivine with 20 $\mum$ radius 88\% vacuum porous grains of amorphous
olivine.  Style and color convention of model components same as for Figure 4.}
\end{figure}

\clearpage

\begin{table}[h,t]
{\tiny
\caption[Significance of Opacities for Forsterite Exemplars]{Significance 
(mass/uncertainty) of Opacities for Forsterite Exemplars 
\label{table1v}}
\begin{tabular}{lcccccccl}
\hline \hline
          Opacity
          & DK Tau
          & GN Tau
          & HBC 656
          & IS Tau
          & ROXs 42C
          & V836 Tau
          & V955 Tau
          & \\
\hline
CfoJ98 & 0 & 4.3 & 5.0 & 1.4 & 0.7 & 0 & 0 & $\longleftarrow$\\
ColJ98 & 4.5 & 0 & 0 & 0 & 0 & 0 & 0 & \\
ChoJ98 & 0 & 0 & 0 & 0 & 0 & 0 & 0.1 & \\
CfaJ98 & 0 & 0 & 0 & 0 & 0.9 & 0 & 0 & \\
CFo91.4 & 0 & 0 & 0 & 0 & 0 & 0 & 0 & \\
CFo89.3 & 0 & 0 & 0 & 0 & 0 & 1.6 & 0 & \\
CFo86.5 & 0 & 0 & 0 & 0 & 0 & 0 & 1.8 & \\
CFo84.4 & 0.4 & 0 & 0 & 0 & 0 & 0 & 0 & \\
CFo80 & 0 & 0 & 0 & 0 & 0 & 0 & 0 & \\
CFo21.8 & 0 & 0 & 0.6 & 0 & 0 & 0 & 0 & \\
Cannfor & 0 & 0 & 0 & 0 & 0 & 0 & 0 & \\
Cfohyb & 0 & 0 & 1.8 & 1.0 & 0 & 0 & 0.3 & $\longleftarrow$\\
CS6CDE & 0 & 0 & 0.9 & 0.1 & 0 & 0 & 0.8 & $\Longleftarrow$\\
CS6CDE2 & 1.3 & 0 & 2.6 & 2.7 & 3.2 & 1.1 & 2.2 & $\Longleftarrow$\\
WfoJ98 & 0 & 0 & 0 & 0 & 0 & 0 & 0 & \\
WolJ98 & 0 & 0 & 0 & 0 & 0 & 0 & 0 & \\
WhoJ98 & 0 & 0 & 0 & 0 & 0.3 & 0 & 0 & \\
WfaJ98 & 3.1 & 0 & 0.9 & 2.5 & 2.4 & 0.4 & 1.7 & $\longleftarrow$\\
WFo91.4 & 0 & 0 & 0 & 0 & 0 & 0 & 1.2 & \\
WFo89.3 & 0 & 0 & 0 & 0 & 0 & 0 & 0 & \\
WFo86.5 & 0 & 0 & 0 & 1.3 & 0 & 0 & 0 & \\
WFo84.4 & 0.1 & 0 & 0 & 0 & 0 & 0 & 0 & \\
WFo80 & 0 & 1.2 & 1.1 & 0 & 0.6 & 0 & 0 & \\
WFo21.8 & 0 & 0 & 2.1 & 0 & 0 & 0 & 0 & \\
Wannfor & 4.0 & 0 & 2.6 & 1.2 & 0 & 3.3 & 0 & $\longleftarrow$\\
Wfohyb & 0 & 0 & 0 & 0 & 0 & 0.6 & 0 & \\
WS6CDE & 0 & 0.1 & 0.9 & 0 & 0 & 1.2 & 0 & $\Longleftarrow$\\
WS6CDE2 & 0.6 & 0.4 & 2.6 & 0.5 & 0.9 & 0 & 0.4 & $\Longleftarrow$\\

\hline
\end{tabular}
\tablecomments{\footnotesize A given entry in a column is the 
mass weight of the component named in the first entry on the same 
row divided by its uncertainty, giving the significance of a particular 
opacity used in a given model.  The naming convention is as described in Section 
3.2.  Arrows in the final column indicate recurring opacities as discussed in 
the text.  The opacity used most at both temperatures is the CDE2 opacity using 
\citet{sog06} optical constants, with the CDE opacity from the same optical constants 
also frequently included.  Both of these are indicated by double-lined arrows.}
}
\end{table}

\clearpage

\begin{table}[h,t]
{\tiny
\caption[Significance of Opacities for Enstatite and Forsterite Exemplars, 
Jointly]{Significance (mass/uncertainty) of Opacities for Enstatite and Forsterite 
Exemplars, Jointly\label{table2v}}
\begin{tabular}{lcccccccccccl}
\hline \hline
          
          & DH
          & FN
          & Haro
          & HK
          & DK
          & GN
          & HBC
          & IS
          & ROXs
          & V836
          & V955
          & \\
          Opacity
          & Tau
          & Tau
          & 6-37
          & Tau
          & Tau
          & Tau
          & 656
          & Tau
          & 42C
          & Tau
          & Tau
          & \\
\hline
Cen90 & 0 & 0 & 1.6 & 0 & 1.9 & 2.1 & 0 & 0 & 0 & 0.9 & 0 & $\Longleftarrow$\\
Cen80 & 0 & 0 & 0 & 0 & 0 & 0 & 0 & 0 & 0 & 0 & 0 & \\
Cen70 & 0 & 0 & 0 & 0 & 0 & 0 & 1.0 & 0 & 0 & 0 & 0 & \\
Cen50 & 0 & 0 & 0 & 0 & 0 & 0 & 0.1 & 0 & 0 & 0 & 0 & \\
Cen00 & 0 & 0 & 0 & 0.4 & 0 & 0 & 0 & 1.0 & 2.1 & 0 & 1.4 & \\
CJ98CDE2 & 2.7 & 2.7 & 0 & 0.7 & 0 & 1.5 & 2.5 & 0.7 & 0.9 & 0 & 0 & $\longleftarrow$\\
CK0so & 0 & 0 & 0 & 0 & 0 & 0 & 0 & 0 & 0 & 0 & 0 & \\
CJ98b & 0 & 0 & 0 & 0 & 1.4 & 0 & 0 & 0 & 0 & 0 & 0 & \\
CJ98c & 0.1 & 0 & 0 & 0 & 0 & 0 & 3.4 & 0 & 2.7 & 0.4 & 0 & \\
CS6CDE2 & 0 & 0 & 0 & 0 & 0 & 0 & 0 & 0 & 2.3 & 0 & 0 & \\
CS6tCDE0.10 & 0 & 0 & 0.9 & 2.6 & 2.2 & 0 & 0 & 4.6 & 0 & 2.4 & 3.6 &
$\Longleftarrow$\\
CS6tCDE0.08 & 0 & 0 & 0 & 0 & 0 & 0 & 0 & 0 & 1.5 & 0 & 0 & \\
CS6tCDE0.06 & 0 & 0 & 0 & 0 & 0 & 2.6 & 0 & 0 & 0 & 0 & 0 & \\
CS6tCDE0.04 & 4.0 & 3.3 & 0 & 0 & 0 & 0 & 8.1 & 0 & 0 & 0 & 0 & \\
Wen90 & 6.2 & 3.5 & 3.6 & 3.6 & 3.5 & 0.2 & 4.9 & 2.1 & 0.3 & 2.5 & 1.3 &
$\Longleftarrow$\\
Wen80 & 0 & 0 & 0 & 0 & 0 & 0 & 0 & 0 & 0.3 & 0 & 0.5 & \\
Wen70 & 0.7 & 0 & 0 & 0 & 0 & 0 & 0 & 0 & 0 & 0 & 0 & \\
Wen50 & 0 & 0 & 0 & 0 & 0 & 0 & 0 & 0 & 0 & 0 & 0 & \\
Wen00 & 1.8 & 0 & 2.6 & 1.9 & 2.4 & 0 & 3.2 & 1.0 & 0 & 4.3 & 0 & $\longleftarrow$\\
WJ98CDE2 & 0 & 0 & 0.3 & 0.2 & 0 & 0 & 0.5 & 0 & 0 & 0 & 0 & \\
WK0so & 0 & 2.0 & 0 & 0 & 0 & 0 & 0 & 0 & 0 & 0 & 0 & \\
WJ98b & 0 & 0 & 0 & 0 & 0 & 0 & 0 & 0 & 0 & 0 & 0 & \\
WJ98c & 0 & 0 & 0 & 0 & 0 & 0 & 0 & 0 & 0 & 0 & 0 & \\
WS6CDE2 & 0 & 0 & 0 & 0 & 0 & 0 & 0 & 0 & 0 & 0 & 0 & \\
WS6tCDE0.10 & 4.2 & 0.9 & 1.2 & 0 & 2.9 & 0 & 3.2 & 1.4 & 0 & 2.0 & 0 &
$\Longleftarrow$\\
WS6tCDE0.08 & 0 & 0 & 0 & 0 & 0 & 0 & 2.4 & 0 & 1.3 & 0 & 0 & \\
WS6tCDE0.06 & 0 & 0 & 0 & 0 & 0 & 0 & 0 & 0 & 0 & 0 & 1.2 & \\
WS6tCDE0.04 & 0 & 0 & 0 & 0.9 & 0 & 1.3 & 0 & 0 & 0 & 0 & 0 & \\
\hline
\end{tabular}
\tablecomments{\footnotesize Entries in the table are mass weight of a 
component divided by uncertainty, as in Table 1.  Arrows in the final column 
indicate recurring opacities as 
discussed in the text.  The enstatite opacity used most at both temperatures is the 
``en90'' opacity from \citet{chi02}, indicated by double-lined arrows.  The optimal 
forsterite opacity is the tCDE opacity using optical constants by \citet{sog06} 
with bounding parameter t=0.10, indicated by double-lined arrows.}
}
\end{table}

\clearpage

\begin{table}[h,t]
{\tiny
\caption[Significance of Opacities for Amorphous Silicate Exemplars]{Significance 
(mass/uncertainty) of Opacities for Amorphous Silicate Exemplars \label{table3v}}
\begin{tabular}{lcccccccl}
\hline \hline
          Opacity
          & CoKu Tau/4
          & DM Tau
          & FM Tau
          & GM Aur
          & LkCa 15
          & TW Cha
          & UY Aur
          & \\
\hline
CD95OlSm & 0 & 0 & 0 & 0 & 0 & 0 & 0 & $\Longleftarrow$\\
CD95OlLg & 0 & 0 & 0 & 0 & 0 & 0 & 0 & $\Longleftarrow$\\
CD79OlSm & 5.5 & 0 & 0 & 0 & 0 & 0 & 0 & \\
CD79OlLg & 0 & 0 & 0 & 0 & 0 & 0 & 0 & \\
CD95Py5Sm & 12.4 & 0 & 0 & 0 & 0 & 0 & 1.4 & \\
CD95Py5Lg & 0 & 0 & 0 & 0 & 0 & 0 & 0 & \\
CD95Py10Sm & 6.6 & 0 & 7.3 & 14.1 & 0 & 0 & 0 & $\longleftarrow$\\
CD95Py10Lg & 0 & 0 & 0 & 0 & 0 & 0 & 0 & \\
CJ94PySm & 0 & 0 & 0 & 4.2 & 12.0 & 10.6 & 0 & $\Longleftarrow$\\
CJ94PyLg & 0 & 0 & 0 & 0 & 0 & 0 & 0 & $\Longleftarrow$\\
CD79PySm & 0.8 & 3.0 & 0 & 7.5 & 0 & 0.4 & 5.4 & $\longleftarrow$\\
CD79PyLg & 0 & 0 & 0 & 0 & 0 & 0 & 0 & \\
CSD96PySm & 0 & 0 & 0 & 0 & 0 & 7.0 & 0 & \\
CSD96PyLg & 0 & 0 & 0 & 0 & 0 & 0 & 0 & \\
WD95OlSm & 8.9 & 5.8 & 15.0 & 16.2 & 23.2 & 22.4 & 9.3 & $\Longleftarrow$\\
WD95OlLg & 0 & 0 & 0 & 0 & 0 & 0 & 0 & $\Longleftarrow$\\
WD79OlSm & 0 & 4.3 & 0 & 0.4 & 1.1 & 0 & 1.7 & $\longleftarrow$\\
WD79OlLg & 0 & 1.0 & 0 & 0 & 0 & 2.5 & 0.1 & $\longleftarrow$\\
WD95Py5Sm & 0 & 3.0 & 0 & 0 & 0 & 0 & 0 & \\
WD95Py5Lg & 0 & 15.1 & 0.3 & 0 & 0 & 0 & 0 & \\
WD95Py10Sm & 0 & 0.5 & 0.5 & 0 & 0 & 0 & 2.0 & $\longleftarrow$\\
WD95Py10Lg & 0 & 0 & 3.2 & 0 & 0 & 0 & 0 & $\longleftarrow$\\
WJ94PySm & 0 & 0 & 0 & 0 & 0 & 0 & 0 & $\Longleftarrow$\\
WJ94PyLg & 0 & 0 & 0 & 0 & 2.5 & 4.7 & 0 & $\Longleftarrow$\\
WD79PySm & 0 & 0 & 0 & 0 & 0 & 0 & 0 & \\
WD79PyLg & 0 & 1.7 & 1.1 & 0 & 0 & 0 & 1.8 & $\longleftarrow$\\
WSD96PySm & 0 & 0 & 0 & 7.7 & 3.5 & 0 & 1.8 & $\longleftarrow$\\
WSD96PyLg & 0 & 0 & 0 & 0 & 0 & 0 & 0 & \\
\hline
\end{tabular}
\tablecomments{\footnotesize Entries in the table are mass weight of a 
component divided by uncertainty, as in Table 1.  Arrows in the final column 
indicate recurring opacities as discussed 
in the text.  Of the amorphous olivines, the one included in the most 
models at the highest significance is the amorphous olivine from \citet{dor95}, 
indicated by double-lined arrows.  Of the amorphous pyroxenes, the one that shows 
up in the most models at the highest significance is the amorphous pyroxene of 
cosmic composition by \citet{jag94}, indicated by double-lined arrows.}
}
\end{table}

\clearpage

\begin{table}[h,t]
{
\caption[Trapezium Dust Model Parameters]{Trapezium Dust Model 
Parameters\label{table4v}}
\begin{tabular}{lcccccc}
\hline \hline
          Dust
          & 281\,K
          & 
          & 
          & 145\,K
          & 
          & \\
          Species
          & Value
          & Unc.
          & Sgnfcnc.
          & Value
          & Unc.
          & Sgnfcnc.\\
          (1)
          & (2)
          & (3)
          & (4)
          & (5)
          & (6)
          & (7)\\
\hline
Blackbody & 59.6 & 8.8 & 6.8 & \nodata & \nodata & \nodata\\
Sm. Am. Py. & 1.90 & 0.63 & 3.0 & \nodata & \nodata & \nodata\\
Sm. Am. Ol. & 2.05 & 0.75 & 2.7 & 358 & 47 & 7.6\\
Silica & 0.367 & 0.292 & 1.3 & 23.0 & 29.9 & 0.8\\       
\hline
\end{tabular}
\tablecomments{\footnotesize Col. (1): Dust component.  ``Sm. Am. Py.'' 
and ``Sm Am. Ol.'' denote small amorphous pyroxene and small amorphous 
olivine, respectively.  Cols. (2) and (5): 
Parameter value for warm and cool dust, respectively.  Cols. (3) and (6): 
Uncertainties on parameter values for warm and cool dust, respectively.  
Cols. (4) and (7): Significances of component, being the parameter value 
divided by its uncertainty, for warm and cool dust, respectively.  
Parameter values and uncertainties for blackbody are in units of 
10$^{-14}$ steradians, while parameter values and uncertainties for other 
dust components are in units of 10$^{-16}$ $\frac{gm}{cm^{2}}$.  ``...'' 
indicates the component at the temperature indicated by placement in the 
columns was not included in the model, so its mass is zero.  Other 
components not included in the model (also having zero mass) are large 
amorphous pyroxene, large amorphous olivine, enstatite, and forsterite.  
$\chi^{2}$ per degree of freedom for this model is 1.6.}
}
\end{table}

\clearpage

\begin{table}[h,t]
{
\caption[Degeneracies between Dust Components]{Degeneracies between Dust 
Components\label{table5}}
\begin{tabular}{ccc}
\hline \hline
          
          & 
          & Correl.\\
          Component
          & Component
          & Coeff.\\
          (1)
          & (2)
          & (3)\\
\hline
Wm. Sm. Am. Py. & Wm. Sm. Am. Ol. & -0.73\\
Wm. Lg. Am. Py. & Wm. Lg. Am. Ol. & -0.72\\ 
Wm. Enstatite & Wm. Silica & -0.50\\
Cl. Sm. Am. Py. & Cl. Sm. Am. Ol. & -0.82$>$r$>$-0.89\\
Cl. Lg. Am. Py. & Cl. Lg. Am. Ol. & -0.87$\geq$r$\geq$-0.89\\
Cl. Enstatite & Cl. Forsterite & -0.46\\
Cl. Enstatite & Cl. Silica & -0.29$\geq$r$\geq$-0.50\\
\hline
\end{tabular}
\tablecomments{\footnotesize Cols. (1) and (2): pair of dust components 
with significant degeneracy.  ``Wm.'' denotes warm, ``Cl.'' denotes cool, 
``Sm.'' denotes small, ``Lg.'' denotes large, ``Am.'' denotes amorphous, 
``Py.'' denotes pyroxene, and ``Ol.'' denotes olivine.  Col. (3): 
Correlation coefficients between pair of dust components obtained from 
covariance matrices.  The probability, P, of the correlation coefficient 
between each of the seven pairs listed in the table being drawn from a 
random distribution of data is less than or equal to 0.1\%, assuming 
there are 133 independent data points in each spectrum for which the a 
covariance matrix is computed.}
}
\end{table}

\clearpage

\begin{landscape}
\begin{deluxetable}{lccccccccccc}
\tabletypesize{\scriptsize}
\tablewidth{680pt}
\tablecaption{Dust Model Parameters \label{table1vi}}
\tablehead{\colhead{} & \colhead{} & \colhead{} & \colhead{small} &
\colhead{small} & \colhead{large} & \colhead{large} & \colhead{} & \colhead{} &
\colhead{} & \colhead{Total} & \colhead{}\\
\colhead{} & \colhead{Temp} & \colhead{} & \colhead{Amorphous} &
\colhead{Amorph.} & \colhead{Amorph.} & \colhead{Amorph.} & \colhead{Crystalline} &
\colhead{Crystalline} & \colhead{Crystalline} & \colhead{Dust} & \colhead{}\\
\colhead{Object} & \colhead{(K)} & \colhead{$\Omega_{BB}$} & 
\colhead{Pyroxene\tablenotemark{a}} & \colhead{Olivine\tablenotemark{b}} & 
\colhead{Pyroxene\tablenotemark{c}} & \colhead{Olivine\tablenotemark{d}} & 
\colhead{Enstatite\tablenotemark{e}} & \colhead{Forsterite\tablenotemark{f}} & 
\colhead{Silica\tablenotemark{g}} & \colhead{Mass} & \colhead{$\frac{\chi^2}{\rm
d.o.f.}$}\\
\colhead{(1)} & \colhead{(2)} & \colhead{(3)} & \colhead{(4)} & \colhead{(5)} &
\colhead{(6)} & \colhead{(7)} & \colhead{(8)} & \colhead{(9)} & \colhead{(10)} &
\colhead{(11)} & \colhead{(12)}}
\startdata
AA Tau & 127 & 123.8 $\pm$ 3.9 & 75.1 $\pm$ 30.1 & 0.0 $\pm$ 18.8 & 0.0 $\pm$ 21.6 &
0.0 $\pm$ 14.3 & 6.2 $\pm$ 11.7 & 7.6 $\pm$ 10.6 & 11.1 $\pm$ 14.4 & 59.1 & 2.9\\
``'' & 545 & 1.22 $\pm$ 0.02 & 9.7 $\pm$ 4.4 & 28.9 $\pm$ 5.0 & 34.2 $\pm$ 6.1 &
22.6 $\pm$ 5.8 & 4.5 $\pm$ 3.5 & 0.0 $\pm$ 3.2 & 0.0 $\pm$ 2.5 & 0.937 & \nodata\\
BP Tau & 103 & 64.3 $\pm$ 10.9 & 15.5 $\pm$ 4.9 & 24.4 $\pm$ 3.8 & 45.2 $\pm$ 4.6 &
0.0 $\pm$ 2.6 & 8.4 $\pm$ 2.5 & 3.0 $\pm$ 2.1 & 3.5 $\pm$ 3.7 & 981 & 3.6\\
``'' & 431 & 2.73 $\pm$ 0.05 & 45.6 $\pm$ 6.4 & 27.7 $\pm$ 5.2 & 9.3 $\pm$ 6.9 &
0.0 $\pm$ 6.4 & 12.9 $\pm$ 3.8 & 4.6 $\pm$ 3.4 & 0.0 $\pm$ 2.8 & 1.95 & \nodata\\
CI Tau & 115 & 312.4 $\pm$ 9.93272 & 80.2 $\pm$ 17.9 & 11.8 $\pm$ 10.0 & 0.0 $\pm$
11.6 & 0.0 $\pm$ 7.6 & 0.0 $\pm$ 6.9 & 8.0 $\pm$ 6.3 & 0.0 $\pm$ 9.2 & 278 & 3.3\\
``'' & 545 & 1.50 $\pm$ 0.03 & 0.0 $\pm$ 3.5 & 35.1 $\pm$ 4.1 & 47.6 $\pm$ 4.9 &
11.4 $\pm$ 4.6 & 4.7 $\pm$ 2.7 & 0.7 $\pm$ 2.5 & 0.5 $\pm$ 1.9 & 1.66 & \nodata\\
CoKu Tau/3 & 103 & 0.0 $\pm$ 0.0 & 0.0 $\pm$ 7.8 & 3.0 $\pm$ 5.6 & 24.9 $\pm$ 5.5 &
54.0 $\pm$ 7.6 & 3.9 $\pm$ 3.6 & 9.1 $\pm$ 3.4 & 5.2 $\pm$ 5.4 & 488 & 4.7\\
``'' & 602 & 0.975 $\pm$ 0.019 & 11.1 $\pm$ 3.2 & 13.2 $\pm$ 3.4 & 28.0 $\pm$
4.4 & 36.6 $\pm$ 4.3 & 7.0 $\pm$ 2.6 & 4.1 $\pm$ 2.4 & 0.0 $\pm$ 1.8 & 1.06 & \\
CoKu Tau/4 & 114 & 354.2 $\pm$ 8.2 & 32.8 $\pm$ 3.9 & 47.6 $\pm$ 4.2 & 13.8 $\pm$
3.7 & 0.0 $\pm$ 2.8 & 0.0 $\pm$ 2.3 & 2.3 $\pm$ 2.0 & 3.4 $\pm$ 2.7 & 677 & 2.2\\
``'' & 200 & 0.420 $\pm$ 0.230 & 8.4 $\pm$ 10.0 & 91.6 $\pm$ 22.4 & 0.0 $\pm$
14.1 & 0.0 $\pm$ 12.5 & 0.0 $\pm$ 6.9 & 0.0 $\pm$ 5.1 & 0.0 $\pm$ 7.6 & 4.64 & \nodata\\
CW Tau & 151 & 142.2 $\pm$ 3.9 & 0.0 $\pm$ 26.8 & 24.8 $\pm$ 17.8 & 0.0 $\pm$ 21.1 &
0.0 $\pm$ 14.0 & 15.5 $\pm$ 12.7 & 29.8 $\pm$ 15.7 & 29.9 $\pm$ 17.3 & 59.1 & 5.3\\
``'' & 659 & 1.52 $\pm$ 0.03 & 0.0 $\pm$ 7.5 & 51.6 $\pm$ 11.7 & 42.1 $\pm$ 10.8 &
0.0 $\pm$ 12.4 & 2.9 $\pm$ 5.9 & 0.0 $\pm$ 5.2 & 3.4 $\pm$ 4.2 & 0.643 & \nodata\\
CX Tau & 151 & 58.3 $\pm$ 1.4 & 27.2 $\pm$ 15.1 & 62.7 $\pm$ 19.8 & 0.0 $\pm$ 16.0 &
0.0 $\pm$ 11.5 & 0.0 $\pm$ 8.8 & 3.4 $\pm$ 7.5 & 6.8 $\pm$ 9.4 & 27.0 & 2.6\\
``'' & 602 & 0.400 $\pm$ 0.013 & 7.3 $\pm$ 6.8 & 15.2 $\pm$ 7.4 & 24.4 $\pm$ 9.9 &
38.9 $\pm$ 9.9 & 7.6 $\pm$ 5.4 & 3.2 $\pm$ 5.1 & 3.4 $\pm$ 4.2 & 0.268 & \nodata\\
CY Tau & 127 & 22.4 $\pm$ 1.0 & 22.6 $\pm$ 37.7 & 0.0 $\pm$ 30.1 & 0.0 $\pm$ 34.0 &
0.0 $\pm$ 21.9 & 0.0 $\pm$ 20.2 & 48.2 $\pm$ 38.1 & 29.2 $\pm$ 28.6 & 9.50 & 4.0\\
``'' & 602 & 0.319 $\pm$ 0.006 & 0.0 $\pm$ 22.1 & 69.9 $\pm$ 39.9 & 0.0 $\pm$ 33.1
& 0.0 $\pm$ 30.1 & 12.8 $\pm$ 17.0 & 0.0 $\pm$ 13.839 & 17.3 $\pm$ 15.1 & 0.0460 & \nodata\\
DD Tau & 127 & 220.2 $\pm$ 8.4 & 80.5 $\pm$ 18.1 & 0.0 $\pm$ 10.7 & 0.0 $\pm$ 11.8 &
0.0 $\pm$ 7.9 & 10.1 $\pm$ 6.8 & 9.3 $\pm$ 6.4 & 0.0 $\pm$ 9.0 & 233 & 3.5\\
``'' & 488 & 4.72 $\pm$ 0.07 & 26.3 $\pm$ 9.1 & 0.0 $\pm$ 9.7 & 7.9 $\pm$ 13.0 &
46.2 $\pm$ 12.3 & 13.7 $\pm$ 6.8 & 5.8 $\pm$ 6.1 & 0.0 $\pm$ 4.8 & 1.42 & \nodata\\
DE Tau & 115 & 226.5 $\pm$ 7.7 & 84.7 $\pm$ 9.9 & 5.2 $\pm$ 5.6 & 0.0 $\pm$ 6.2 &
0.0 $\pm$ 4.0 & 0.0 $\pm$ 3.5 & 10.1 $\pm$ 3.4 & 0.0 $\pm$ 4.6 & 390 & 4.7\\
``'' & 545 & 1.17 $\pm$ 0.02 & 0.0 $\pm$ 6.1 & 59.2 $\pm$ 10.4 & 0.0 $\pm$ 10.1 &
28.0 $\pm$ 8.2 & 5.0 $\pm$ 4.8 & 7.8 $\pm$ 4.3 & 0.0 $\pm$ 3.4 & 0.600 & \nodata\\
DF Tau & 163 & 46.6 $\pm$ 2.2 & 0.0 $\pm$ 14.1 & 0.0 $\pm$ 10.2 & 0.0 $\pm$ 12.3 &
86.4 $\pm$ 21.3 & 0.0 $\pm$ 6.8 & 4.6 $\pm$ 5.8 & 9.0 $\pm$ 8.0 & 56.2 & 5.3\\
``'' & 716 & 1.34 $\pm$ 0.02 & 63.6 $\pm$ 27.0 & 0.0 $\pm$ 17.4 & 0.0 $\pm$ 23.8 &
0.0 $\pm$ 23.4 & 18.5 $\pm$ 12.5 & 17.9 $\pm$ 11.6 & 0.0 $\pm$ 7.6 & 0.250 & \nodata\\
DG Tau & 127 & 2810.9 $\pm$ 56.2 & 86.3 $\pm$ 18.4 & 0.0 $\pm$ 11.0 & 0.0 $\pm$ 11.3
& 0.0 $\pm$ 7.3 & 0.0 $\pm$ 5.8 & 11.7 $\pm$ 6.4 & 1.9 $\pm$ 8.2 & 1350 & 6.4\\
``'' & 431 & 23.1 $\pm$ 0.4 & 0.0 $\pm$ 335.1 & 0.0 $\pm$ 531.6 & 0.0 $\pm$ 745.9
& 0.0 $\pm$ 855.8 & 0.0 $\pm$ 326.0 & 100.0 $\pm$ 1347.6 & 0.0 $\pm$ 162.1 & 0.113 & \nodata\\
DH Tau & 127 & 89.4 $\pm$ 2.8 & 6.7 $\pm$ 14.5 & 48.9 $\pm$ 13.4 & 0.0 $\pm$ 12.4 &
0.0 $\pm$ 8.5 & 17.9 $\pm$ 7.7 & 26.6 $\pm$ 8.9 & 0.0 $\pm$ 7.0 & 68.3 & 8.4\\
``'' & 944 & 0.0953 $\pm$ 0.0032 & 0.0 $\pm$ 4.1 & 7.5 $\pm$ 4.9 & 0.0 $\pm$ 7.4 &
47.3 $\pm$ 6.7 & 26.0 $\pm$ 4.5 & 14.8 $\pm$ 3.8 & 4.3 $\pm$ 2.1 & 0.115 & \nodata\\
DK Tau & 175 & 74.0 $\pm$ 2.1 & 0.0 $\pm$ 14.0 & 9.1 $\pm$ 9.4 & 19.0 $\pm$ 11.3 &
15.9 $\pm$ 8.5 & 17.1 $\pm$ 7.4 & 15.4 $\pm$ 6.8 & 23.4 $\pm$ 9.1 & 54.3 & 5.7\\
``'' & 830 & 0.631 $\pm$ 0.020 & 0.0 $\pm$ 2.2 & 1.5 $\pm$ 2.6 & 30.3 $\pm$ 3.1 &
54.2 $\pm$ 3.6 & 7.1 $\pm$ 1.9 & 4.8 $\pm$ 1.8 & 2.0 $\pm$ 1.1 & 1.58 & \nodata\\
DL Tau & 139 & 147.2 $\pm$ 5.2 & 0.0 $\pm$ 25.1 & 0.0 $\pm$ 17.3 & 36.0 $\pm$ 20.3 &
41.0 $\pm$ 20.0 & 10.0 $\pm$ 11.8 & 6.8 $\pm$ 10.6 & 6.2 $\pm$ 14.3 & 75.1 & 3.8\\
``'' & 488 & 3.63 $\pm$ 0.06 & 16.2 $\pm$ 22.6 & 0.0 $\pm$ 25.3 & 0.0 $\pm$ 38.2 &
0.0 $\pm$ 34.9 & 34.4 $\pm$ 25.8 & 26.8 $\pm$ 21.7 & 22.6 $\pm$ 18.7 & 0.385 & \nodata\\
DM Tau & 121 & 119.1 $\pm$ 2.7 & 0.0 $\pm$ 15.5 & 70.6 $\pm$ 18.5 & 0.0 $\pm$ 13.6 &
0.0 $\pm$ 9.1 & 0.0 $\pm$ 6.1 & 12.6 $\pm$ 7.2 & 16.9 $\pm$ 8.3 & 62.2 & 4.5\\
``'' & 260 & 0.295 $\pm$ 0.040 & 0.0 $\pm$ 2.4 & 25.3 $\pm$ 2.7 & 34.0 $\pm$ 3.2 &
40.4 $\pm$ 3.3 & 0.0 $\pm$ 1.7 & 0.2 $\pm$ 1.7 & 0.0 $\pm$ 1.2 & 2.90 & \nodata\\
DN Tau & 127 & 100.1 $\pm$ 3.6 & 77.1 $\pm$ 23.7 & 16.4 $\pm$ 13.5 & 0.0 $\pm$ 16.2
& 0.0 $\pm$ 10.9 & 0.0 $\pm$ 9.1 & 5.9 $\pm$ 8.1 & 0.6 $\pm$ 12.5 & 68.4 & 2.9\\
``'' & 431 & 1.99 $\pm$ 0.03 & 0.0 $\pm$ 20.5 & 49.7 $\pm$ 28.9 & 0.0 $\pm$ 31.9 &
0.0 $\pm$ 29.2 & 20.6 $\pm$ 17.0 & 7.7 $\pm$ 13.3 & 22.0 $\pm$ 15.8 & 0.283 & \nodata\\
DO Tau & 139 & 565.9 $\pm$ 11.4 & 79.4 $\pm$ 35.5 & 0.0 $\pm$ 20.1 & 0.0 $\pm$ 25.1
& 0.0 $\pm$ 16.9 & 0.0 $\pm$ 13.4 & 12.7 $\pm$ 12.8 & 7.9 $\pm$ 15.9 & 136 & 2.3\\
``'' & 545 & 3.59 $\pm$ 0.07 & 0.0 $\pm$ 5.3 & 1.3 $\pm$ 6.3 & 12.1 $\pm$ 8.2 &
76.8 $\pm$ 11.0 & 5.6 $\pm$ 4.3 & 4.2 $\pm$ 4.0 & 0.0 $\pm$ 2.9 & 2.03 & \nodata\\
DP Tau & 139 & 145.5 $\pm$ 4.8 & 34.2 $\pm$ 6.7 & 50.1 $\pm$ 7.3 & 0.0 $\pm$ 6.6 &
0.0 $\pm$ 4.4 & 4.0 $\pm$ 3.6 & 4.7 $\pm$ 3.2 & 6.9 $\pm$ 4.9 & 236 & 2.9\\
``'' & 488 & 2.40 $\pm$ 0.05 & 0.0 $\pm$ 6.0 & 64.7 $\pm$ 11.8 & 0.0 $\pm$ 10.5 &
29.9 $\pm$ 8.7 & 0.0 $\pm$ 5.1 & 5.4 $\pm$ 4.8 & 0.0 $\pm$ 2.9 & 1.16 & \nodata\\
DQ Tau & 127 & 236.4 $\pm$ 7.1 & 93.1 $\pm$ 30.6 & 0.0 $\pm$ 15.0 & 0.0 $\pm$ 18.0 &
0.0 $\pm$ 11.9 & 0.0 $\pm$ 10.6 & 6.9 $\pm$ 9.9 & 0.0 $\pm$ 13.2 & 122 & 3.1\\
``'' & 431 & 4.43 $\pm$ 0.07 & 8.1 $\pm$ 16.9 & 42.8 $\pm$ 22.8 & 0.0 $\pm$ 27.6 &
0.0 $\pm$ 25.4 & 41.1 $\pm$ 21.6 & 7.9 $\pm$ 12.1 & 0.0 $\pm$ 9.3 & 0.642 & \nodata\\
DR Tau & 127 & 683.4 $\pm$ 17.9 & 47.6 $\pm$ 10.8 & 41.4 $\pm$ 9.8 & 0.0 $\pm$ 10.0
& 0.0 $\pm$ 6.7 & 0.0 $\pm$ 5.5 & 3.9 $\pm$ 4.8 & 7.0 $\pm$ 6.8 & 594 & 2.1\\
``'' & 545 & 6.24 $\pm$ 0.10 & 0.0 $\pm$ 7.4 & 59.0 $\pm$ 12.5 & 5.5 $\pm$ 11.6 &
34.5 $\pm$ 9.9 & 0.5 $\pm$ 6.0 & 0.5 $\pm$ 5.4 & 0.0 $\pm$ 4.0 & 2.32 & \nodata\\
DS Tau & 139 & 31.1 $\pm$ 1.2 & 0.0 $\pm$ 25.1 & 21.9 $\pm$ 16.6 & 40.8 $\pm$ 20.2 &
0.0 $\pm$ 13.3 & 21.1 $\pm$ 13.0 & 10.9 $\pm$ 10.6 & 5.3 $\pm$ 13.5 & 19.1 & 5.1\\
``'' & 545 & 0.674 $\pm$ 0.014 & 0.0 $\pm$ 3.7 & 45.3 $\pm$ 5.0 & 44.7 $\pm$ 5.1 &
0.0 $\pm$ 5.2 & 4.1 $\pm$ 2.9 & 3.8 $\pm$ 2.7 & 2.1 $\pm$ 1.9 & 0.682 & \nodata\\
F04147+2822 & 163 & 9.70 $\pm$ 0.54 & 0.0 $\pm$ 8.1 & 18.9 $\pm$ 5.4 & 61.1 $\pm$
8.5 & 0.0 $\pm$ 4.9 & 4.2 $\pm$ 3.8 & 15.8 $\pm$ 3.8 & 0.0 $\pm$ 4.7 & 24.1 & 2.2\\
``'' & 602 & 0.361 $\pm$ 0.009 & 3.2 $\pm$ 3.5 & 36.2 $\pm$ 4.1 & 60.6 $\pm$
5.4 & 0.0 $\pm$ 4.9 & 0.0 $\pm$ 2.6 & 0.0 $\pm$ 2.4 & 0.0 $\pm$ 1.9 & 0.485 & \nodata\\
FM Tau & 127 & 93.4 $\pm$ 3.0 & 60.2 $\pm$ 11.1 & 27.1 $\pm$ 7.9 & 0.0 $\pm$ 9.2 &
0.0 $\pm$ 6.2 & 5.6 $\pm$ 5.2 & 3.5 $\pm$ 4.6 & 3.6 $\pm$ 6.2 & 107 & 2.4\\
``'' & 431 & 1.32 $\pm$ 0.03 & 7.6 $\pm$ 4.4 & 78.3 $\pm$ 9.3 & 10.5 $\pm$ 6.4 &
0.0 $\pm$ 6.3 & 1.6 $\pm$ 3.4 & 2.0 $\pm$ 3.1 & 0.0 $\pm$ 2.4 & 1.19 & \nodata\\
FN Tau & 139 & 254.4 $\pm$ 5.3 & 74.6 $\pm$ 19.7 & 0.0 $\pm$ 12.9 & 0.0 $\pm$ 14.3 &
0.0 $\pm$ 10.2 & 0.1 $\pm$ 8.6 & 25.3 $\pm$ 9.6 & 0.0 $\pm$ 7.0 & 105 & 5.1\\
``'' & 488 & 1.80 $\pm$ 0.04 & 0.0 $\pm$ 5.8 & 38.2 $\pm$ 7.8 & 0.0 $\pm$ 10.0 &
30.6 $\pm$ 8.3 & 28.8 $\pm$ 6.2 & 2.4 $\pm$ 4.6 & 0.0 $\pm$ 3.2 & 1.11 & \nodata\\
FO Tau & 115 & 133.8 $\pm$ 6.0 & 43.5 $\pm$ 6.6 & 23.8 $\pm$ 5.2 & 19.2 $\pm$ 5.5 &
0.0 $\pm$ 4.2 & 8.9 $\pm$ 3.5 & 4.6 $\pm$ 3.2 & 0.0 $\pm$ 4.5 & 332 & 2.7\\
``'' & 488 & 1.57 $\pm$ 0.03 & 42.6 $\pm$ 11.8 & 0.0 $\pm$ 9.6 & 40.2 $\pm$ 12.5 &
0.0 $\pm$ 12.9 & 8.1 $\pm$ 7.0 & 2.6 $\pm$ 6.7 & 6.5 $\pm$ 5.5 & 0.537 & \nodata\\
FP Tau & 115 & 39.4 $\pm$ 2.0 & 13.2 $\pm$ 6.4 & 0.0 $\pm$ 5.0 & 82.2 $\pm$ 9.4 &
0.0 $\pm$ 3.9 & 1.4 $\pm$ 3.1 & 3.2 $\pm$ 2.7 & 0.0 $\pm$ 4.3 & 110 & 5.3\\
``'' & 488 & 0.400 $\pm$ 0.008 & 90.2 $\pm$ 41.6 & 0.0 $\pm$ 18.4 & 0.0 $\pm$ 27.8
& 0.0 $\pm$ 24.7 & 0.9 $\pm$ 13.1 & 8.9 $\pm$ 11.7 & 0.0 $\pm$ 9.2 & 0.0743 & \nodata\\
FQ Tau & 151 & 12.7 $\pm$ 0.4 & 47.2 $\pm$ 119.2 & 0.0 $\pm$ 95.8 & 0.0 $\pm$ 116.3
& 0.0 $\pm$ 80.9 & 0.0 $\pm$ 64.0 & 51.7 $\pm$ 125.2 & 1.1 $\pm$ 60.0 & 1.02 & 3.9\\
``'' & 830 & 0.0635 $\pm$ 0.0016 & 10.0 $\pm$ 6.0 & 0.0 $\pm$ 7.3 & 4.6 $\pm$ 10.5
& 71.4 $\pm$ 12.3 & 11.9 $\pm$ 5.2 & 1.5 $\pm$ 4.7 & 0.6 $\pm$ 3.7 & 0.0345 & \nodata\\
FS Tau & 127 & 445.3 $\pm$ 12.2 & 27.2 $\pm$ 9.6 & 18.9 $\pm$ 7.5 & 45.4 $\pm$ 9.8 &
0.0 $\pm$ 6.5 & 0.0 $\pm$ 5.2 & 3.6 $\pm$ 4.5 & 4.9 $\pm$ 6.8 & 415 & 2.8\\
``'' & 488 & 3.05 $\pm$ 0.06 & 0.0 $\pm$ 6.1 & 44.8 $\pm$ 9.0 & 8.2 $\pm$ 9.9 &
40.1 $\pm$ 9.1 & 2.7 $\pm$ 5.1 & 4.3 $\pm$ 4.6 & 0.0 $\pm$ 3.2 & 1.45 & \nodata\\
FT Tau & 115 & 101.0 $\pm$ 4.2 & 74.8 $\pm$ 18.6 & 10.5 $\pm$ 11.1 & 0.0 $\pm$ 12.6
& 0.0 $\pm$ 8.2 & 0.0 $\pm$ 7.2 & 13.3 $\pm$ 7.0 & 1.4 $\pm$ 10.5 & 108 & 5.9\\
``'' & 431 & 1.29 $\pm$ 0.03 & 7.5 $\pm$ 6.5 & 60.6 $\pm$ 11.3 & 15.4 $\pm$ 9.4 &
0.0 $\pm$ 9.7 & 7.4 $\pm$ 5.2 & 4.6 $\pm$ 4.7 & 4.4 $\pm$ 3.8 & 0.675 & \nodata\\
FV Tau & 127 & 359.6 $\pm$ 13.4 & 35.7 $\pm$ 7.2 & 58.7 $\pm$ 8.6 & 0.0 $\pm$ 6.8 &
0.0 $\pm$ 4.5 & 1.8 $\pm$ 3.9 & 1.1 $\pm$ 3.5 & 2.7 $\pm$ 4.8 & 652 & 3.6\\
``'' & 545 & 5.24 $\pm$ 0.09 & 0.0 $\pm$ 3.5 & 20.0 $\pm$ 5.0 & 0.0 $\pm$ 6.8 &
72.6 $\pm$ 8.1 & 0.0 $\pm$ 3.6 & 7.5 $\pm$ 3.4 & 0.0 $\pm$ 1.5 & 3.30 & \nodata\\
FX Tau & 115 & 58.3 $\pm$ 5.4 & 0.0 $\pm$ 6.5 & 55.3 $\pm$ 6.3 & 0.0 $\pm$ 5.2 &
29.8 $\pm$ 4.1 & 1.3 $\pm$ 3.0 & 6.8 $\pm$ 2.7 & 6.7 $\pm$ 3.8 & 358 & 5.2\\
``'' & 545 & 0.788 $\pm$ 0.019 & 6.4 $\pm$ 3.2 & 48.1 $\pm$ 4.6 & 30.3 $\pm$ 4.3 &
0.7 $\pm$ 4.7 & 8.9 $\pm$ 2.5 & 5.5 $\pm$ 2.4 & 0.0 $\pm$ 1.6 & 1.10 & \nodata\\
FZ Tau & 139 & 135.1 $\pm$ 4.5 & 0.0 $\pm$ 32.6 & 50.9 $\pm$ 29.3 & 0.0 $\pm$ 26.4 &
0.0 $\pm$ 18.0 & 21.6 $\pm$ 16.8 & 14.7 $\pm$ 13.6 & 12.8 $\pm$ 17.5 & 55.7 & 7.5\\
``'' & 545 & 2.68 $\pm$ 0.05 & 0.0 $\pm$ 6.0 & 0.0 $\pm$ 6.5 & 51.2 $\pm$ 9.0 &
27.2 $\pm$ 7.9 & 6.7 $\pm$ 4.7 & 4.9 $\pm$ 4.3 & 10.0 $\pm$ 3.9 & 1.29 & \nodata\\
GG Tau A & 139 & 192.2 $\pm$ 5.0 & 73.2 $\pm$ 16.0 & 8.6 $\pm$ 9.7 & 0.0 $\pm$ 11.9
& 0.0 $\pm$ 8.1 & 0.0 $\pm$ 6.5 & 12.8 $\pm$ 6.6 & 5.4 $\pm$ 6.9 & 130 & 2.4\\
``'' & 545 & 1.65 $\pm$ 0.04 & 0.0 $\pm$ 2.9 & 35.7 $\pm$ 3.5 & 49.3 $\pm$ 4.2 &
12.5 $\pm$ 3.9 & 0.3 $\pm$ 2.3 & 0.9 $\pm$ 2.2 & 1.3 $\pm$ 1.5 & 2.44 & \nodata\\
GG Tau B & 139 & 46.1 $\pm$ 1.0 & 51.6 $\pm$ 8.7 & 37.1 $\pm$ 7.2 & 0.0 $\pm$ 8.0 &
0.0 $\pm$ 5.5 & 0.0 $\pm$ 4.3 & 8.7 $\pm$ 4.2 & 2.7 $\pm$ 4.8 & 38.2 & 7.5\\
``'' & 1172 & 0.0224 $\pm$ 0.0009 & 0.0 $\pm$ 3.4 & 12.6 $\pm$ 4.2 & 25.5 $\pm$
5.3 & 43.1 $\pm$ 5.6 & 12.7 $\pm$ 3.1 & 6.2 $\pm$ 2.9 & 0.0 $\pm$ 1.5 & 0.0438 & \nodata\\
GH Tau & 139 & 65.8 $\pm$ 2.0 & 80.8 $\pm$ 15.9 & 0.0 $\pm$ 9.1 & 0.0 $\pm$ 10.7 &
0.0 $\pm$ 7.2 & 0.0 $\pm$ 5.9 & 14.0 $\pm$ 5.8 & 5.2 $\pm$ 7.5 & 56.5 & 4.3\\
``'' & 716 & 0.349 $\pm$ 0.007 & 0.0 $\pm$ 6.3 & 1.3 $\pm$ 7.3 & 10.5 $\pm$ 9.6 &
76.8 $\pm$ 12.8 & 7.7 $\pm$ 5.0 & 0.6 $\pm$ 4.6 & 3.2 $\pm$ 3.5 & 0.182 & \nodata\\
GI Tau & 115 & 171.6 $\pm$ 10.1 & 0.0 $\pm$ 7.6 & 0.0 $\pm$ 5.7 & 87.1 $\pm$ 11.0 &
0.0 $\pm$ 4.0 & 9.2 $\pm$ 3.5 & 3.3 $\pm$ 3.2 & 0.4 $\pm$ 5.4 & 610 & 3.9\\
``'' & 488 & 2.99 $\pm$ 0.06 & 19.9 $\pm$ 3.5 & 38.5 $\pm$ 4.3 & 9.2 $\pm$ 5.1 &
29.3 $\pm$ 4.3 & 1.5 $\pm$ 2.7 & 1.6 $\pm$ 2.4 & 0.0 $\pm$ 1.8 & 3.22 & \nodata\\
GK Tau & 175 & 115.6 $\pm$ 2.6 & 0.0 $\pm$ 10.0 & 0.0 $\pm$ 7.2 & 38.7 $\pm$ 8.6 &
35.1 $\pm$ 7.6 & 10.5 $\pm$ 4.9 & 9.8 $\pm$ 4.5 & 6.0 $\pm$ 5.8 & 90.4 & 4.2\\
``'' & 1001 & 0.283 $\pm$ 0.013 & 0.0 $\pm$ 2.3 & 40.3 $\pm$ 3.3 & 45.6 $\pm$ 3.6
& 6.6 $\pm$ 3.9 & 3.8 $\pm$ 2.0 & 3.4 $\pm$ 2.0 & 0.3 $\pm$ 1.0 & 0.992 & \nodata\\
GM Aur & 91 & 810.0 $\pm$ 32.8 & 58.4 $\pm$ 3.8 & 37.7 $\pm$ 3.0 & 0.0 $\pm$ 3.1 &
0.0 $\pm$ 2.0 & 0.0 $\pm$ 1.7 & 2.1 $\pm$ 1.7 & 1.9 $\pm$ 2.5 & 3610 & 4.9\\
``'' & 488 & 0.326 $\pm$ 0.014 & 1.0 $\pm$ 3.5 & 98.1 $\pm$ 8.8 & 0.0 $\pm$ 5.3 &
0.0 $\pm$ 4.9 & 0.0 $\pm$ 2.5 & 0.9 $\pm$ 2.6 & 0.0 $\pm$ 1.6 & 0.714 & \nodata\\
GN Tau & 139 & 48.9 $\pm$ 2.2 & 0.0 $\pm$ 18.2 & 36.4 $\pm$ 14.0 & 0.0 $\pm$ 14.3 &
0.0 $\pm$ 9.8 & 33.6 $\pm$ 12.3 & 19.9 $\pm$ 9.3 & 10.1 $\pm$ 11.4 & 50.7 & 5.6\\
``'' & 545 & 1.06 $\pm$ 0.02 & 0.0 $\pm$ 3.1 & 12.2 $\pm$ 3.1 & 53.7 $\pm$ 4.5 &
22.7 $\pm$ 3.8 & 1.1 $\pm$ 2.4 & 2.8 $\pm$ 2.1 & 7.6 $\pm$ 1.8 & 1.42 & \nodata\\
GO Tau & 115 & 59.1 $\pm$ 1.9 & 86.5 $\pm$ 17.5 & 0.0 $\pm$ 9.9 & 0.0 $\pm$ 10.9 &
0.0 $\pm$ 7.0 & 0.0 $\pm$ 6.1 & 11.2 $\pm$ 5.8 & 2.3 $\pm$ 8.1 & 53.2 & 5.3\\
``'' & 545 & 0.213 $\pm$ 0.006 & 0.0 $\pm$ 5.1 & 34.2 $\pm$ 6.0 & 42.1 $\pm$ 7.1 &
14.3 $\pm$ 7.1 & 4.3 $\pm$ 3.9 & 3.1 $\pm$ 3.3 & 2.0 $\pm$ 2.8 & 0.184 & \nodata\\
Haro 6-28 & 163 & 14.9 $\pm$ 0.6 & 0.0 $\pm$ 13.5 & 0.0 $\pm$ 9.6 & 0.0 $\pm$ 12.7 &
82.4 $\pm$ 19.9 & 0.0 $\pm$ 6.7 & 6.0 $\pm$ 5.7 & 11.6 $\pm$ 7.4 & 15.1 & 11.8\\
``'' & 659 & 0.197 $\pm$ 0.006 & 11.3 $\pm$ 9.5 & 0.0 $\pm$ 11.1 & 44.2 $\pm$
14.8 & 0.0 $\pm$ 16.4 & 21.4 $\pm$ 9.0 & 14.5 $\pm$ 7.8 & 8.6 $\pm$ 6.1 & 0.0825 & \nodata\\
Haro 6-37 & 163 & 74.0 $\pm$ 2.0 & 61.0 $\pm$ 26.4 & 1.2 $\pm$ 18.9 & 0.0 $\pm$ 22.5
& 0.0 $\pm$ 15.5 & 23.3 $\pm$ 14.4 & 11.1 $\pm$ 11.1 & 3.4 $\pm$ 13.5 & 26.5 & 4.9\\
``'' & 716 & 0.939 $\pm$ 0.022 & 6.9 $\pm$ 6.5 & 28.3 $\pm$ 8.0 & 0.0 $\pm$
11.6 & 36.9 $\pm$ 9.8 & 18.8 $\pm$ 6.0 & 3.9 $\pm$ 4.9 & 5.0 $\pm$ 3.8 & 0.464 & \nodata\\
HK Tau & 115 & 360.7 $\pm$ 9.4 & 91.6 $\pm$ 10.0 & 0.0 $\pm$ 5.1 & 0.0 $\pm$ 5.7 &
0.0 $\pm$ 3.7 & 0.0 $\pm$ 3.4 & 8.3 $\pm$ 3.1 & 0.1 $\pm$ 5.0 & 506 & 6.2\\
``'' & 488 & 0.809 $\pm$ 0.020 & 0.0 $\pm$ 6.2 & 14.3 $\pm$ 7.1 & 13.1 $\pm$ 9.9 &
48.3 $\pm$ 9.9 & 20.8 $\pm$ 5.9 & 3.5 $\pm$ 4.8 & 0.0 $\pm$ 3.5 & 0.483 & \nodata\\
HN Tau & 139 & 171.0 $\pm$ 5.1 & 66.6 $\pm$ 12.6 & 0.0 $\pm$ 8.8 & 0.0 $\pm$ 9.7 &
0.0 $\pm$ 6.6 & 0.0 $\pm$ 5.5 & 20.2 $\pm$ 5.9 & 13.2 $\pm$ 6.6 & 165 & 4.5\\
``'' & 545 & 2.01 $\pm$ 0.04 & 0.0 $\pm$ 3.6 & 44.6 $\pm$ 5.0 & 22.8 $\pm$ 5.0 &
28.1 $\pm$ 4.8 & 0.0 $\pm$ 2.9 & 3.0 $\pm$ 2.7 & 1.5 $\pm$ 1.9 & 2.07 & \nodata\\
HO Tau & 139 & 11.7 $\pm$ 0.6 & 82.7 $\pm$ 30.1 & 0.0 $\pm$ 15.9 & 0.0 $\pm$ 19.0 &
0.0 $\pm$ 12.1 & 0.0 $\pm$ 10.2 & 16.9 $\pm$ 11.3 & 0.4 $\pm$ 17.9 & 9.81 & 5.6\\
``'' & 488 & 0.211 $\pm$ 0.006 & 0.1 $\pm$ 5.8 & 53.0 $\pm$ 8.6 & 32.9 $\pm$ 7.6 &
0.0 $\pm$ 8.2 & 8.3 $\pm$ 4.4 & 3.9 $\pm$ 3.8 & 1.8 $\pm$ 3.1 & 0.178 & \nodata\\
HP Tau & 139 & 386.8 $\pm$ 9.8 & 88.5 $\pm$ 31.8 & 0.0 $\pm$ 15.5 & 0.0 $\pm$ 18.5 &
0.0 $\pm$ 12.2 & 0.0 $\pm$ 10.5 & 11.5 $\pm$ 10.3 & 0.0 $\pm$ 18.4 & 178 & 2.1\\
``'' & 488 & 2.98 $\pm$ 0.07 & 0.0 $\pm$ 3.5 & 52.8 $\pm$ 5.6 & 28.4 $\pm$ 4.9 &
14.6 $\pm$ 5.0 & 0.0 $\pm$ 2.9 & 1.8 $\pm$ 2.6 & 2.3 $\pm$ 1.8 & 3.55 & \nodata\\
HQ Tau & 187 & 77.3 $\pm$ 2.1 & 23.2 $\pm$ 5.9 & 25.7 $\pm$ 4.8 & 46.4 $\pm$ 6.2 &
0.0 $\pm$ 4.3 & 0.0 $\pm$ 3.4 & 3.2 $\pm$ 3.0 & 1.5 $\pm$ 4.1 & 108 & 4.9\\
``'' & 1400 & 0.227 $\pm$ 0.008 & 0.0 $\pm$ 2.9 & 46.5 $\pm$ 4.5 & 24.5 $\pm$ 4.3
& 25.7 $\pm$ 4.4 & 1.8 $\pm$ 2.5 & 1.4 $\pm$ 2.5 & 0.0 $\pm$ 1.4 & 0.480 & \nodata\\
IP Tau & 103 & 88.6 $\pm$ 6.3 & 0.0 $\pm$ 5.4 & 27.7 $\pm$ 3.9 & 64.4 $\pm$ 6.0 &
0.0 $\pm$ 2.7 & 2.6 $\pm$ 2.5 & 0.8 $\pm$ 2.1 & 4.4 $\pm$ 4.1 & 546 & 4.2\\
``'' & 488 & 0.648 $\pm$ 0.016 & 23.7 $\pm$ 3.9 & 64.9 $\pm$ 6.8 & 5.9 $\pm$ 5.8 &
0.0 $\pm$ 5.5 & 1.4 $\pm$ 3.1 & 4.2 $\pm$ 2.7 & 0.0 $\pm$ 1.9 & 0.785 & \nodata\\
IQ Tau & 139 & 51.8 $\pm$ 1.9 & 92.6 $\pm$ 54.1 & 0.0 $\pm$ 29.2 & 0.0 $\pm$ 33.0 &
0.0 $\pm$ 21.9 & 0.0 $\pm$ 17.1 & 7.4 $\pm$ 17.6 & 0.0 $\pm$ 18.9 & 16.9 & 6.3\\
``'' & 602 & 0.782 $\pm$ 0.016 & 0.0 $\pm$ 4.3 & 37.0 $\pm$ 5.4 & 10.7 $\pm$ 6.5 &
43.2 $\pm$ 6.1 & 4.2 $\pm$ 3.5 & 4.8 $\pm$ 3.1 & 0.0 $\pm$ 2.3 & 0.588 & \nodata\\
IS Tau & 175 & 13.7 $\pm$ 0.5 & 0.0 $\pm$ 31.1 & 0.0 $\pm$ 22.0 & 0.0 $\pm$ 26.8 &
0.0 $\pm$ 19.0 & 3.8 $\pm$ 15.2 & 64.3 $\pm$ 36.2 & 31.9 $\pm$ 21.4 & 5.39 & 3.3\\
``'' & 659 & 0.417 $\pm$ 0.010 & 0.0 $\pm$ 3.5 & 0.0 $\pm$ 3.7 & 41.8 $\pm$ 4.8 &
44.0 $\pm$ 4.8 & 7.1 $\pm$ 2.7 & 2.8 $\pm$ 2.5 & 4.2 $\pm$ 2.0 & 0.504 & \nodata\\
IT Tau & 127 & 53.5 $\pm$ 2.3 & 91.6 $\pm$ 17.3 & 0.0 $\pm$ 9.3 & 0.0 $\pm$ 10.3 &
0.0 $\pm$ 7.0 & 3.3 $\pm$ 5.7 & 4.9 $\pm$ 5.1 & 0.2 $\pm$ 7.3 & 72.8 & 6.3\\
``'' & 659 & 0.649 $\pm$ 0.011 & 0.0 $\pm$ 25.3 & 37.1 $\pm$ 30.8 & 0.0 $\pm$ 40.4
& 0.0 $\pm$ 38.7 & 52.6 $\pm$ 38.6 & 10.4 $\pm$ 17.6 & 0.0 $\pm$ 13.8 & 0.0709 & \nodata\\
Lk Ca 15 & 103 & 267.6 $\pm$ 8.7 & 86.5 $\pm$ 8.0 & 0.0 $\pm$ 4.5 & 0.0 $\pm$ 4.9 &
0.0 $\pm$ 2.9 & 0.0 $\pm$ 2.6 & 6.2 $\pm$ 2.8 & 7.4 $\pm$ 3.8 & 562 & 7.5\\
``'' & 716 & 0.182 $\pm$ 0.006 & 0.0 $\pm$ 2.5 & 88.7 $\pm$ 6.2 & 9.7 $\pm$ 3.8
& 0.0 $\pm$ 3.9 & 0.0 $\pm$ 2.0 & 1.6 $\pm$ 2.0 & 0.0 $\pm$ 1.1 & 0.451 & \nodata\\
RW Aur A & 151 & 230.2 $\pm$ 5.5 & 67.2 $\pm$ 25.0 & 0.0 $\pm$ 16.0 & 0.0 $\pm$ 19.5
& 0.0 $\pm$ 13.0 & 0.0 $\pm$ 10.8 & 18.5 $\pm$ 11.1 & 14.2 $\pm$ 14.4 & 88.3 & 3.9\\
``'' & 602 & 2.49 $\pm$ 0.05 & 4.1 $\pm$ 6.2 & 68.1 $\pm$ 12.2 & 10.9 $\pm$ 9.4
& 11.9 $\pm$ 9.0 & 3.8 $\pm$ 5.0 & 1.2 $\pm$ 4.6 & 0.0 $\pm$ 3.5 & 1.29 & \nodata\\
UY Aur & 151 & 912.9 $\pm$ 16.9 & 10.3 $\pm$ 19.5 & 0.0 $\pm$ 15.2 & 72.4 $\pm$ 25.7
& 0.0 $\pm$ 13.1 & 0.0 $\pm$ 10.1 & 4.6 $\pm$ 8.9 & 12.7 $\pm$ 11.9 & 284 & 1.4\\
``'' & 545 & 4.96 $\pm$ 0.13 & 0.0 $\pm$ 3.6 & 59.4 $\pm$ 6.3 & 25.9 $\pm$ 5.1 &
14.7 $\pm$ 5.3 & 0.0 $\pm$ 2.9 & 0.0 $\pm$ 2.6 & 0.0 $\pm$ 1.7 & 6.23 & \nodata\\
UZ Tau E & 175 & 93.0 $\pm$ 2.4 & 47.8 $\pm$ 10.4 & 16.3 $\pm$ 7.5 & 30.6 $\pm$ 9.0
& 0.0 $\pm$ 6.8 & 1.0 $\pm$ 5.3 & 1.6 $\pm$ 4.6 & 2.8 $\pm$ 6.8 & 77.3 & 5.4\\
``'' & 1400 & 0.282 $\pm$ 0.009 & 0.0 $\pm$ 3.1 & 7.0 $\pm$ 3.7 & 4.8 $\pm$ 5.3
& 85.9 $\pm$ 7.3 & 0.1 $\pm$ 2.7 & 2.3 $\pm$ 2.6 & 0.0 $\pm$ 1.6 & 0.401 & \nodata\\
V410 Anon 13 & 103 & 7.28 $\pm$ 0.65 & 54.4 $\pm$ 6.8 & 6.8 $\pm$ 5.0 & 33.0 $\pm$
5.4 & 0.0 $\pm$ 3.6 & 1.0 $\pm$ 3.3 & 4.3 $\pm$ 3.0 & 0.4 $\pm$ 5.1 & 42.5 & 5.6\\
``'' & 488 & 0.101 $\pm$ 0.002 & 46.3 $\pm$ 13.6 & 0.0 $\pm$ 10.5 & 10.4
$\pm$ 14.4 & 0.0 $\pm$ 14.0 & 24.0 $\pm$ 8.9 & 13.2 $\pm$ 7.0 & 6.1 $\pm$ 6.4 & 0.0318 & \nodata\\
V710 Tau & 139 & 35.5 $\pm$ 1.0 & 62.9 $\pm$ 23.1 & 0.0 $\pm$ 15.7 & 0.0 $\pm$ 18.1
& 0.0 $\pm$ 11.6 & 0.0 $\pm$ 10.2 & 14.7 $\pm$ 10.2 & 22.4 $\pm$ 14.8 & 16.0 & 8.7\\
``'' & 602 & 0.427 $\pm$ 0.008 & 0.0 $\pm$ 8.0 & 63.4 $\pm$ 13.8 & 0.0 $\pm$
12.4 & 14.1 $\pm$ 10.3 & 14.5 $\pm$ 6.2 & 2.4 $\pm$ 5.5 & 5.6 $\pm$ 5.3 & 0.176 & \nodata\\
V807 Tau & 127 & 117.4 $\pm$ 3.2 & 54.9 $\pm$ 7.9 & 30.5 $\pm$ 6.0 & 0.0 $\pm$ 7.0 &
0.0 $\pm$ 4.7 & 0.0 $\pm$ 3.7 & 8.6 $\pm$ 3.6 & 6.0 $\pm$ 4.5 & 137 & 3.4\\
``'' & 773 & 0.428 $\pm$ 0.008 & 0.0 $\pm$ 14.6 & 8.9 $\pm$ 16.2 & 83.4 $\pm$
32.5 & 0.0 $\pm$ 25.1 & 2.4 $\pm$ 12.3 & 5.2 $\pm$ 10.6 & 0.0 $\pm$ 8.6 & 0.0752 & \nodata\\
V836 Tau & 127 & 37.1 $\pm$ 1.4 & 58.4 $\pm$ 16.0 & 0.0 $\pm$ 12.1 & 0.0 $\pm$ 13.7
& 0.0 $\pm$ 9.2 & 15.8 $\pm$ 7.9 & 16.5 $\pm$ 7.6 & 9.3 $\pm$ 8.6 & 32.6 & 4.5\\
``'' & 602 & 0.237 $\pm$ 0.007 & 0.0 $\pm$ 3.3 & 24.0 $\pm$ 3.4 & 55.8 $\pm$ 4.8
& 3.7 $\pm$ 4.4 & 7.8 $\pm$ 2.4 & 4.3 $\pm$ 2.2 & 4.5 $\pm$ 1.8 & 0.387 & \nodata\\
V955 Tau & 115 & 81.8 $\pm$ 5.9 & 30.0 $\pm$ 8.4 & 49.4 $\pm$ 9.0 & 0.0 $\pm$ 7.8 &
0.0 $\pm$ 5.2 & 3.5 $\pm$ 4.6 & 15.4 $\pm$ 4.7 & 1.7 $\pm$ 6.5 & 257 & 2.5\\
``'' & 488 & 2.06 $\pm$ 0.03 & 12.3 $\pm$ 6.1 & 0.0 $\pm$ 6.5 & 63.0 $\pm$ 10.0
& 0.0 $\pm$ 8.6 & 11.5 $\pm$ 4.8 & 4.8 $\pm$ 4.3 & 8.4 $\pm$ 3.9 & 0.966 & \nodata\\
VY Tau & 115 & 47.7 $\pm$ 1.7 & 62.9 $\pm$ 38.7 & 0.0 $\pm$ 30.0 & 0.0 $\pm$ 31.3 &
0.0 $\pm$ 22.0 & 0.0 $\pm$ 17.1 & 37.1 $\pm$ 27.4 & 0.0 $\pm$ 15.5 & 15.3 & 7.2\\
``'' & 545 & 0.226 $\pm$ 0.006 & 22.0 $\pm$ 6.2 & 19.7 $\pm$ 6.5 & 43.8 $\pm$ 8.7
& 0.0 $\pm$ 9.1 & 5.0 $\pm$ 4.8 & 8.0 $\pm$ 4.5 & 1.5 $\pm$ 3.6 & 0.153 & \nodata\\
XZ Tau & 151 & 787.4 $\pm$ 18.1 & 75.3 $\pm$ 22.1 & 20.5 $\pm$ 12.5 & 0.0 $\pm$ 16.8
& 0.0 $\pm$ 12.2 & 0.0 $\pm$ 8.3 & 4.2 $\pm$ 7.8 & 0.0 $\pm$ 8.9 & 317 & 2.0\\
``'' & 488 & 12.1 $\pm$ 0.2 & 0.0 $\pm$ 9.6 & 13.4 $\pm$ 10.4 & 68.3 $\pm$ 17.7 &
10.4 $\pm$ 14.4 & 4.8 $\pm$ 7.7 & 3.0 $\pm$ 7.0 & 0.0 $\pm$ 5.5 & 3.54 & \nodata\\
ZZ Tau & 211 & 6.43 $\pm$ 0.15 & 0.0 $\pm$ 33.3 & 49.4 $\pm$ 34.8 & 0.0 $\pm$ 35.2 &
0.0 $\pm$ 28.2 & 0.4 $\pm$ 18.9 & 0.0 $\pm$ 15.6 & 50.1 $\pm$ 35.1 & 1.17 & 3.5\\
``'' & 1400 & 0.0106 $\pm$ 0.0007 & 8.6 $\pm$ 4.3 & 0.0 $\pm$ 5.4 & 78.9 $\pm$
10.1 & 0.0 $\pm$ 8.5 & 7.0 $\pm$ 3.8 & 1.2 $\pm$ 3.7 & 4.3 $\pm$ 2.4 & 0.0218 & \nodata\\
ZZ Tau IRS & 127 & 280.3 $\pm$ 7.5 & 84.4 $\pm$ 11.1 & 11.0 $\pm$ 5.9 & 0.0 $\pm$
7.3 & 0.0 $\pm$ 4.8 & 0.0 $\pm$ 3.7 & 4.6 $\pm$ 3.9 & 0.0 $\pm$ 4.7 & 323 & 4.9\\
``'' & 545 & 1.28 $\pm$ 0.03 & 0.0 $\pm$ 3.1 & 0.0 $\pm$ 4.1 & 0.0 $\pm$ 6.2 &
96.3 $\pm$ 8.8 & 0.0 $\pm$ 2.9 & 3.7 $\pm$ 2.8 & 0.0 $\pm$ 1.5 & 1.18 & \nodata\\
\enddata
\tablecomments{\footnotesize Col. (1): Object name.  Col. (2): 
One of two dust model temperatures (Kelvin).  Col. (3): Solid angle, 
$\Omega_{BB}$, of blackbody of temperature specified in Col. (2) 
representing continuum emission, expressed in units of $10^{-16}$ 
steradians.  Cols. (4)-(10): mass percentages and their propagated 
uncertainties of all dust masses at 
temperature specified in Col. (2).  One dust model is completely 
specified by two adjacent rows - the row following the object$'$s 
name and the row beneath that one.  Propagated uncertainties are 
the square root of the sum of the squares of terms, such that each 
term is the product of the uncertainty in one of seven mass weights 
and the partial derivative of the mass fraction of the dust species 
in question with respect to the same mass weight.  This way, even 
if a species has zero mass, its 
uncertainty still contributes to the propagated uncertainty in 
question.  Col. (11): Total dust mass at
one temperature in $10^{-4}$ lunar masses, computed assuming 140pc 
to Taurus-Auriga.  Col. (12): $\chi^{2}$ per degree of freedom, 
determined over 7.7\,$<$\,$\lambda$\,$<$\,37$\mum$.}
\tablenotetext{a}{\footnotesize Optical constants for amorphous 
pyroxene of cosmic composition from \citet{jag94}, assuming CDE2 
\citep{fab01}}
\tablenotetext{b}{\footnotesize Optical constants for amorphous 
olivine MgFeSiO$_4$ from \citet{dor95}, assuming CDE2}
\tablenotetext{c}{\footnotesize Optical constants for amorphous 
pyroxene of cosmic composition from \citet{jag94}, using the 
Bruggeman EMT and Mie theory (Bohren \& Huffman 1983) with a 
volume fraction of vacuum of $f$\,=\,0.6 for porous spherical 
grains of radius 5$\mum$}
\tablenotetext{d}{\footnotesize Optical constants for amorphous 
olivine MgFeSiO$_4$ from \citet{dor95}, using the Bruggeman EMT 
and Mie theory (Bohren \& Huffman 1983) with a volume fraction of 
vacuum of $f$\,=\,0.6 for porous spherical grains of radius 5$\mum$}
\tablenotetext{e}{\footnotesize Opacities for clinoenstatite 
Mg$_{0.9}$Fe$_{0.1}$SiO$_3$ from \citet{chi02}}
\tablenotetext{f}{\footnotesize Optical constants for 3 
crystallographic axes of forsterite, Mg$_{2}$SiO$_4$, from 
\citet{sog06}, assuming tCDE (see discussion in Section 4.2.3).}
\tablenotetext{g}{\footnotesize Opacity for annealed silica by 
\citet{fab00}.}
\end{deluxetable}
\end{landscape}

\clearpage

\begin{table}[h,t]
{\tiny
\tablenum{A1}
\caption[]{Mispointing Corrections \label{tablea1}}
\begin{tabular}{lcccccccccc}
\hline \hline
          
          & 
          & 
          & 
          & 
          & 
          & LL2nod1 or
          & LL2nod2 or
          & LL1nod1 or
          & LL1nod2 or\\
          Object 
          & modules
          & SL2nod1 
          & SL2nod2 
          & SL1nod1 
          & SL1nod2 
          & SHnod1
          & SHnod2 
          & LHnod1 
          & LHnod2\\
          (1) 
          & (2) 
          & (3) 
          & (4)
          & (5) 
          & (6) 
          & (7) 
          & (8) 
          & (9)
          & (10)\\
\hline
AA Tau & SLLL & 1.00 & 1.00 & 1.00 & 1.00 & 1.02 & 1.00 & 1.02 & 1.04\\
BP Tau & SLLL & 1.01 & 1.00 & 1.05 & 1.00 & 1.00 & 1.01 & 1.00 & 1.01\\
CI Tau & SLLL & 1.00 & 1.00 & 1.00 & 1.01 & 1.00 & 1.03 & 1.00 & 1.01\\
CoKu Tau/3 & SLLL & 1.01 & 1.00 & 1.00 & 1.00 & 0.96 & 1.00 & 1.00 & 1.00\\
CoKu Tau/4 & SLLL & 1.01 & 1.00 & 1.00 & 1.01 & 1.00 & 1.00 & 1.00 & 1.01\\
CW Tau & SLSHLH & 1.13 & 1.14 & 1.10 & 1.09 & 1.25 & 1.00 & 1.00 & 1.00\\
CX Tau & SLLL & 1.12 & 1.00 & 1.24 & 1.20 & 1.21 & 1.20 & 1.20 & 1.21\\
CY Tau & SLLL & 1.01 & 1.00 & 1.00 & 1.01 & 1.01 & 1.00 & 1.00 & 1.00\\
DD Tau & SLLL & 1.00 & 1.00 & 1.01 & 1.00 & 1.00 & 1.01 & 1.00 & 1.00\\
DE Tau & SLLL & 1.00 & 1.00 & 1.00 & 1.00 & 1.00 & 1.00 & 1.00 & 1.01\\
DF Tau & SLSHLH & 1.00 & 1.00 & 1.00 & 1.00 & 1.17 & 0.96 & 1.02 & 1.00\\
DG Tau & SLSHLH & 1.00 & 1.00 & 1.05 & 1.00 & 1.12 & 1.00 & 1.05 & 1.06\\
DH Tau & SLLL & 1.00 & 1.00 & 1.00 & 1.00 & 1.00 & 1.01 & 1.00 & 1.03\\
DK Tau & SLSHLH & 1.01 & 1.00 & 1.00 & 1.01 & 1.00 & 1.01 & 1.00 & 1.00\\
DL Tau & SLSHLH & 1.01 & 1.00 & 1.00 & 1.00 & 1.00 & 1.04 & 1.00 & 1.01\\
DM Tau & SLLL & 1.15 & 1.15 & 1.15 & 1.15 & 1.00 & 1.00 & 1.00 & 1.01\\
DN Tau & SLLL & 1.00 & 1.00 & 1.00 & 1.00 & 1.02 & 1.00 & 1.00 & 1.01\\
DO Tau & SLSHLH & 1.01 & 1.00 & 1.00 & 1.00 & 1.00 & 1.07 & 1.05 & 1.06\\
DP Tau & SLLL & 1.00 & 1.00 & 1.01 & 1.00 & 1.01 & 1.00 & 1.00 & 1.01\\
DQ Tau & SLLL & 1.05 & 1.05 & 1.05 & 1.05 & 1.00 & 1.00 & 1.00 & 1.00\\
DR Tau & SLSHLH & 1.00 & 1.00 & 1.00 & 1.00 & 1.05 & 0.95 & 1.01 & 1.00\\
DS Tau & SLLL & 1.00 & 1.00 & 1.00 & 1.00 & 1.00 & 1.00 & 1.00 & 1.00\\
F04147+2822 & SLLL & 1.00 & 1.00 & 1.00 & 1.01 & 1.00 & 1.01 & 1.01 & 1.00\\
FM Tau & SLLL & 1.02 & 1.00 & 1.00 & 1.00 & 1.00 & 1.00 & 1.00 & 1.00\\
FN Tau & SLLL & 1.02 & 1.00 & 1.10 & 1.00 & 1.10 & 1.10 & 1.10 & 1.10\\
FO Tau & SLLL & 1.03 & 1.00 & 1.15 & 0.92 & 1.00 & 1.01 & 1.00 & 1.00\\
FP Tau & SLLL & 1.01 & 1.00 & 1.00 & 1.01 & 1.00 & 1.00 & 1.00 & 1.00\\
FQ Tau & SLLL & 1.00 & 1.01 & 1.00 & 1.00 & 1.01 & 1.00 & 1.00 & 1.00\\
FS Tau & SLSHLH & 1.00 & 1.00 & 1.02 & 1.00 & 1.09 & 0.93 & 1.00 & 1.03\\
FT Tau & SLLL & 1.00 & 1.00 & 1.00 & 1.00 & 1.00 & 1.00 & 1.00 & 1.00\\
FV Tau & SLSHLH & 1.01 & 1.00 & 1.04 & 1.00 & 1.12 & 0.97 & 1.00 & 1.00\\
FX Tau & SLLL & 1.01 & 1.00 & 1.02 & 1.00 & 1.00 & 1.00 & 1.05 & 1.06\\
FZ Tau & SLSHLH & 1.01 & 1.00 & 1.03 & 1.00 & 1.19 & 0.97 & 1.01 & 1.00\\
GG Tau A & SLLL & 1.04 & 1.03 & 1.04 & 1.03 & 1.00 & 1.00 & 1.00 & 1.00\\
GG Tau B & SLLL & 1.10 & 1.15 & 1.10 & 1.15 & 1.02 & 1.00 & 1.01 & 1.00\\
GH Tau & SLLL & 1.00 & 1.00 & 1.00 & 1.01 & 1.00 & 1.01 & 1.00 & 1.01\\
GI Tau & SLLL & 1.02 & 1.00 & 1.04 & 1.00 & 1.00 & 1.00 & 1.00 & 1.01\\
GK Tau & SLSHLH & 1.02 & 1.00 & 1.03 & 1.00 & 1.06 & 1.05 & 1.05 & 1.06\\
GM Aur & SLLL & 1.00 & 1.00 & 1.00 & 1.02 & 1.00 & 1.01 & 1.00 & 1.00\\
GN Tau & SLLL & 1.01 & 1.00 & 1.00 & 1.00 & 1.00 & 1.00 & 1.00 & 1.00\\
GO Tau & SLLL & 1.03 & 1.00 & 1.00 & 1.00 & 1.00 & 1.01 & 1.00 & 1.01\\
Haro 6-28 & SLSHLH & 1.03 & 1.00 & 1.00 & 1.00 & 1.00 & 1.00 & 1.00 & 1.00\\
Haro 6-37 & SLSHLH & 1.00 & 1.00 & 1.03 & 1.00 & 1.06 & 1.00 & 1.00 & 1.00\\
HK Tau & SLLL & 1.00 & 1.00 & 1.00 & 1.01 & 1.00 & 1.01 & 1.00 & 1.01\\
HN Tau & SLSHLH & 1.01 & 1.00 & 1.00 & 1.00 & 1.00 & 1.00 & 1.00 & 1.00\\
HO Tau & SLLL & 1.01 & 1.00 & 1.00 & 1.00 & 1.06 & 1.00 & 1.02 & 1.00\\
HP Tau & SLSHLH & 1.02 & 1.00 & 1.00 & 1.01 & 1.07 & 1.13 & 1.06 & 1.10\\
HQ Tau & SLSHLH & 1.01 & 1.00 & 1.00 & 1.01 & 1.00 & 1.11 & 1.03 & 1.03\\
IP Tau & SLLL & 1.01 & 1.00 & 1.01 & 1.00 & 1.00 & 1.00 & 1.00 & 1.01\\
IQ Tau & SLLL & 1.01 & 1.00 & 1.00 & 1.00 & 1.02 & 1.04 & 1.02 & 1.04\\
IS Tau & SLLL & 1.00 & 1.02 & 1.00 & 1.02 & 1.00 & 1.00 & 1.00 & 1.00\\
IT Tau & SLLL & 1.02 & 1.00 & 1.00 & 1.04 & 1.00 & 1.01 & 1.00 & 1.01\\
LkCa 15 & SLLL & 1.01 & 1.00 & 1.00 & 1.02 & 1.00 & 1.02 & 1.00 & 1.00\\
RW Aur A & SLSHLH & 1.00 & 1.00 & 1.00 & 1.00 & 1.01 & 1.00 & 1.01 & 1.00\\
UY Aur & SLSHLH & 1.00 & 1.00 & 1.00 & 1.00 & 1.11 & 1.01 & 1.08 & 1.07\\
UZ Tau E & SLSHLH & 1.03 & 1.00 & 1.00 & 1.02 & 1.37 & 0.87 & 1.10 & 1.11\\
V410 Anon 13 & SLLL & 1.02 & 1.00 & 1.00 & 1.02 & 1.08 & 1.00 & 1.05 & 1.00\\
V710 Tau & SLLL & 1.02 & 1.00 & 1.12 & 1.07 & 1.04 & 1.00 & 1.05 & 1.06\\
V807 Tau & SLLL & 1.01 & 1.00 & 1.00 & 1.01 & 1.01 & 1.00 & 1.00 & 1.00\\
V836 Tau & SLLL & 1.00 & 1.00 & 1.00 & 1.01 & 1.00 & 1.00 & 1.00 & 1.00\\
V955 Tau & SLLL & 1.00 & 1.00 & 1.00 & 1.01 & 1.00 & 1.00 & 1.00 & 1.01\\
VY Tau & SLLL & 1.00 & 1.00 & 1.01 & 1.00 & 1.01 & 1.00 & 1.00 & 1.01\\
XZ Tau & SLSHLH & 1.00 & 1.01 & 1.01 & 1.00 & 1.00 & 1.01 & 1.00 & 1.02\\
ZZ Tau$^a$ & SLLL & \nodata & \nodata & \nodata & \nodata & \nodata & \nodata &
\nodata & \nodata\\
ZZ Tau IRS & SLLL & 1.00 & 1.00 & 1.03 & 1.00 & 1.00 & 1.00 & 1.00 & 1.00\\
\hline
\end{tabular}
\tablecomments{\footnotesize Col. (1): Object name.  Cols. (2)-(9): Multiplicative 
scalars applied to the spectrum of one order of one
nod for a given object to match the flux density of the other nod as described in
the text.  SL2 means Short-Low order 2, SL1 is Short-Low order 1, LL2 is Long-Low
order 2, LL1 is Long-Low order 1.}
\tablenotetext{a}{\footnotesize For ZZ Tau, there were 6 DCEs per spectral order.  
Each DCE of each order was normalized to the DCE with the highest average flux 
density over that order, except for Long-Low Bonus order.  In that case, the DCEs 
were normalized to the average flux level of the DCE with the second-highest flux 
density level \citep[see discussion by][]{sarg08}.}
}
\end{table}

\end{document}